\documentclass[11pt,letterpaper]{article}

\usepackage[margin=1in]{geometry}
\usepackage[T1]{fontenc}
\usepackage[utf8]{inputenc}
\usepackage{lmodern}
\usepackage{microtype}
\usepackage{amsmath,amssymb}
\usepackage{graphicx}
\usepackage{booktabs}
\usepackage{array}
\usepackage{enumitem}
\usepackage{xcolor}
\usepackage{authblk}
\usepackage{siunitx}
\DeclareSIUnit{\nothing}{\relax}                          % \SI{8.0}{\mega\nothing} -> 8.0 M
\DeclareSIUnit[number-unit-product={}]{\percent}{\%}       % 26.4% not 26.4 %
\usepackage[colorlinks,linkcolor=blue!60!black,citecolor=blue!60!black,urlcolor=blue!60!black]{hyperref}
\usepackage[capitalise,noabbrev]{cleveref}

\definecolor{todoblue}{RGB}{0,90,200}
\newif\ifnotes
\notesfalse                                      % \notesfalse hides every WH / PS / to-PS note (reader copies)

\newif\ifprovenance
\provenancetrue
\newcommand{\provtag}[1]{\ifprovenance\textsuperscript{\tiny\color{gray}[#1]}\fi}
\newcommand{\meashw}{\provtag{meas-hw}}   % measured after place-and-route
\newcommand{\meassim}{\provtag{meas-sim}} % measured in RTL simulation
\newcommand{\measbd}{\provtag{meas-board}} % measured on the board
\newcommand{\csynth}{\provtag{csynth}}    % HLS C-synthesis estimate, not place-and-route
\newcommand{\simu}{\provtag{sim}}         % imaging / detector simulator
\newcommand{\phys}{\provtag{phys}}        % physics (atom-loss) model
\newcommand{\lit}{\provtag{lit}}          % literature value
\newcommand{\roof}{\provtag{roof}}        % roofline / projection
\newcommand{\algo}{\provtag{algo}}        % algorithmic model or fit

\newcommand{\stImpl}{implemented}
\newcommand{\stSpec}{specified}
\newcommand{\stOpen}{open}

\newcolumntype{L}[1]{>{\raggedright\arraybackslash}p{#1}}

\begin{document}

\title{Quantum Data Processing Unit for Neutral-Atom Quantum Computers}
\author[1]{Pranav Srinivas Murali}
\author[1,2]{Russell Marroquin Solares}
\author[1]{Warren Zhou}
\author[1]{Wanda Hou}
\author[1]{Anil Bilgin}
\author[1,2]{Yi-Zhuang You}

\affil[1]{EdenCode Inc., San Jose, CA 95134, USA}
\affil[2]{Department of Physics, University of California, San Diego, CA 92093, USA}
\date{\today}

\maketitle

\begin{abstract}
Quantum computers need real-time classical processing to convert measurement
data into feedback within device-specific timing constraints. We introduce
the quantum data processing unit (QDPU), an architecture between a quantum
device and its controller. Its core is a machine-learned quantum data model
(QDM), adapted to the measurement response and noise characteristics of the device.
We implement an FPGA pipeline for neutral-atom array assembly that performs
occupancy detection and rearrangement scheduling in FPGA logic. One calibrated
matched-filter lane classifies a $50\times50$ array in \SI{12.5}{\micro\second}.
An atom-loss model using a software reference planner evaluates a $35\times35$
target with a \SI{60}{s} trap lifetime and \SI{0.1}{\percent} loss per moved atom.
Removing an assumed \SI{150}{ms} host round trip raises the predicted assembly
success from \SI{8}{\percent} to \SI{85}{\percent} and reduces the wall-clock
time per successful assembly from \SI{19}{s} to \SI{0.16}{s}.
A GPU-based outer loop is proposed for refitting model parameters between runs.
Its NVQLink transport is measured on the board: over \SI{10}{GbE} to an NVIDIA
DGX Spark, the median FPGA-to-host-to-FPGA round trip is \SI{2.74}{\micro\second}
for 128-byte messages forwarded by a CPU core. A separate 12-bit learned qLDPC
kernel is placed and routed for the same FPGA target, with a derived latency
of \SI{0.84}{ms} per round for a $[[1122,148]]$ lifted-product code under
code-capacity noise. A quantised GraphMP ranking kernel for the same code runs
on the board with a median host-to-FPGA-to-host latency of
\SI{50.5}{\micro\second} with one shot in flight, excluding host-side BP+OSD
post-processing. Together, these implementations illustrate a shared
architecture with workload-specific models and timing requirements.
\end{abstract}

\clearpage
\setcounter{tocdepth}{2}\tableofcontents
\clearpage

\section{Introduction}\label{sec:intro}

Quantum computation requires a classical processing layer that converts quantum
measurement data into control decisions within the timing constraints of the
device. Between a measurement and the next operation, the measurement record
must be acquired, interpreted, and translated into an action that the
controller can execute. The data may consist of fluorescence images, qubit
readout signals, or parity measurement outcomes, while the resulting decisions
may specify conditional operations or error corrections.
These tasks differ in their computational structure and timing requirements,
but share a common architectural role: processing quantum measurement data in
real time to close the feedback loop between the quantum device and its
controller.

We introduce the \emph{quantum data processing unit} (QDPU) as a general
architecture for this layer. The QDPU separates the computational task of
interpreting measurement data from the interfaces through which measurements
are acquired and control decisions are delivered. Its core computational
capability is provided by \emph{quantum data models} (QDMs): machine-learned
models trained on quantum measurement data to produce estimates and decisions
used in subsequent device operations. A QDM can be trained and refit to the
measurement response and noise characteristics of the device it serves, while
the QDPU provides the real-time execution environment and interfaces required
to use that model in a feedback loop. This distinction allows the processing
architecture to be specified and evaluated across workloads, while the models
remain adapted to the task and the device. In the proposed feedback path,
measurement inference and rearrangement scheduling run on the QDPU within
the experiment's latency budget.

We perform board-level tests on an AMD Zynq UltraScale+ RFSoC
(\texttt{xczu49dr}, ZCU216 board), connected to an NVIDIA DGX Spark
through NVQLink over \SI{10}{GbE}. These tests measure FPGA--host
round-trip latency and the performance of a quantised GraphMP ranking
kernel running on the FPGA (\cref{sec:platform,sec:workloads:decoding}).

Measurement processing is needed in both noisy intermediate-scale
quantum (NISQ) systems~\cite{preskill2018nisq,bharti2022nisq} and
fault-tolerant processors. In NISQ experiments, repeated executions produce
measurement records that must be interpreted under changing device conditions.
When these results determine subsequent operations, processing latency also
affects the experimental cycle time. Fault-tolerant operation adds sustained
syndrome processing and decoding to these tasks.

Quantum measurement data also impose requirements beyond low latency alone.
Measurement records are intrinsically noisy, and the information available for
inference depends on the time allocated to acquiring them. In neutral-atom
systems, for example, short fluorescence exposures can yield noisy occupancy
measurements.
Measurements implemented through imperfect gates introduce a further
complication: their records contain errors arising from the measurement
circuit as well as the errors they are intended to reveal. Extracting useful
information from these records under a fixed latency constraint is therefore
an inference problem. Learned methods have demonstrated increasingly strong
performance on quantum measurement tasks involving device
data~\cite{bausch2024alphaqubit,phuttitarn2024ml}, motivating QDMs as a central
capability of the QDPU. Their advantage nevertheless depends on the task and
operating conditions; for simpler inference tasks, the benefit over
conventional methods may be limited or conditional, as discussed in
\cref{sec:unit:why}.

\begin{figure}[htbp]
  \centering
  \includegraphics[width=\textwidth]{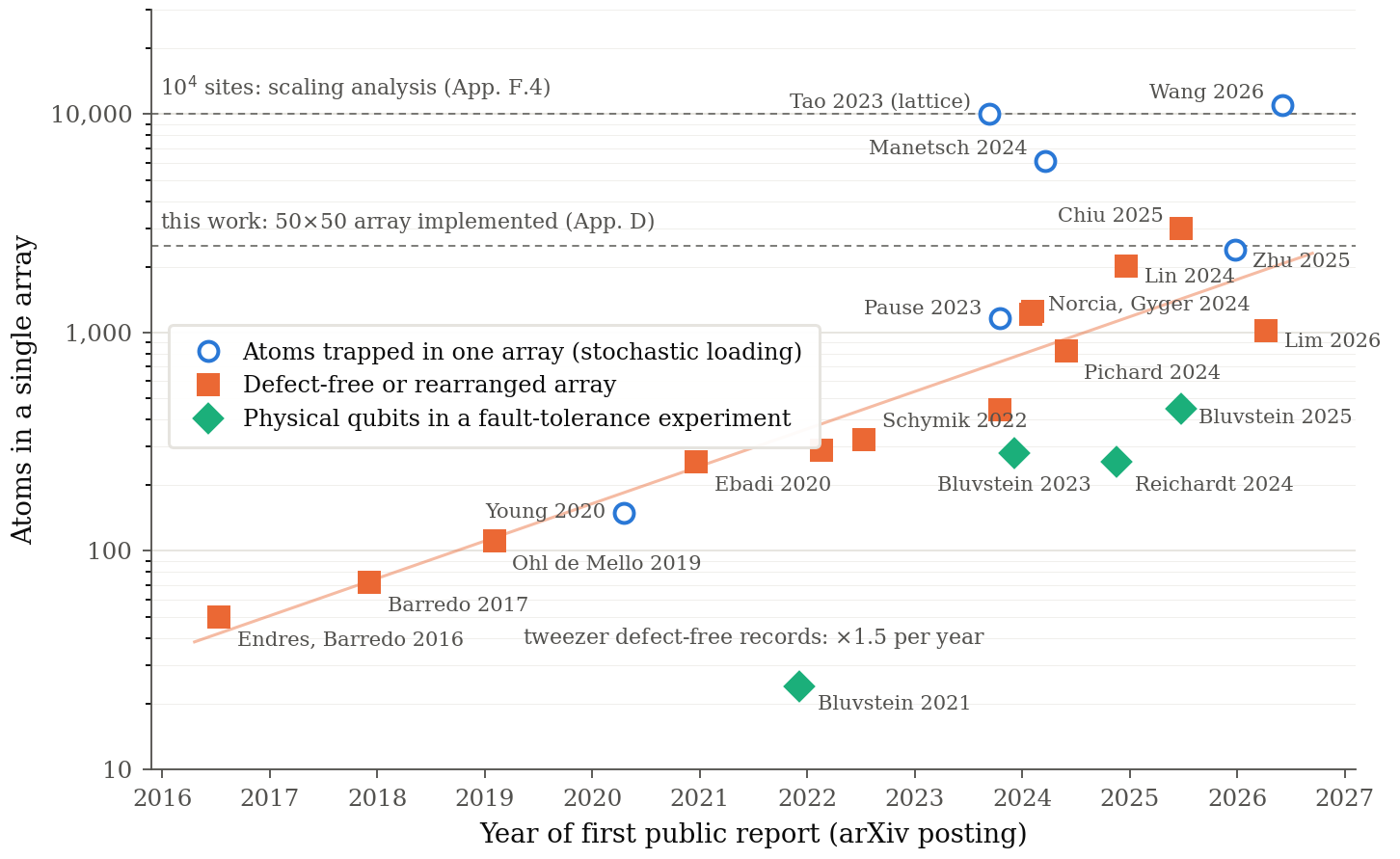}
  \caption{Growth of neutral-atom arrays, 2016 to 2026. Each point is a
  published record, placed at the date of its first public report. Open
  circles: atoms trapped in one array under stochastic loading, without
  rearrangement~\cite{young2020clock,tao2024detection,pause2024supercharged,manetsch2025tweezer6100,zhu2026ytterbium,wang2026metasurface11000}.
  Filled squares: defect-free or rearranged
  arrays~\cite{endres2016assembly,barredo2016assembler,barredo2018threed,ohldemello2019cluster,ebadi2021simulator,scholl2021antiferro,ebadi2022mis,schymik2022cryo,pause2024supercharged,norcia2024iterative,gyger2024continuous,pichard2024rearrangement2000,lin2025aiassembly,chiu2025continuous,lim2026defectfree}.
  Diamonds: physical qubits in a fault-tolerance or logical-qubit
  experiment~\cite{bluvstein2022transport,bluvstein2024logical,reichardt2024faulttolerant,bluvstein2026universal}.
  The line is a fit to the tweezer defect-free records (optical-lattice results
  excluded), a factor of 1.5 per year; the community strategic plan reports
  about 1.8 per year for physical qubit number~\cite{menssen2026strategic}.
  Dashed lines mark the $50\times50$ array implemented in this work and the
  $10^4$-site scale of the scaling analysis in \cref{app:platform:scaling}.}
  \label{fig:growth}
\end{figure}

Neutral-atom arrays provide a concrete setting in which to develop and
evaluate this architecture. Their scale has increased by approximately a
factor of 1.5 per year over the past decade~\cite{saffman2026naqc}, as shown in
\cref{fig:growth}. The size of defect-free arrays prepared through atom
rearrangement has grown from approximately 50 atoms in
2016~\cite{endres2016assembly,barredo2016assembler} to 2,024
atoms~\cite{lin2025aiassembly}; arrays containing more than 3,000 atoms have
also been maintained continuously for hours~\cite{chiu2025continuous} in 2025.
More recently, experiments have reported trapping
6,100~\cite{manetsch2025tweezer6100} and
11,000~\cite{wang2026metasurface11000} atoms. The community strategic plan
identifies the ``10,000 to 100,000 qubit level'' as the next stage of
development~\cite{menssen2026strategic}.

Larger arrays increase both the volume of imaging data and the work required
for occupancy detection and rearrangement scheduling. The available processing
time depends on the atom lifetime, which is affected by vacuum conditions,
trap depth, and imaging light. Reported lifetimes range from minutes in
room-temperature systems~\cite{manetsch2025tweezer6100} to hours in cryogenic
systems~\cite{schymik2021cryo,lim2026defectfree}. Under a fixed total
laser-power budget, increasing the number of traps reduces the power per trap
and hence the trap depth, which can further limit atom
lifetime~\cite{menssen2026strategic}. Scaling the array therefore requires
more measurement data and control decisions to be processed within the time
available for preparation and feedback.

The number of trapped atoms should also be distinguished from the number of
physical qubits used in fault-tolerance or logical-qubit experiments. The
latter remains at the scale of a few
hundred~\cite{bluvstein2024logical,reichardt2024faulttolerant,bluvstein2026universal}.
In \cref{fig:growth}, these counts differ from the largest trapped arrays by
more than an order of magnitude within a given year. Trapping capacity alone
does not establish the scale at which atoms can be operated as computational
qubits. This work addresses the classical processing needed for array
preparation, evaluating occupancy detection and rearrangement scheduling
through their latency and their modelled effect on atom retention.

Existing work on quantum control, interconnects, and real-time decoding
provides important architectural foundations for this processing layer,
particularly for superconducting processors. These devices operate with
measurement cycles of approximately one
microsecond~\cite{acharya2025willow} and are expected to generate measurement
data at terabits per second at utility
scale~\cite{battistel2023realtime}. NVIDIA's NVQLink proposes a vendor-neutral
interconnect between the control system and an HPC
resource~\cite{caldwell2025nvqlink}. Complementary architectural descriptions
include a quantum microarchitecture~\cite{fu2017quma}, a decoder
coprocessor~\cite{battistel2023realtime}, and a six-layer QEC
stack~\cite{chen2026qecstack}. These approaches
provide terminology, interfaces, and implementations for different aspects
of the control and processing stack. In a decoder-oriented specification,
for example, the processing interface accepts a stream of binary measurement
outcomes at a fixed cycle rate and returns corrections to the controller.
Such a specification captures the requirements of a particular workload;
a general processing architecture must also accommodate measurement and
control tasks with different data representations and execution schedules.

The neutral-atom assembly loop illustrates these additional requirements.
Its input is a fluorescence image, and image acquisition consumes part of
the physical time budget: atoms are lost at the vacuum-limited rate during
the exposure, with a smaller additional contribution from the imaging light
under the model in \cref{app:loop:model}. Inference estimates occupancy,
position, and loss from the detected photons, and the resulting estimates
are used to construct a move schedule. Rather than following the fixed
measurement-cycle rate of the superconducting decoding example, the assembly
cycle depends on the time allocated to imaging, inference, and rearrangement.
The allowed inference latency is therefore part of the operating-cycle design.
Despite these differences, image-based array assembly and syndrome decoding
occupy the same architectural position: both interpret measurement records
within a device-imposed latency budget and return decisions that the
controller translates into operations. This shared position motivates a
processing-unit abstraction above any individual algorithm or its
workload-specific input and output formats.

Real-time feedback without host software in the critical path has already
been demonstrated for a hand-designed algorithm on a superconducting device.
A clustering decoder implemented on an FPGA within the control system of an
eight-qubit experiment enabled a conditional gate
\SI{9.6}{\micro\second} after the last of nine measurement
rounds~\cite{caune2026realtime}. This result establishes an important precedent
for integrating real-time classical inference with quantum control. It does
not, however, demonstrate a learned QDM in that loop or address the
image-based measurement data produced by a neutral-atom array. These
requirements motivate an implementation that combines device-adapted
inference with the measurement and control interfaces needed for array
assembly, while retaining an architectural description applicable to other
workloads.

This work develops that description and realizes a neutral-atom QDPU on a
single FPGA. The array-assembly pipeline combines occupancy detection
with rearrangement scheduling, and its performance is evaluated in terms of
both processing latency and modelled atom retention. To examine a
distinct inference workload within the same framework, we also consider the
learned qLDPC decoder developed in the companion
paper~\cite{companion2026graphmp}. Sampling workloads for quantum machine
learning identify further tasks that the unit is specified to support;
these prospective capabilities are distinguished from the array-assembly
pipeline implemented here. The implementation thus serves as a concrete
instance of the broader architecture, rather than defining that architecture
through a single algorithm or qubit modality.

\Cref{sec:requirements} derives the processing requirements from physical
timing and measurement constraints; \cref{sec:unit} develops the QDPU
architecture and its three interfaces; \cref{sec:platform} presents the
single-FPGA implementation for neutral atoms; and \cref{sec:workloads}
examines array assembly, learned qLDPC decoding, and the specified sampling
workloads for quantum machine learning. \Cref{sec:limits} states the
limitations of the present implementation and the remaining work, while
\crefrange{app:frontend}{app:open} report the implementation in full.

\section{Requirements for the Classical Layer}\label{sec:requirements}

The classical layer processes quantum measurement records and uses the results
to guide subsequent operations. Its design must account for measurement noise,
the time available for feedback, and the rate at which measurement data are
generated. These considerations lead to the following requirements.

\paragraph{R1. Noise-aware and adaptive processing.}
In quantum experiments, measurement records are affected by noise in both the
quantum device and the readout process. In neutral-atom systems, for example,
fluorescence images are subject to photon-counting fluctuations and background
light. In quantum error correction, finite gate fidelities and decoherence
introduce errors that must be detected through repeated syndrome extraction.
The extraction circuits are themselves imperfect: faulty gates can introduce
additional data-qubit errors, while readout errors can corrupt the measured
syndrome~\cite{terhal2015qec}.

The statistics of these errors can also change during operation. In
neutral-atom systems, changes in optical alignment, trap conditions, and
background light affect fluorescence measurements. In error-corrected
processors, changes in gate and readout performance alter the noise
distribution relevant to decoding. The classical layer must therefore
adaptively process noisy measurement records, with estimates or model
parameters updated as device conditions change. Learned models provide one
approach to this task, with applications to fluorescence
classification~\cite{phuttitarn2024ml} and quantum error
correction~\cite{bausch2024alphaqubit,varbanov2025nnd}.

\paragraph{R2. Feedback latency.}
Classical processing forms part of the experimental feedback loop.
An occupancy estimate determines which atoms are moved during rearrangement,
and a mid-circuit measurement can determine a subsequent
pulse~\cite{bluvstein2024logical,huie2023repetitive,norcia2023midcircuit}.
In quantum error correction, decoding results determine updates to the
Pauli frame and may be required before a measurement-dependent operation
can proceed~\cite{terhal2015qec,gidney2019autoccz}.

These dependencies constrain the time available for classical processing.
In neutral-atom preparation, longer processing times extend the preparation
cycle and increase the probability of atom loss (\cref{app:loop}).
During quantum computation, delays in conditional operations extend the
time over which quantum information must be preserved. When such delays
determine the rate of computation, execution is
reaction-limited~\cite{gidney2019autoccz}.

The relevant feedback latency includes measurement acquisition, data
transfer, classical processing, and delivery of the resulting control
instruction. Reported latencies may include different subsets of these
stages, so a latency specification must identify both the available time
and the stages included in its measurement.

\paragraph{R3. Keeping pace with measurement.}
The classical layer must process measurement data at least as fast as they
are generated during sustained operation. Otherwise, unprocessed records
accumulate and delay subsequent decisions. In repeated syndrome extraction,
the required processing rate scales with the number of measured qubits and
the extraction frequency~\cite{holmes2020nisqplus,skoric2023parallelwindow}.
Estimates for utility-scale superconducting systems reach tens of megabits
per second per logical qubit and multiple terabits per second in
aggregate~\cite{battistel2023realtime}. This sustained processing requirement
is distinct from the feedback latency of an individual result.

Imaging-based readout also produces substantial data volumes. A single
$2048\times2048$ sixteen-bit image contains $6.7\times10^{7}$ bits, whereas
the occupancy map needed for rearrangement requires only one bit per atom
site (\cref{app:platform}). Extracting this information close to the acquisition
hardware reduces the amount of data transferred to subsequent control
stages. The cost of data transfer is evident in a continuously operated
neutral-atom array, where image transfer and rearrangement limit an
\SI{80}{ms} reloading cycle, and increasing the camera region adds
\SIrange{10}{20}{ms}~\cite{chiu2025continuous}.

These requirements concern inference from noisy and changing measurement
records, feedback latency, and sustained processing capacity. They must be
considered across the path from measurement acquisition to control.

\section{The QDPU}\label{sec:unit}

\subsection{Definition}\label{sec:unit:definition}

A \emph{quantum data processing unit} (QDPU) is a real-time processing unit
between a quantum device and its controller. It interprets quantum measurement
data and produces feedback decisions within the latency budget of the
experiment. Its learned inference model is the \emph{quantum data model}
(QDM), trained on measurement data and refitted as device conditions change
(\cref{sec:unit:loops}). The CNN occupancy detector of
\cref{sec:workloads:assembly} and the decoder of the companion
paper~\cite{companion2026graphmp} are the two QDMs considered here. The QDPU
also includes the real-time blocks that supply their inputs and use their
outputs, such as the rearrangement engine.

The QDPU is defined by its processing role and interfaces. Its implementation
may use an FPGA, an FPGA coupled to a GPU, or an application-specific integrated
circuit (ASIC), depending on the latency and throughput requirements of the
workload (\cref{sec:workloads:decoding}). The implementation considered here
uses a single FPGA (\cref{sec:platform}).

Related functions appear in existing quantum-control architectures: a
measurement discrimination unit within a quantum
microarchitecture~\cite{fu2017quma}, a decoder
coprocessor~\cite{battistel2023realtime}, and a separation between a decoder
engine and a feedforward controller~\cite{chen2026qecstack}. The
QDPU abstraction groups measurement interpretation and feedback computation
within a common processing stage, while allowing the data representations,
models, and timing requirements to vary by workload.

Comparable processing stages occur in classical systems. Solid-state drive
controllers interpret noisy NAND-flash measurements and manage device-dependent
changes~\cite{cai2017flash}. Coherent optical receivers use digital signal
processing to compensate for channel impairments and recover transmitted
data~\cite{savory2010coherent,zhao2019coherentdsp}. These examples illustrate
the separation of a physical device from the processing needed to interpret
its output (\cref{tab:analogy}).

\begin{table}[htbp]
  \centering
  \caption{Comparison of the proposed QDPU with two classical processing
  systems. Each combines interpretation of noisy device output with
  device-dependent adaptation and additional control or management tasks.}
  \label{tab:analogy}
  \small
  \begin{tabular}{@{}L{2.6cm}L{4.1cm}L{4.1cm}L{4.1cm}@{}}
    \toprule
     & \textbf{QDPU (this work)} & \textbf{Solid-state drive controller}~\cite{cai2017flash} & \textbf{Coherent optical receiver DSP}~\cite{savory2010coherent,zhao2019coherentdsp} \\
    \midrule
    Medium & Atom array, read out by fluorescence imaging & NAND flash cell array & Optical fibre span \\[2pt]
    Measurement impairments & Photon-counting fluctuations, background light, and atom loss during acquisition & Overlapping charge-level distributions and changes with wear, retention time, and temperature & Dispersion, polarisation rotation, laser phase noise, and amplifier noise \\[2pt]
    Output & Site occupancy, a move schedule, gate-level feedback & Corrected user data & Corrected client bits \\[2pt]
    Jobs besides error correction & Site detection, array assembly, state discrimination; later calibration and mid-circuit control & Read-voltage optimisation, wear levelling, bad-block management, data placement & Timing recovery, dispersion compensation, polarisation demultiplexing, carrier phase and frequency recovery \\[2pt]
    Adapts to the individual device & Model retrained as optics, background and alignment drift (\cref{sec:unit:loops}) & Read reference voltages re-optimised per block as cells wear & Equalisers and phase estimators adapt continuously \\
    \bottomrule
  \end{tabular}
\end{table}

In these classical systems, error correction is one function within a broader
processing stage that also performs detection, compensation, or calibration.
This organisation provides a useful analogy for the QDPU. For neutral-atom
feedback, the processing latency has an additional consequence: it extends
the interval during which atoms can be lost before the next operation
(\cref{app:loop}).

The controller generates waveforms and determines pulse timing; the QDPU
supplies the decisions that condition those operations. The processing stage
can support state discrimination, array assembly, and decoding, with calibration
and mid-circuit control as further applications. Training and model updates
belong to the outer loop, while inference runs in the inner loop under the
workload's latency and throughput constraints.

\subsection{Two loops}\label{sec:unit:loops}

\begin{figure}[htbp]
  \centering
  \includegraphics[width=0.55\textwidth]{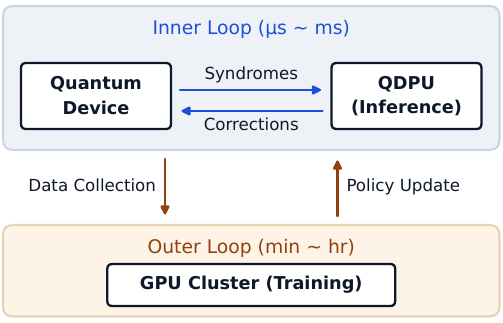}
  \caption{Proposed inner and outer loops of a QDPU. The inner loop processes
  measurement records and returns feedback decisions within the time available
  for subsequent operations. The outer loop collects measurement data and
  labels and supplies updated model parameters as device conditions change.}
  \label{fig:loops}
\end{figure}

The requirements in \cref{sec:requirements} distinguish the execution of
inference from adaptation to changing device conditions. In the proposed
architecture (\cref{fig:loops}), a trained QDM processes measurement records
in the inner feedback loop, subject to the latency and sustained processing
requirements of R2 and R3. An outer loop collects measurement data and labels
and returns updated model parameters, addressing the device-adaptation
requirement of R1. Its update interval depends on device drift and the
availability of training data. Separating these tasks allows model training
to proceed without delaying individual feedback decisions.

\subsection{Three interfaces}\label{sec:unit:interfaces}

\begin{figure}[htbp]
  \centering
  \includegraphics[width=\textwidth]{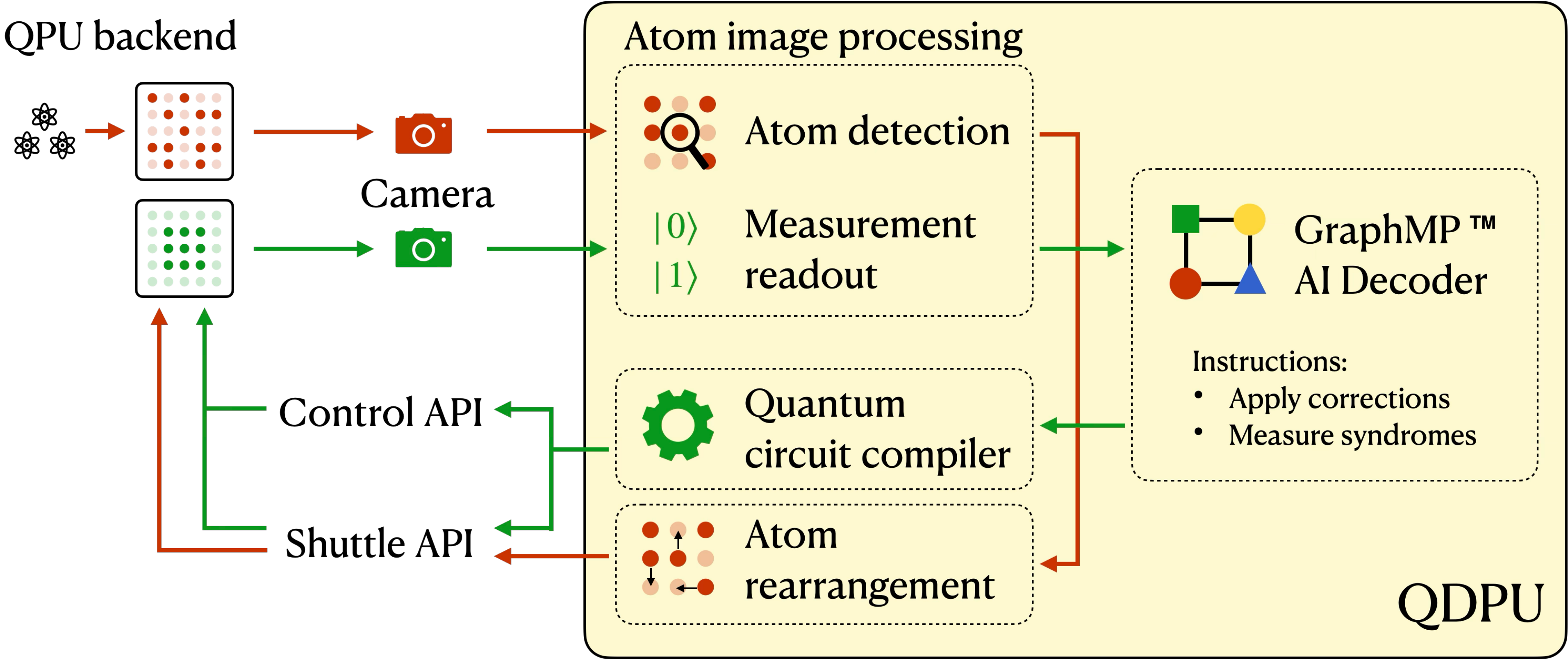}
  \caption{The QDPU between the QPU back-end and its controller. The atom array
  and its camera are drawn twice, once per loop. Red, array loading: atoms
  from the source (grey) fill sites at random (dark red, occupied; pale,
  empty); the frame goes to atom detection, then to atom rearrangement, whose
  move schedule returns through the Shuttle API. Green, computation: the
  assembled array (dark green, qubits in use; pale, empty) is read out as
  qubit states; the GraphMP decoder infers corrections from the syndromes.
  The circuit compiler uses the decoding results to generate instructions
  issued through the Control API (gates) and the Shuttle API (moves).
  Arrows follow the data in each loop; both pass through the QDPU.
  The outer loop to the training cluster is drawn in \cref{fig:loops}.}
  \label{fig:stack}
\end{figure}

A QDPU has three external interfaces (\cref{fig:stack}). The device-side
\emph{ingress} receives measurement records, which may be camera frames,
photon counts, digitised readout signals, or syndrome bits. The controller-side
\emph{egress} delivers results used in subsequent operations, such as an
occupancy map and move schedule, a Pauli-frame update, or a branch selection.
The \emph{outer-loop} interface transfers measurement data and labels for
training and receives updated model parameters. Unlike the feedback path,
model updates need not be completed within an individual measurement cycle
(\cref{app:platform}).

\subsection{Workloads}\label{sec:unit:workloads}

The two workloads evaluated here are array assembly and error decoding
(\cref{sec:workloads:assembly,sec:workloads:decoding}). Array assembly combines
occupancy detection with rearrangement scheduling. Decoding interprets
syndrome records to determine corrections or Pauli-frame updates. Both require
measurement interpretation, but their input formats and timing constraints
differ.

Further applications include state discrimination, calibration, and mid-circuit
control. Mid-circuit readout and conditional pulses have been implemented using
FPGA-based control~\cite{bluvstein2024logical,lee2026cavity,norcia2023midcircuit},
while automated calibration can run between experimental
shots~\cite{kelly2018calibration}. Task-specific models for these applications
could share the QDPU's execution environment and data interfaces. Quantum
sampling for machine learning is a prospective workload, discussed in
\cref{sec:workloads:qml}.

\subsection{Why learned inference}\label{sec:unit:why}

The benefit of learned inference depends on the workload and operating
conditions. For atom readout, matched filters can achieve CNN-level readout
error with substantially lower computational cost~\cite{kent2025matchedfilter}.
The array-assembly pipeline supports either a calibrated matched filter or
a small CNN. The closed-loop evaluation uses the matched filter at the
selected operating point; detector accuracy and implementation costs are
reported in \cref{app:detector,app:impl}.

For decoding, the companion paper evaluates learned models across several
qLDPC code families and noise models (\cref{sec:workloads:decoding}). Device
drift also motivates recalibration or model updates, as discussed in R1.
Learned models provide one approach to adapting inference to these changes.
The array-assembly evaluation relates detection accuracy and processing latency
to modelled atom loss, allowing the inference benefit to be assessed alongside
its implementation cost.

\section{A Neutral-Atom QDPU: Platform and Interfaces}\label{sec:platform}

Neutral-atom array preparation provides an initial application for the QDPU.
Increasing array sizes expand the number of sites that must be imaged and
classified (\cref{fig:growth}). The preparation cycle also has a clear
measurement-to-control sequence: a camera images the array, classical processing
estimates occupancy and plans rearrangement, and acousto-optic deflectors move
the atoms~\cite{chiu2025continuous,dadpour2025lowlatency}. The millisecond
timescales of array preparation allow this sequence to be studied with an
FPGA-based implementation.

Stochastic loading requires rearrangement to prepare a defect-free target, and
subsequent atom loss can require further preparation. Zoned architectures also
transport atoms between storage, interaction, and readout regions during
computation~\cite{bluvstein2022transport,bluvstein2024logical}. The present
work addresses the initial array-assembly pipeline.

\paragraph{Platform.}
The FPGA target is an AMD Zynq UltraScale+ RFSoC \texttt{xczu49dr} on the
ZCU216 evaluation board, a device class used in open-source quantum-control
platforms~\cite{stefanazzi2022qick}. The array-assembly pipeline meets a
\SI{200}{MHz} timing target after place-and-route with either detector, and
with the matched filter it also closes as a complete ZCU216 bitstream with an
on-chip self-test (\cref{app:impl:board}). With
the matched filter used in the loop it uses \SI{20.7}{\percent} of the
device's LUTs and \SI{1.5}{\percent} of its DSP slices; with the $8\times8$
CNN, \SI{23.2}{\percent} and \SI{13.7}{\percent}. Forty fixed-template matched-filter lanes
route on their own at \SI{668}{MHz} (\cref{app:impl}). The learned qLDPC
kernel described in \cref{app:decoder} is included in a co-residency resource
estimate. In that estimate, one matched-filter lane and the decoder
use \SI{40}{\percent} of the DSP slices; four CNN cores and the decoder would
use \SI{93}{\percent}, making DSP capacity the limiting resource for the CNN
(\cref{app:impl:coresident}). Joint implementation is needed to validate this
budget.

\paragraph{Interfaces.}
\Cref{tab:main_ifaces} maps the three interfaces onto a neutral-atom system.
The proposed ingress receives a $2048\times2048$ region of interest from a
Hamamatsu ORCA-Quest~2 qCMOS camera over
CoaXPress~\cite{hamamatsu2024orcaquest2}. The imaging simulator used for the
detector evaluation is configured for this camera
family~\cite{winklmann2023imagesim,capstum2024simulator}.

The egress connects to a Quantum Machines OPX1000 through an OPNIC interface
card~\cite{qm_opx1000,qm_opnic}. The proposed topology places the QDPU, OPNIC,
CPU, and GPU on one server's PCIe fabric. Each cycle, the QDPU writes occupancy
and a move schedule to server memory. Software using the OPNIC SDK forwards
this record to the OPX1000 over the vendor's NVQLink connection; the controller
generates the AOD and gate waveforms (\cref{app:platform:controller}).

The outer loop can use a GPU in the same server over PCIe, or a remote GPU
cluster over NVQLink. A round-trip FPGA-to-GPU-to-FPGA latency of
\SI{3.96}{\micro\second} and link rates up to \SI{400}{Gb/s} have been
reported for NVQLink~\cite{caldwell2025nvqlink,nvidia_nvqlink}. On the
ZCU216, an NVQLink link to an NVIDIA DGX Spark over \SI{10}{GbE}, built on
NVIDIA's Holoscan Sensor Bridge IP~\cite{nvidia_holoscan_hsb}, has a median
FPGA-to-host-to-FPGA round trip of \SI{2.74}{\micro\second} for 128-byte
messages forwarded by a CPU core and \SI{6.40}{\micro\second} when forwarded
by the GPU. The timestamp-to-MAC and MAC-to-timestamp logic contributes
\SI{0.52}{\micro\second}, excluding the transceivers
(\cref{app:if:nvqlink}). Camera ingest and controller egress are specified
but not implemented. The outer-loop transport has been measured with test
traffic, but the traffic it is to carry has not.

\begin{table}[htbp]
  \caption{The three interfaces of the neutral-atom QDPU. Status describes
  the application interface: \stSpec{} denotes a defined interface and
  \stOpen{} an interface still to be specified. NVQLink transport performance
  is measured independently (\cref{app:if:nvqlink}).
  \Cref{app:platform:ifaces} gives the channel-by-channel specification.}
  \label{tab:main_ifaces}
  \centering
  \small
  \begin{tabular}{@{}L{2.2cm}L{3.6cm}L{4.2cm}L{3.0cm}l@{}}
    \toprule
    Interface & Instrument & Carries & Rate or size & Status \\
    \midrule
    Ingress (device side) & Hamamatsu ORCA-Quest~2 qCMOS~\cite{hamamatsu2024orcaquest2} over CoaXPress & $2048\times2048\times16$-bit frames; $8\times8$-pixel site windows extracted on chip & quad CXP-6, $4 \times \SI{6.25}{Gb/s}$ (\SI{20}{Gb/s} payload); $6.7\times10^{7}$ bits per frame & \stSpec \\
    Egress (controller side) & Quantum Machines OPX1000~\cite{qm_opx1000} via its OPNIC~\cite{qm_opnic}: PCIe into server memory, then NVQLink to the controller & one occupancy bit per site; move stream as (from, to) site pairs & \num{2500} bits $+$ moves per cycle & \stSpec \\
    Outer loop & GPU in the same server over PCIe, or a GPU cluster over NVQLink~\cite{caldwell2025nvqlink,nvidia_nvqlink}; NVQLink transport measured at \SI{10}{GbE} (\cref{app:if:nvqlink}) & site windows and labels out; QDM parameters in (detector template or CNN weights, \cref{tab:cnn}; decoder) & minutes to hours; matched filter: 64 \texttt{int8} coefficients and one integer threshold; CNN: about \SI{7.2}{kB} of 12-bit weights and biases; decoder: about \SI{9.8}{kB} (\cref{app:decoder}) & \stOpen \\
    \bottomrule
  \end{tabular}
\end{table}

\paragraph{Processing near acquisition.}
A $2048\times2048$ sixteen-bit frame contains $6.7\times10^{7}$ bits, whereas
an occupancy map requires one bit per site. Extracting occupancy at the camera
interface reduces the amount of image data transferred to subsequent control
stages. Image transfer is a reported limitation in a continuously operated
neutral-atom array~\cite{chiu2025continuous}. The array-assembly evaluation in
\cref{sec:workloads:assembly} examines how processing latency affects
modelled atom loss.

\section{Workloads}\label{sec:workloads}

This section evaluates array assembly and learned qLDPC decoding as two QDPU
workloads. Array assembly converts images into occupancy estimates and move
schedules; decoding converts syndrome records into correction information.
Their implementations are assessed within a shared FPGA resource budget.
The final subsection discusses quantum sampling for machine learning as a
prospective extension.

\subsection{Array assembly}\label{sec:workloads:assembly}

The pipeline converts cropped site images into a stream of atom moves
(\cref{app:frontend}). A detector classifies each $8\times8$-pixel site
window, and a row assembler packs the resulting occupancy bits into grid rows.
The rearrangement engine reads the occupancy memory and target pattern and
emits source--destination pairs until the pattern is complete or the array is
declared infeasible.

Two detectors use the same one-bit-per-site interface: a matched filter, a
template and threshold calibrated from labelled windows, with 64
multiply--accumulates per site, and a convolutional network with \num{226944}
multiply--accumulates per site, quantised to 12 bits
(\cref{app:detector,app:quant}). On simulator data corrected for a readout
defect (\cref{app:detector}), the network is more accurate only where photons
are scarcest. At \SI{2}{ms} exposure, eight detected photons per atom, the
simulated per-site imaging fidelities are \SI{91.6}{\percent} for the matched filter
and \SI{95.4}{\percent} for the CNN; at \SI{5}{ms} they are
\SI{99.77}{\percent} and \SI{99.89}{\percent}, and from \SI{10}{ms} both
exceed \SI{99.99}{\percent} (\cref{tab:floor}). The integrated pipeline meets
a \SI{200}{MHz} timing target on the \texttt{xczu49dr} with either detector at
the $8\times8$ window. Detection of a $50\times50$ array takes
\SI{12.5}{\micro\second} with one matched-filter lane and \SI{22.5}{ms} with
one CNN core; forty matched-filter lanes route on their own at
\SI{668}{MHz} (\cref{app:impl}).

Previous work has accelerated several stages of this loop. An FPGA
rearrangement engine computes moves for a $30\times30$ target in a
$50\times50$ array in approximately
\SI{1}{\micro\second}~\cite{guo2025rearrange}, while an FPGA detection
accelerator processes a $256\times256$-pixel frame in
\SI{115}{\micro\second}~\cite{winklmann2026atomdetection}. A GPU system
with direct memory access reports a \SI{55.8}{ms} cycle for a $32\times32$
array~\cite{dadpour2025lowlatency}. Other reported feedback times include
\SI{282}{\micro\second} for 24 tweezers~\cite{hu2026lowlatency} and
\SI{25.3}{ms} for a $16\times16$ array~\cite{guo2026atomflow}. A host-based
system classifies \num{2400} sites in \SI{85}{\micro\second} using 24 CPU
cores and completes a \SI{460}{\micro\second} cycle through a dedicated
controller link~\cite{qm2026readout}.

These timings cover different parts of the measurement-to-control path.
In a reported GPU system, image acquisition accounts for \SI{46}{ms} of the
\SI{55.8}{ms} cycle~\cite{dadpour2025lowlatency}. Here, detection and
rearrangement are integrated through place-and-route, and their timing
provides inputs to the atom-loss model of \cref{app:loop}. This model uses a
corrected software reference planner; the column-step correction has not yet
been incorporated into the RTL. The model assumes a
\SI{150}{ms} host round trip (\cref{tab:params}). This exceeds the
approximately \SI{10}{ms} host contribution reported for the GPU system with
direct memory access~\cite{dadpour2025lowlatency}, as well as the
sub-millisecond feedback times of dedicated controller
links~\cite{hu2026lowlatency,qm2026readout}. The predicted benefit of
removing the host delay depends on this assumption;
\cref{app:loop:lifetime} shows that the conclusion holds for any host round
trip above a few milliseconds: at a trap lifetime of \SI{60}{s}, \SI{10}{ms}
of host round trip already lowers the assembly success probability from 0.85
to 0.78.

\begin{table}[htbp]
    \caption{Latency, resource, and data-volume comparisons for the neutral-atom
  array-assembly workload. Atom-loss predictions use the model of
  \cref{app:loop} and the parameters of \cref{tab:params}: $50\times50$
  array with a centred $35\times35$ target, $\tau_{\mathrm{vac}} =
  \SI{60}{s}$ and \SI{0.1}{\percent} loss per moved atom unless stated, each
  configuration at the exposure that minimises its wall-clock per assembly.
  The host round trip is assumed to be \SI{150}{ms}.}
  \label{tab:exchange}
  \centering
  \small
  \begin{tabular}{@{}L{3.4cm}L{7.6cm}l@{}}
    \toprule
    Comparison & Result & Reference \\
    \midrule
    Wall-clock $\to$ assembly & Removing the assumed \SI{150}{ms} host round trip raises the assembly success probability from \SI{8}{\percent} to \SI{85}{\percent} and cuts the wall-clock per successful assembly from \SI{19}{s} to \SI{0.16}{s}; with lossless transport, success rises from \SI{23}{\percent} to \SI{100}{\percent}. Dead time is \SIrange{2}{84}{\percent} of all atom loss with the QDPU in the loop and \SIrange{11}{96}{\percent} with a host, for trap lifetimes of \SIrange{5}{1380}{s}. & \cref{app:loop:lifetime} \\
    Exposure and inference latency & At the loop's \SI{10}{ms} exposure neither detector produced errors in the evaluated simulated samples, so the learned detector saves no exposure, and its \SI{22.5}{ms} of inference on one core is dead time: success falls from 0.85 to 0.69 and the wall-clock per assembly rises from \SI{0.16}{s} to \SI{0.43}{s}. With no inference time the two give the same result. & \cref{app:loop:latency} \\
    Inference cost $\to$ resources & One matched-filter lane takes \SI{1.5}{\percent} of the device's DSP slices and keeps pace with the camera; one CNN core takes \SI{13.7}{\percent}. With the decoder resident, the totals are \SI{40}{\percent} and \SI{93}{\percent} (four CNN cores). & \cref{app:impl:coresident} \\
    Image $\to$ occupancy map & $6.7\times10^{7}$ image bits in, $2.5\times10^{3}$ occupancy bits out per frame; move descriptors are additional. & \cref{app:platform} \\
    \bottomrule
  \end{tabular}
\end{table}

Under the assumed \SI{150}{ms} host round trip, the host loop completes a
$35\times35$ assembly in \SI{8}{\percent} of attempts at a trap lifetime of
\SI{60}{s}; with the QDPU in the loop it completes \SI{85}{\percent}, and the
wall-clock per successful assembly falls from \SI{19}{s} to \SI{0.16}{s}
(\cref{tab:exchange}). At this array size, the assumed host delay substantially
reduces the probability of completing an assembly. At longer trap lifetimes,
the host loop still takes approximately $11$ times as long per successful
assembly as the QDPU loop at \SI{300}{s}, and $7.5$ times as long at
\SI{1380}{s}. This comparison depends on the host delay, the trap lifetime
and the transport loss; at \SI{1}{\percent} loss per moved atom, no
configuration completes the assembly in the simulated trials.

The reference loop uses the matched filter at a \SI{10}{ms} exposure,
with \SI{12.5}{\micro\second} of detection time. Detector choices and their
inference costs are compared in \cref{app:loop:latency}. At this operating
point, camera readout and atom transport account for \SI{64}{\percent} of
the \SI{27.5}{ms} modelled cycle, including about \SI{10}{ms} of transport.

\subsection{Decoding}\label{sec:workloads:decoding}

Belief propagation is widely used for classical low-density parity-check
(LDPC) decoding, but short cycles and degeneracy can impair its performance
on quantum LDPC codes~\cite{poulin2008iterative}. Symmetric trapping sets
provide one characterisation of failures in syndrome-based iterative
decoding~\cite{raveendran2021trapping}. Ordered-statistics post-processing can
substantially improve performance~\cite{panteleev2021bposd,roffe2020landscape}.
For surface codes, learned decoders have also been evaluated on experimental
data and have outperformed matching and tensor-network decoders in the
reported benchmarks~\cite{bausch2024alphaqubit,varbanov2025nnd}.

The companion paper~\cite{companion2026graphmp} introduces GraphMP, a QDM for
qLDPC decoding: a graph message-passing decoder in
which a single trained model applies across decoding graphs of different size
and structure. Its messages are computed by local, parallel operations on the
Tanner graph, and it runs in two modes: a low-latency single pass, and a
bounded-width search guided by the model's probabilistic output that trades
latency for stronger logical-error suppression without changing the model.

The model is trained through an iterative procedure called a decoder factory.
In each iteration, the current model guides a wide search for corrections to
difficult decoding cases. These corrections provide supervised training data
for the next model, so that it learns to make better single-pass predictions
from the results of the search.

Across qLDPC code families, sizes, physical error rates and noise models, the resulting
decoder transfers to graphs it was not trained on and is competitive with
strong classical decoders, and the same cycle refits it as the hardware noise
evolves. Before device data exist, the training distribution can come from an
emulator that models the processor, its atom-loss channels and the classical
control flow together, such as the PPVM framework~\cite{luo2026ppvm}.
On the companion paper's benchmark of twelve decoding tasks spanning
bivariate-bicycle, generalised-bicycle, lifted-product and mitten codes, the
search-enhanced decoder reaches a lower logical error rate than BP+OSD on all
nine code-capacity tasks and is broadly competitive with Relay-BP; the
single-pass model overtakes the belief-propagation teacher it was distilled
from within one generation~\cite{companion2026graphmp}. Separately, for the
LP20 $[[1122,148,\le 20]]$ lifted-product code under code-capacity noise,
the \texttt{xczu49dr} kernel configuration described
in \cref{app:decoder} has a post-route clock period of
\SI{3.514}{ns}, corresponding to a kernel latency of
\SI{0.84}{ms} per decoding instance. A partition across programmable
logic and the AI Engine array of a Versal device has an estimated
latency of approximately \SI{0.2}{ms} per decoding instance.
These timings apply to the stated decoding task and hardware
configurations. Decoder hardware must satisfy both a sustained
throughput requirement and the latency budget of dependent operations.
Pauli-frame updates can often be deferred, whereas a conditional operation
requires the relevant decoded result before it can proceed. Both uses can
be supported by real-time decoding.
Reported timings depend on the code, the syndrome history and
the decoder implementation. Batched inference time per shot
must also be distinguished from the full response time of a
feedback operation.

\paragraph{Pauli-frame tracking.}
A Pauli correction need not be applied physically to the qubits: it can be
recorded in a classical frame and incorporated into the interpretation of
later measurements~\cite{knill2005pauliframe,riesebos2017pauliframes}.
The decoder must keep pace with syndrome generation on average, and the
relevant frame information must be available before an operation that
depends on it. Conditional operations in teleportation-based non-Clifford
gates impose such dependencies~\cite{gidney2019autoccz}.
If decoding persistently falls behind syndrome generation, the resulting
backlog can cause an exponential slowdown of the
computation~\cite{terhal2015qec,skoric2023parallelwindow}.
Parallel decoding of syndrome-history windows can increase
throughput~\cite{skoric2023parallelwindow,tan2023scalable}.
In a neutral-atom experiment, fully connected neural decoders were
used for distance-3 and distance-5 surface codes, with $3\times3$
and $5\times5$ data-qubit arrays, respectively, and four rounds of
syndrome extraction. The Methods report an inference time of
\SI{0.33}{\micro\second} per shot on an NVIDIA A100 GPU for batches
of approximately $10^4$ shots~\cite{bluvstein2026universal}.
This batch-amortized inference time characterizes throughput;
it does not establish single-shot feedback latency or scaling
to larger codes.
A GPU belief-propagation decoder reports
latencies below \SI{50}{\micro\second} for a $[[784,24,24]]$
code~\cite{ferraz2025gpuqldpc}.
Syndrome records can also support post-processing. Virtual error cancellation
combines the results of a low-latency in-loop decoder with those of a more
computationally intensive decoder applied after execution, reducing bias in
logical observables without modifying the quantum
circuits~\cite{shtanko2026measurenow}.
A distance-5 surface-code experiment on a superconducting processor
reported an average decoding latency of \SI{63}{\micro\second} while
keeping pace with a \SI{1.1}{\micro\second} cycle for a million cycles.
The decoder ran on a workstation, and its output was not fed back into
the logical circuit~\cite{acharya2025willow}.

\paragraph{Real-time decoding.}
Real-time decoding supports frame updates and, when required, supplies
results for conditional operations during circuit execution. The decoder
must sustain the syndrome rate and return the relevant result before
each dependent operation. In teleportation-based implementations of
non-Clifford gates, for example, a decoded measurement outcome can
determine a conditional correction or a subsequent measurement
basis~\cite{gidney2019autoccz}.
Heralded erasures provide location information that can be incorporated
into decoding~\cite{wu2022erasure}. In non-Abelian topological codes,
recovery can require successive rounds of measurement and correction
because fusion outcomes are nondeterministic. Intrinsic heralding
provides additional syndrome information about the error
path~\cite{jing2026heralding}.
The feedback deadline is set by the next dependent operation; it does
not require each syndrome round to be decoded within one QEC cycle.
An FPGA Collision Clustering decoder integrated into a
superconducting control system was tested on an eight-qubit
stability experiment with four weight-two $ZZ$ checks.
Its syndrome-extraction circuit is equivalent to that of a
circular repetition code. For the nine-round feedback experiment,
the full response time from the final syndrome-extraction round
to a conditional gate was \SI{9.6}{\micro\second}, comprising
\SI{6.5}{\micro\second} of decoding and
\SI{3.1}{\micro\second} of communication and
control~\cite{caune2026realtime}.
Standalone hardware implementations report \SI{810}{ns} per round for a
distance-21 surface code on simulated syndromes~\cite{barber2025realtime}
and \SI{11.5}{ns} per round for a distributed union-find decoder at
distance 21 under phenomenological noise~\cite{liyanage2023fpga}.
FPGA decoding of the gross code has also been
reported~\cite{maurer2025grossdecode}.

\begin{table}[htbp]
  \caption{Classical processing requirements for Pauli-frame tracking and
  real-time decoding with feedforward. The two uses can share a decoder and
  can both be supported within the QDPU processing architecture.}
  \label{tab:regimes}
  \centering
  \small
  \begin{tabular}{@{}L{2.9cm}L{5.4cm}L{5.4cm}@{}}
    \toprule
    & Pauli-frame tracking & Real-time decoding \\
    \midrule
    Decoder output is & a Pauli correction recorded in a classical frame~\cite{knill2005pauliframe} & a decoded outcome used to select a conditional operation~\cite{caune2026realtime} \\
    Must meet & sustained throughput sufficient to avoid a backlog~\cite{terhal2015qec} & the throughput requirement and the deadline of each dependent operation~\cite{gidney2019autoccz,caune2026realtime} \\
    Failure mode & a growing backlog that delays operations requiring the frame~\cite{terhal2015qec,skoric2023parallelwindow} & a delayed conditional operation or a missed feedback deadline \\
    Tolerates & deferred frame updates, batching, and parallel decoding windows, subject to future dependencies~\cite{skoric2023parallelwindow,tan2023scalable} & batching or windowing that respects the feedback deadline \\
    Typical hardware emphasis & throughput through parallel processing on CPUs or GPUs~\cite{bluvstein2026universal,ferraz2025gpuqldpc,acharya2025willow} & short response times through hardware acceleration and close controller integration~\cite{barber2025realtime,liyanage2023fpga,maurer2025grossdecode,caune2026realtime} \\
    \bottomrule
  \end{tabular}
\end{table}

The requirements in \cref{tab:regimes} motivate different hardware
configurations. Low-latency processing supports conditional operations,
while parallel and batched decoding can sustain the throughput needed for
frame tracking. A heterogeneous QDPU could combine these capabilities
through an FPGA--GPU interface. The reported
\SI{3.96}{\micro\second} FPGA-to-GPU-to-FPGA round trip of
NVQLink~\cite{caldwell2025nvqlink} provides a reference point for the
communication cost of such a configuration. On the ZCU216, NVQLink to a DGX
Spark at \SI{10}{GbE} measures \SI{2.74}{\micro\second} through a CPU core
and \SI{6.40}{\micro\second} through the GPU at the median. With one shot
in flight, a quantised GraphMP ranking kernel for the LP20 code achieves a
median host-to-FPGA-to-host latency of \SI{50.5}{\micro\second}, with
bit-exact agreement against the PyTorch reference. This interval includes
about \SIrange{4}{5}{\micro\second} of link overhead and excludes host-side
BP+OSD post-processing (\cref{app:if:nvqlink}).

The relevant timing constraint depends on the experiment. The
\SI{0.84}{ms} decoder latency reported here exceeds both the
\SI{1.1}{\micro\second} superconducting measurement
cycle~\cite{acharya2025willow} and the time available in reported
sub-millisecond neutral-atom mid-circuit feedback
loops~\cite{bluvstein2024logical,lee2026cavity}. Other neutral-atom operations
have longer cycles. Preparation and reloading take tens of milliseconds in a
continuously operated array~\cite{chiu2025continuous}. In the repeated
toric-code experiment of Atom Computing, syndrome-extraction cycles take
approximately \SI{80}{ms} and \SI{150}{ms} for the two code sizes studied,
or approximately \SI{30}{ms} when measured ancillas are replaced rather than
reloaded~\cite{atomcomputing2026toric}. The decoder latency is shorter than
these cycle times, although integration with a particular experiment requires
matching its decoding task and feedback schedule. FPGA execution is therefore
considered here for workloads with compatible timing budgets, with GPU-based
frame tracking as a further option.

\subsection{Quantum machine learning}\label{sec:workloads:qml}

Quantum sampling provides another potential workload. In several quantum
machine-learning approaches, measurements from a quantum processor supply
samples to a classical training procedure. Sampling tasks also underlie
experimental demonstrations of quantum computational
advantage~\cite{arute2019supremacy}, although these demonstrations do not by
themselves establish an advantage for machine learning.

Proposed generative models include quantum circuit Born machines trained on
ion-trap hardware~\cite{benedetti2019generative}, Born machines with
worst-case classical-hardness results~\cite{coyle2020born}, quantum Boltzmann
machines~\cite{amin2018boltzmann}, and quantum generative adversarial
networks~\cite{zoufal2019qgan}. Generative quantum models without barren
plateaus have also been trained and sampled on a 68-qubit processor for
distributions believed to be classically intractable~\cite{huang2025generative}.
A quantum Markov chain Monte Carlo method has shown convergence in fewer steps
than its classical comparator in simulation and on a superconducting
processor~\cite{layden2023mcmc}. Neutral-atom demonstrations include quantum
reservoir learning with up to 108 qubits~\cite{kornjaca2024reservoir} and a
quantum GAN generator~\cite{dibrita2024recon}.

The practical advantage of these methods remains open. Classical algorithms
have matched some proposed quantum speedups~\cite{tang2019recommendation},
and barren plateaus can make training difficult at
scale~\cite{mcclean2018barren}. Reviews discuss these limitations and the
evidence needed to establish useful
advantage~\cite{biamonte2017qml,cerezo2022challenges,schuld2022advantage}.

Such workloads require classical processing at several timescales. A classical
optimiser updates circuit parameters using accumulated
samples~\cite{cerezo2021variational,bharti2022nisq}. Changes in device
conditions may also require recalibration or readout-model updates. Dynamic
circuits impose a separate feedback constraint when mid-circuit measurements
determine later gates within the coherence
time~\cite{corcoles2021dynamic,baumer2024entanglement,fossfeig2023adaptive,iqbal2024nonabelian}.
The QDPU interfaces could support measurement acquisition and conditional
control in such workflows, alongside a classical training
system~\cite{alexeev2024quantumcentric}. Application-model training and
adaptation of the QDPU's measurement models remain distinct tasks. These
sampling workloads are prospective extensions of the architecture and are not
implemented here.

\section{Discussion and Outlook}\label{sec:limits}

We introduced the QDPU as a processing architecture between quantum
measurement acquisition and feedback control, with models fitted to the
device and hardware chosen for the latency and throughput requirements
of each workload. For neutral-atom array assembly, we implemented a
pipeline combining occupancy detection, by a calibrated matched filter or a
small CNN, with rearrangement scheduling and evaluated its FPGA timing and resource use after
place-and-route. The atom-loss model relates detector performance and
processing time to predicted atom retention, allowing inference gains
to be assessed alongside their hardware cost. A separate implementation
of learned qLDPC decoding provides a second workload for evaluating the
architecture.

The detector evaluation uses simulated images rather than real camera frames.
The simulator is configured for the ORCA-Quest camera family but has not been
validated against the intended camera (\cref{app:detector}). We corrected a
readout defect in the simulator; the uncorrected data had led earlier drafts of
this work to the opposite conclusion about the detector.

The CNN pipeline is implemented out of context and excludes board-level I/O
constraints. The matched-filter pipeline has also been built as a board
bitstream with an on-chip self-test, but has not yet run on the board
(\cref{app:impl:board}); the build found and fixed a move-loss defect in the
rearrangement engine under controller backpressure. Camera ingest and the controller link are specified but not
implemented (\cref{app:platform}). The outer-loop NVQLink transport has been
measured on the board at \SI{10}{GbE} with test traffic and a GraphMP
ranking kernel, but has not been joined to the pipeline (\cref{app:if:nvqlink}). The frame-to-schedule latency is therefore
derived from implementation and simulation results, and the hand-off from
server memory through the OPNIC software has not been measured. The vendor's
planning figure is approximately \SI{1}{\micro\second} per PCIe transfer
inside the OPC, including memory-to-OPNIC transfer. The decoder has been placed
and routed separately (\cref{app:decoder}); its co-residency with the
pipeline is a synthesis-based resource estimate pending joint place-and-route
(\cref{app:impl:coresident}). The interface-integration evaluation covers two
occupancy cases (\cref{app:frontend:verification}).

The evaluations also use different geometries: a $14\times14$ array in a
256-pixel frame for imaging, and a $50\times50$ array in a 2048-pixel region
of interest, with a $35\times35$ target, for the loop model and the FPGA
design (\cref{tab:params,app:detector}). The site pitch of the target apparatus
is not yet fixed. The loop model uses detector error rates from the simulated
\SI{4}{\micro\metre} sparse array.

The loop results use a corrected software reference planner whose column
step has not yet been incorporated into the RTL. Its full-array replanning
on repair cycles increases transport loss, so the reported success
probabilities and latency gains are conditional on this policy
(\cref{app:loop:lifetime}).

The next implementation step is to integrate the interfaces in
\cref{tab:main_ifaces} on the ZCU216 board and measure camera ingest with the
ORCA-Quest~2 and feedback delivery to the OPX1000. Recorded frames will be used
to evaluate the detector under experimental imaging conditions. Joint
place-and-route of the pipeline and decoder is needed to validate the
co-residency budget. Further work includes calibration, mid-circuit control,
heralded-loss handling, and the sampling applications discussed in
\cref{sec:workloads:qml}. Their hardware configurations will depend on the
latency, throughput, and resource requirements of each workload.

For neutral-atom QEC, the QDPU framework makes measurement inference
and decoding explicit components of the system design. In particular,
qLDPC implementations require code size and the length of syndrome
histories to be evaluated alongside decoder accuracy, compute and
memory resources, and feedback timing. The separate decoder
implementation reported here provides an initial hardware reference
for that assessment. Extending it to noisy syndrome extraction and
atom-loss information, and measuring the complete feedback response,
would help determine which code, decoder and hardware combinations
are suitable for specific neutral-atom QEC protocols.

\section*{Acknowledgments}
We thank Luis Fernandez Martinez, Wenchao Xu, Joonhee Choi,
Yaniv Kurman, Brian Kain and Javier Duarte for helpful discussions and feedback.

This research was supported in part by grant NSF PHY-2309135 to the Kavli
Institute for Theoretical Physics (KITP).

\appendix

\noindent\emph{Provenance tags.} Quantitative claims in the appendices carry a
superscript tag naming their source: \meashw{} measured after place-and-route;
\measbd{} measured on the board; \meassim{} measured in RTL simulation; \csynth{} a C-synthesis estimate;
\simu{} the imaging simulator; \phys{} the atom-loss model; \lit{} a
literature value; \roof{} a projection from datasheets or standards;
\algo{} an algorithmic model or fit. \Cref{app:open} lists what remains
open.

\section{Rearrangement Pipeline: Architecture and Interface Contract}\label{app:frontend}

\subsection{Scope}\label{app:frontend:scope}

\begin{figure}[htbp]
  \centering
  \includegraphics[width=\textwidth]{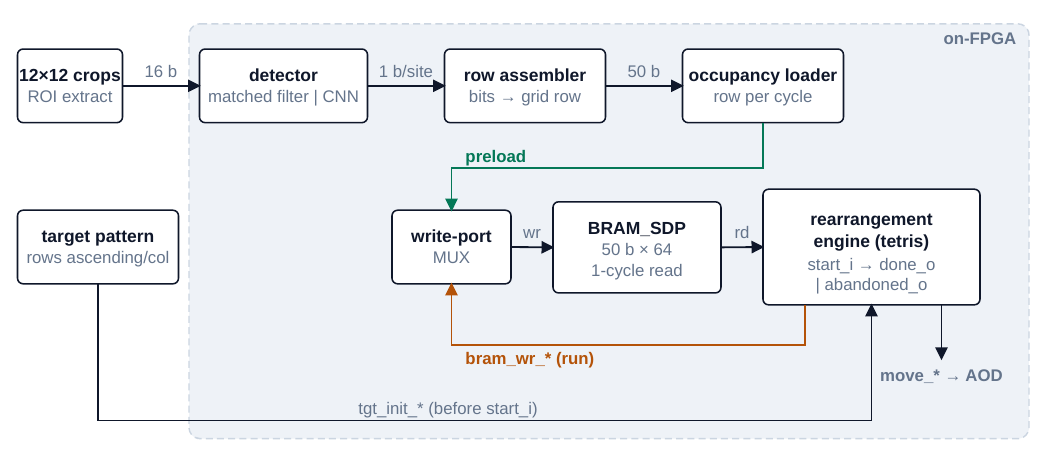}
  \caption{Rearrangement pipeline block diagram. Shaded blocks are FPGA-resident. Edge
  labels give bus width and sustained rate.}
  \label{fig:arch}
\end{figure}

Throughout the appendices, the \emph{QDPU} (quantum data processing unit,
\cref{sec:unit}) is the real-time unit between the camera and the controller
that converts measurement data into gate-level feedback by learned inference;
in this work it is realised
on a single FPGA, and the \emph{rearrangement pipeline} (hereafter the
pipeline) is its first workload, the one
implemented and evaluated here. The pipeline is the chain from cropped camera
pixels to a stream of atom moves (\cref{fig:arch}). It has four components:

\begin{itemize}[leftmargin=1.2em,itemsep=3pt,topsep=3pt]\sloppy
  \item \textbf{Site detector} (matched filter or CNN, \cref{app:detector}):
        takes the $8\times8$-pixel window around one site and emits one bit,
        occupied or empty.
  \item \textbf{Row assembler}: packs the $N^2$ occupancy bits into $N$ words
        of $N$ bits each, one word per grid row.
  \item \textbf{Occupancy memory}, with its loader and write-port multiplexer:
        a dual-port block RAM holding the current grid; the loader writes it
        one row per cycle before a run, and the multiplexer hands the write
        port to the engine once the run starts.
  \item \textbf{Rearrangement engine} (\texttt{tetris}): reads the occupancy
        memory and the target pattern and emits a stream of moves, each a
        (from, to) pair of sites, ending in \texttt{done\_o} or, if the array
        cannot be completed, \texttt{abandoned\_o}.
\end{itemize}

The engine is a pre-existing block. The out-of-context builds of
\cref{tab:impl} use its RTL unchanged, and the board build adds one fix to
its move FIFO (\cref{app:impl:board}); it implements a tetris-style compression
algorithm~\cite{shuai2023tetris} modified to minimise move
distance~\cite{cheng2021mindist}. The minimum distance can potentially reduce the atom loss associated with physically moving the atoms. The algorithm picks the shortest move distance from the list of possible moves for each row. The other three components are new to this work and were built to
write the memory the engine already reads, so the engine's behaviour is
independent of how its occupancy arrives (\cref{app:frontend:verification}).

Rearrangement is latency-critical for two reasons. Loading of the tweezer array
is stochastic, and atoms are lost during a computation at stochastic times and
sites, so a defect-free register has to be assembled before a circuit runs and
re-assembled whenever loss is detected. Beyond initialisation, QEC (quantum
error correction) cycles and the logical operations built on them require
repeated rearrangement of atoms between zones: entangled atoms are shuttled
coherently during the computation~\cite{bluvstein2022transport}, and zoned
architectures separate storage, entangling and readout regions that atoms
move between every cycle~\cite{bluvstein2024logical}. Every rearrangement adds
idle time in which the register accumulates decoherence and loss, so the
throughput of the shuttling schedule and the circuit that consumes it
bounds the useful clock rate of the machine. Therefore, both the shuttling
algorithm and the circuit-level scheduling that generates the
moves~\cite{tan2024compiling} have to be efficient. This work addresses the initialisation loop as
the first step.

The chain is implemented for a $50\times50$ site array and taken through
synthesis, placement and routing on the \texttt{xczu49dr} (\cref{app:impl}). The camera ingest and controller link of \cref{app:platform} are not part
of it. Board-level integration, and the tests
of the upstream (camera) and downstream (controller) interfaces specified in \cref{app:platform}, are the subject of a follow-on paper. 

\subsection{Occupancy contract}\label{app:frontend:contract}

Detection and rearrangement communicate through one bit per site and one
memory. The detector emits occupancy one grid row at a time as an $N$-bit word
(LSB $=$ column~0) into an external simple dual-port block RAM
(\texttt{WIDTH}$=N$, \texttt{DEPTH}$=N_{\mathrm{PAD}}$, one-cycle read) that
the engine owns during a run. Only the write port is arbitrated: the loader
drives it during preload, one row per cycle, and the engine during the run as
it commits moves; a registered \texttt{running} flag selects the writer, so the
two are never active in the same cycle. The engine owns the read port
throughout. The target pattern enters through a separate port as (column, row)
pairs, rows ascending within each column, before the start pulse; moves leave
over a ready/valid handshake as from/to row and column descriptors.
\Cref{tab:ports} lists the port groups.

\begin{table}[htbp]
  \caption{Port groups of the rearrangement pipeline at $N=50$. Widths follow from
  $N$ and $N_{\mathrm{PAD}}$, the width of the outer loading zone, set to twice
  the packed grid ($N_{\mathrm{PAD}} = 2N$) in the experiments reported here;
  per-signal names follow the reference top. All
  port groups are implemented and exercised in RTL simulation
  (\cref{app:frontend:verification}).}
  \label{tab:ports}
  \centering
  \small
  \begin{tabular}{@{}L{3.3cm}lL{2.8cm}L{6.6cm}@{}}
    \toprule
    Port group & Dir. & Width & Semantics \\
    \midrule
    Crop input to detector & in & $P\times8\times8\times16$ b & one $8\times8$ crop per lane per cycle, as built; $P=1$ for the CNN path \\
    Occupancy write (\texttt{bram\_wr\_*}) & in & $1 + \lceil\log_2 N_{\mathrm{PAD}}\rceil + N$ & enable, row address, $N$-bit row word (LSB $=$ col.~0); loader during preload, engine during run \\
    Write-port select (\texttt{running}) & int. & 1 & registered; 0 $=$ loader owns write port, 1 $=$ engine owns it \\
    Occupancy read & int. & $\lceil\log_2 N_{\mathrm{PAD}}\rceil + N$ & engine-owned; one-cycle read latency \\
    Target pattern (\texttt{tgt\_init\_*}) & in & $2\lceil\log_2 N\rceil$ & (column, row) pairs, rows ascending per column; accepted before \texttt{start\_i} only \\
    Control (\texttt{start\_i}, \texttt{done\_o}, \texttt{abandoned\_o}) & in/out & 1 each & start pulse; assembly complete; array infeasible \\
    Move stream (\texttt{move\_*}) & out & $4\lceil\log_2 N\rceil$ $+$ ready/valid & from/to row and column per accepted beat \\
    \bottomrule
  \end{tabular}
\end{table}

\subsection{Module inventory}\label{app:rtl:inventory}

\Cref{tab:rtl:modules} lists the implemented blocks. Three categories appear:
the rearrangement engine and its submodules; the front-end blocks written to
produce the occupancy the engine reads; and a small library of register,
memory and arithmetic primitives shared across the project, instantiated
explicitly rather than inferred.

\begin{table}[htbp]
  \caption{RTL inventory; \cref{tab:impl} gives the resources of the active
  integrated configurations. Both detector configurations have been placed
  and routed, and parallel matched-filter variants have also been evaluated
  separately (\cref{app:impl:pnr}).}
  \label{tab:rtl:modules}
  \centering
  \small
  \begin{tabular}{@{}L{5.4cm}L{10.4cm}@{}}
    \toprule
    Block & Role \\
    \midrule
    \texttt{tetris} & Two-phase engine top: per-row sweep, phase-2 handoff, shared move FIFO\\
    \texttt{row\_assign} & Compile-time selection of the phase-1 assignment policy\\
    \texttt{row\_assign\_min\_displacement} & Minimum-travel assignment; order-preserving dynamic program\\
    \texttt{row\_assign\_earliest\_defect} & Urgency-ordered assignment; alternative policy\\
    \texttt{column\_compress} & Phase~2: vertical compression, one column at a time \\
    \texttt{bitonic\_sort} & 64-wide, 21-stage pipelined sorter on the urgency key \\
    \texttt{source\_col\_encoder} & Row bitmap $\to$ ascending atom column indices\\
    \texttt{target\_queue\_store} & $N$ per-column target-row queues in LUTRAM\\
    \midrule
    \texttt{site\_row\_assembler} & Per-site detector decisions $\to$ $N$-bit occupancy rows \\
    \texttt{atom\_readout} & Occupancy rows $\to$ grid-memory write port \\
    \texttt{tgt\_pattern\_loader} & Target pattern from ROM into the queue preload port \\
    \texttt{target\_pattern\_gen} & Generated target pattern, per-column ascending \\
    \texttt{atom\_rearrangement\_cnn\_top} & System top: detector, assembler, loader, arbiter, memory, engine \\
    \texttt{matched\_filter\_detect} / \texttt{\_array} & Single-site and $P$-lane matched-filter detector (\cref{app:detector:mf}) \\
    \midrule
    \texttt{BRAM\_SDP}, \texttt{FIFO}, \texttt{MUX}, \texttt{DFF}, \texttt{DFFR}, \texttt{COUNTONES}, \texttt{COUNTER} & Memory, queue, arbitration, register and population-count primitives\\
    \bottomrule
  \end{tabular}
\end{table}

\subsection{What the engine computes}\label{app:rtl:algorithm}

The engine implements tetris-style compaction in two
phases~\cite{shuai2023tetris}. Phase~1 processes the grid row by row, assigning
atoms to columns that still require them and moving the selected atoms
horizontally. After a successful sweep, each target column contains the required
number of atoms. Phase~2 moves atoms vertically within each column onto the
target rows. Multiple atoms in one row or column can be transported in a single
acousto-optic sweep. The number of sequential sweeps is therefore bounded by
the rows and columns processed, rather than by the total number of atoms moved.

Both phases reduce to the same one-dimensional problem: pair an ascending list
of source indices with an ascending list of destination indices so as to
minimise total travel. By the rearrangement inequality the optimum for two
lists of equal length is index-by-index matching, so no search is needed once
the two lists are ordered; phase~2 therefore contains no sort at all, because
the bit scan that reads the source and target masks produces both lists in
ascending order by construction. Phase~1 is the harder case only because the
two lists have different lengths: with $K$ atoms in the row and $M$ columns
still needing atoms, one side must be subsetted, and which subset is chosen is
the policy question of \cref{app:rtl:policy}.

\begin{figure}[htbp]
  \centering
  \includegraphics[width=0.78\textwidth]{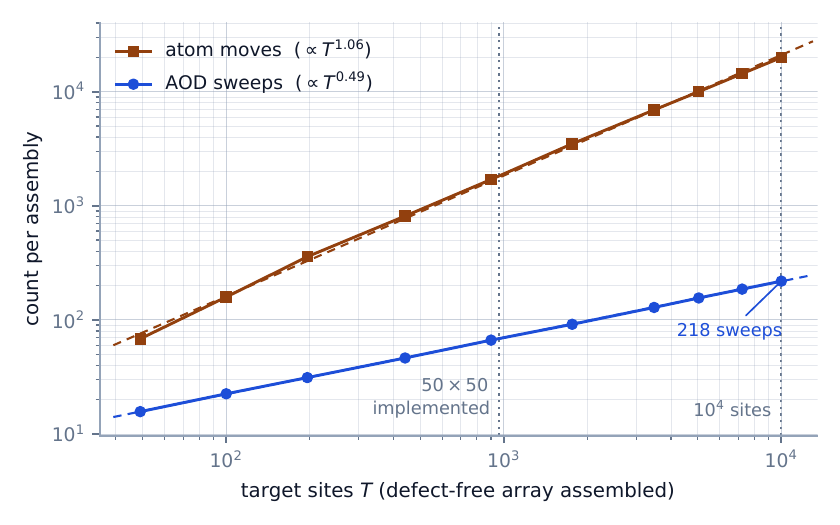}
  \caption{Cost of one assembly pass against target size for the algorithm the
  engine implements: acousto-optic sweeps (the sequential operations the
  controller must play) and total atom moves (the transport charged in atoms).
  Points are means over 20 seeds up to $T=\num{1764}$ and 8 above, at trap
  fill \num{0.6}, central region of interest, every point defect-free; dashed
  lines are power-law fits\algo{}. The implemented $50\times50$ design and the
  $10^{4}$-site scale of \cref{app:platform:scaling} are marked.}
  \label{fig:rtl:sweeps}
\end{figure}

\Cref{fig:rtl:sweeps} gives the cost of this decomposition as a function of
array size, measured on the reference implementation of the same algorithm.
Sweeps grow as $T^{0.49}$ in the number of target sites $T$ and moves as
$T^{1.06}$: a defect-free $10^{4}$-atom array is assembled in one pass with
about \num{218} sweeps and about $2T$ moves, every seed defect-free at a trap
fill of \num{0.6}\algo{}. The sweep count is the latency-critical figure,
because each sweep is one physical operation on the controller
(\cref{app:platform:aod}); the move count is the atom-loss exposure charged in
\cref{app:loop:lifetime}.

\subsection{Phase-1 assignment policies}\label{app:rtl:policy}

Two policies are implemented behind one port list and selected at elaboration
(\cref{tab:rtl:policy}). \emph{Minimum displacement} is deployed. It orders the
needy columns by urgency, pairs them with the row's atoms so as to minimise
$\sum_i |\mathrm{src}_i - \mathrm{dst}_i|$, and when the row holds more atoms
than there are needy columns it runs an order-preserving dynamic program to
choose which $M$ of the $K$ atoms to move. \emph{Earliest defect} keeps the
urgency order and pairs greedily. It needs no dynamic program, so its cost per
row is the fixed $K + N + 27$ cycles: about twice the deployed policy's fast
path, and a third of its dynamic-program path, at the operating point of
\cref{fig:rtl:cycles}. It does not minimise travel.

\begin{table}[htbp]
  \caption{The two phase-1 assignment policies. $K$ is the number of atoms in
  the row and $M$ the number of columns whose target queue is still
  non-empty; $N=50$ and the sort pipeline is $\mathrm{GS}=21$ stages deep.}
  \label{tab:rtl:policy}
  \centering
  \small
  \begin{tabular}{@{}L{3.4cm}L{5.4cm}L{5.4cm}@{}}
    \toprule
     & \textbf{Minimum displacement} (deployed) & \textbf{Earliest defect} \\
    \midrule
    Objective & Minimum total horizontal travel for the row & Fill the column whose next unfilled row is highest up, first \\
    Destination order & Urgency key (next needed row, then column index), from the bitonic sort & The same key and the same sorter \\
    Subsetting when $K > M$ & Order-preserving dynamic program over the $M \times K$ table, then traceback & None: the first $K$ sort outputs are taken \\
    Cycles per row & $K + 27$ when $K \le M$; $MK + K + M + N + 28$ when $K > M$ & $K + N + 27$ \\
    Extra state & $M \times K$ take-matrix, $2 \times N$ cost buffer & $N$-bit chosen mask \\
    Verified against & Move-for-move equality with the software model (\cref{app:frontend:verification}) & Co-simulation of \texttt{done\_o} / \texttt{abandoned\_o} only \\
    \bottomrule
  \end{tabular}
\end{table}

The deployed policy limits the number of pairs to $\min(K,M)$, so a row does
not emit more assignments than there are columns requiring atoms. The
earliest-defect variant instead pairs all $K$ atoms with the first $K$ entries
of the sorted destination list, including padded entries for columns whose
queues are empty. In the feasible co-simulation case of
\cref{app:frontend:verification}, a $50\times50$ array at a per-site loading
probability of \num{0.7} contains approximately \num{1745} atoms, while the
central $6\times6$ target requires \num{36}. The variant emits \num{1584}
moves before asserting \texttt{done\_o}. A software model of the same rule
emits \num{1637} moves, within \SI{4}{\percent} of the co-simulation
count\algo{}. Limiting assignments avoids these unnecessary moves and their
associated transport loss (\cref{app:loop:lifetime}).

\subsection{Datapath blocks}\label{app:rtl:blocks}

\begin{figure}[htbp]
  \centering
  \includegraphics[width=\textwidth]{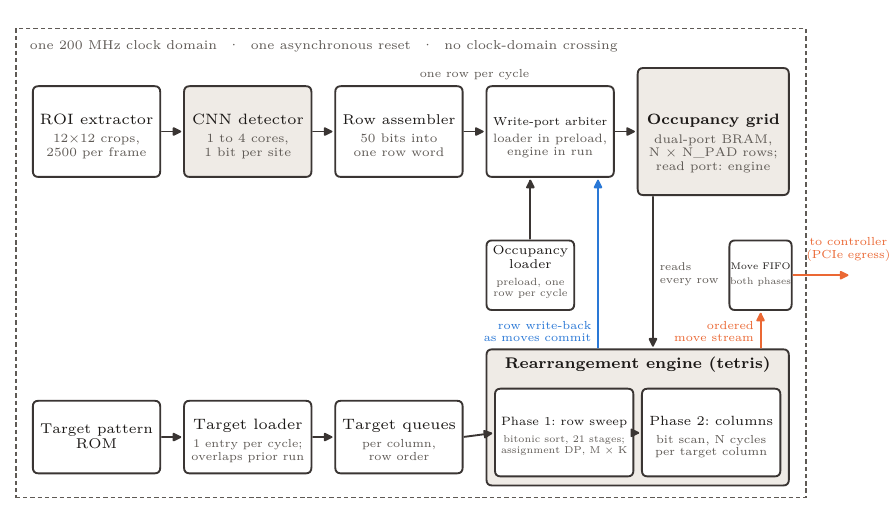}
  \caption{The implemented pipeline. Detection writes the occupancy grid one
  row at a time through an arbitrated write port; the engine owns the read
  port throughout a run and writes each row back as its moves commit. Phase~1
  and phase~2 share one move FIFO, so the controller sees a single ordered
  move stream. }
  \label{fig:rtl:arch}
\end{figure}

\Cref{fig:rtl:arch} shows the implemented blocks and what flows between them.

\paragraph{Source column encoder.} Combinational. It turns the $N$-bit row
bitmap into the ascending list of columns holding atoms, as a prefix
population count followed by a scatter, and reports $K$. It is the smallest
block in the engine and the one that sets its clock period
(\cref{app:impl:timing}).

\paragraph{Target queue store.} $N$ independent circular queues in LUTRAM, one
per column, holding that column's still-unfilled target rows in ascending
order. The pattern is pushed one entry per cycle before the start pulse; during
a run the registered head of each queue is the column's urgency, and a dequeue
mask advances the heads of the columns that were assigned an atom. Holding the
pattern per column, rather than as a bitmap, is what makes urgency a
single-cycle read.

\paragraph{Bitonic sorter.} A 64-wide, 21-stage pipelined sorting network on
the 12-bit key formed from the head row and the column index, so columns
needing an atom high in the array sort first and ties break by column. Columns
with an empty queue take a sentinel key and sink to the tail. The network is
generated by nested elaboration-time loops over the phase and sub-stage
dimensions, and every per-element constant resolves at elaboration, so each
cell synthesises to a comparator, a fixed-polarity multiplexer and a register,
with no runtime control. The sorter is fed a new candidate set every row and
costs the depth of its pipeline, \num{21} cycles, once per row.

\paragraph{Assignment.} On the deployed path, the dynamic program fills one
cell of the $M \times K$ table per cycle from two alternating cost buffers,
records a take bit per cell, and walks the take matrix back to select the
sources. The cost width is 12 bits, which covers the maximum cumulative travel
$N(N-1) = \num{2450}$ for $N = 50$. A final scan returns the chosen sources to
ascending column order before pairing, which is what makes the emitted moves
non-crossing.

\paragraph{Column compression.} Phase~2 reads, for each column, the rows
phase~1 assigned to it and the rows the target asks for, scans both masks
together to produce two ascending lists, and emits one vertical move per
differing pair. Columns with no target rows cost two cycles. Phase~1
guarantees the two lists have equal length when it completes without
abandoning, and the block takes the shorter of the two defensively.

\paragraph{Move FIFO and grid memory.} Both phases write one shared FIFO,
\num{128} entries deep and \num{24} bits wide, one descriptor being the source
and destination row and column; the consumer dequeues over ready/valid, and the
depth absorbs about two and a half rows of moves before back pressure reaches
the state machine. The occupancy grid is an external simple dual-port block RAM
of \num{50}-bit words and \num{64} entries, one-cycle read. Only its write port
is arbitrated: a registered \texttt{running} flag hands it from the loader to
the engine at the start pulse, and the read port belongs to the engine
throughout, which is the mechanism behind the contract of
\cref{app:frontend:contract}.

\subsection{Front-end blocks and the target path}\label{app:rtl:frontend}

Three blocks were written so that the engine reads occupancy that a detector
produced rather than occupancy a testbench injected.

The \emph{site row assembler} packs detector decisions into an $N$-bit
occupancy row and presents each completed row for one cycle. In the
matched-filter build it uses \num{223} LUT and \num{62} FF\meashw{}
(\cref{tab:impl}). The
\emph{occupancy loader} is combinational and drives these rows onto the grid
memory's write port. The \emph{target pattern loader} reads a ROM and supplies
one entry per cycle to the queue preload port, asserting \texttt{done\_o}
after $N_{\mathrm{entries}}+1$ cycles. The default $30\times30$ target uses
\num{900} ROM entries generated from a configuration file; changing the target
requires regenerating this ROM.

Two integration decisions in the system top are worth recording. The detector's
interface widths are parameters rather than constants, because the generated
network's output packing changes between builds: the deployed build gives each
logit its own 64-bit lane, where the earlier one packed both into a single
beat. And an occupancy source multiplexer selects between the detector and an
external injection port, which is what allows the full-grid equivalence test of
\cref{app:frontend:verification} to run in seconds rather than in the twenty minutes a
full grid through the detector would cost.

\subsection{Design decisions}\label{app:frontend:choices}

Three decisions follow from the loop (\cref{app:loop:model}) rather than from
the detector (\cref{tab:decisions}).

\begin{table}[htbp]
  \caption{Architectural decisions of the rearrangement pipeline and the loop constraint
  behind each.}
  \label{tab:decisions}
  \centering
  \small
  \begin{tabular}{@{}L{3.6cm}L{5.4cm}L{5.4cm}@{}}
    \toprule
    Decision & Constraint & Consequence \\
    \midrule
    Detector streams crops; no frame buffer &
    A $2048\times2048\times16$-bit ROI of the qCMOS sensor is \SI{8}{MB}
    (\SI{18.9}{MB} for the full $4096\times2304$ frame), beyond on-chip memory;
    only the neighbourhood of each site ($8\times8$ px in the evaluated
    configuration) carries information &
    Occupancy is emitted incrementally and detection overlaps camera readout \\
    Interface carries one bit per site &
    The engine consumes occupancy alone &
    Matched filter and CNN are drop-in replacements behind the same port \\
    Engine reads an external memory rather than holding the grid &
    Rearrangement mutates the grid as moves are committed, so grid state must
    outlive any single block &
    Loader and engine share one representation with no copy; the next target
    pattern can be preloaded while the previous assembly drains \\
    \bottomrule
  \end{tabular}
\end{table}

\subsection{Clocking and reset}\label{app:frontend:clock}

The implemented block is one synchronous domain and one reset. A single clock port carries a \SI{5.000}{ns}
constraint; the generated detector core's \texttt{ap\_clk} and
\texttt{ap\_rst\_n} are tied to that same clock and reset, and every library
primitive the design instantiates (the grid memory, the move FIFO, the
target queues, every register) is single-clocked. There is no clock-domain
crossing anywhere in the design, and therefore no synchronizer in it.

The blocks operate at different rates within this clock domain. The CNN detector
completes one site every \num{1797} cycles, or
\SI{9.0}{\micro\second}\meassim{}, and the matched-filter lane one site per
cycle, while the row assembler can accept one
site per cycle and the engine reads one grid row per cycle. Handshakes coordinate
these rates: ready/valid on the move stream, \texttt{ap\_start}/\texttt{ap\_done}
for each crop, and the registered \texttt{running} flag for write-port ownership.
These interfaces allow the implemented pipeline to use a single processing clock.

\paragraph{The constraint set.}
The implementation constrains one clock to \SI{5.000}{ns}. It specifies no
additional clock uncertainty, input or output delays, generated clocks, or
false-path and multicycle exceptions\meashw{}. The setup and hold slacks
of the current $8\times8$ matched-filter and CNN builds are reported in
\cref{app:impl:pnr}; the earlier $12\times12$ closure sequence is retained
in \cref{app:impl:timing}. Recovery and removal are not reported.
The \SI{200}{MHz} result therefore applies to the out-of-context block.
Board-level I/O timing, clock distribution and reset-release checks must
be included in the integrated implementation. The board build of
\cref{app:impl:board} adds the board clock through an MMCM and reports
recovery and removal; both are met.

\paragraph{Reset.} The register primitive used throughout asserts
asynchronously on an active-low reset and releases asynchronously as well:
there is no reset synchroniser in the block. Within a single domain this is
benign in simulation and unchecked in implementation, because asynchronous
release is exactly what recovery and removal checks exist to catch and those
checks are unreported. A board-level wrapper should assert reset
asynchronously and release it synchronously, once per domain, and the
implementation run should report recovery and removal alongside setup and hold.
The board build of \cref{app:impl:board} does both: reset asserts when the
MMCM loses lock and releases through a three-stage synchroniser, and recovery
and removal are met with \SI{+0.254}{ns} and \SI{+0.097}{ns} of slack over
\num{15631} endpoints\meashw{}. The memories are deliberately not
reset. The grid memory and target queues have undefined contents until
the preload protocol of \cref{app:frontend:contract} writes every grid row
and target entry before the start pulse. The preload protocol therefore
establishes the initial state.

RTL simulation establishes cycle counts but does not validate board-level
timing or asynchronous reset release. The testbenches use a nominal
\SI{10}{ns} clock; reported latencies are obtained by converting the simulated
cycle counts at the implemented \SI{200}{MHz} clock\meassim{}.

\paragraph{Board-level clock domains.}
The external interfaces introduce additional clock domains
(\cref{app:platform}). Each CXP-6 lane recovers a clock at approximately
\SI{156.25}{MHz}, corresponding to a \SI{6.25}{Gb/s} line rate, 8b/10b
coding, and a 32-bit transceiver interface\roof{}. Elastic buffers align the
lanes before bonding them into a single stream, which crosses into the
processing domain through an asynchronous FIFO. The controller link uses the
PCIe hard block's user clock. Waveform playback and converter clocks belong to
the controller and are outside the QDPU (\cref{app:platform:aod}).

\begin{table}[htbp]
  \caption{Clock structure, implemented and specified. Rates for the unbuilt domains
  follow from the link standards and are not measurements\roof{}.}
  \label{tab:rtl:clock}
  \centering
  \small
  \begin{tabular}{@{}L{2.9cm}L{3.3cm}L{2.5cm}L{4.4cm}l@{}}
    \toprule
    Domain & Source & Rate & What runs in it \\
    \midrule
    Processing clock & Board oscillator through a PLL in the board wrapper & \SI{200}{MHz} constrained; ceiling unmeasured & Detector core, row assembler, loader, write-port arbiter, grid memory, engine, move FIFO \\
    CXP recovered clocks & Per-lane clock recovery in one GTY quad & $\approx$\,\SI{156.25}{MHz} per lane, 4 lanes & Lane alignment, deskew, bonding, link layer\\
    Ingest crossing & Asynchronous FIFO, bonded stream to fabric & --- & Pixel stream into region-of-interest extraction\\
    PCIe user clock & Hard block, fixed by link width and generation & Set by the configuration & Occupancy and move stream to the controller \\
    Outer-loop link & The PCIe user clock when the GPU shares the server; the Ethernet PCS user clock for NVQLink over \SI{10}{GbE} (\cref{app:if:clock,app:if:nvqlink}) & as above; \SI{156.25}{MHz} & Site windows and labels out, model weights in \\
    AOD sample clock & The controller's converters & --- & Waveform playback; outside this device & --- \\
    \bottomrule
  \end{tabular}
\end{table}

\begin{figure}[htbp]
  \centering
  \includegraphics[width=\textwidth]{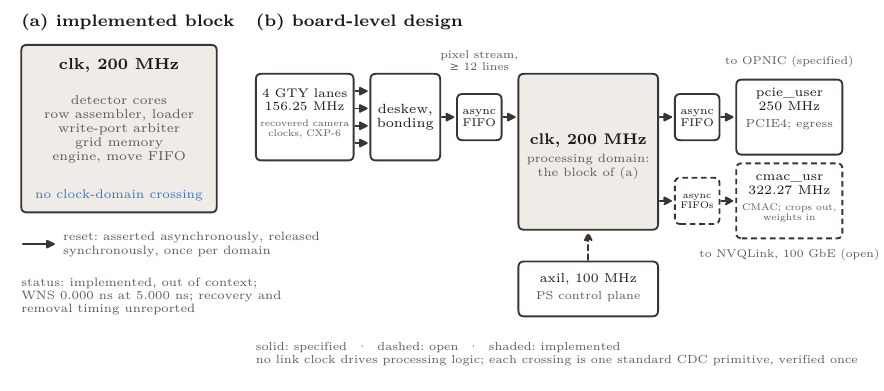}
  \caption{Clock structure. (a)~The implemented block: one \SI{200}{MHz}
  domain, one asynchronously asserted reset, no crossing. (b)~The board-level
  design: four recovered camera clocks aligned and bonded before a single
  crossing into the processing clock, and a second crossing into the
  controller link's user clock. The rates in (b) follow from the link
  standards\roof{}. }
  \label{fig:rtl:clocks}
\end{figure}

\paragraph{Sizing the ingest crossing.} Four CXP-6 lanes carry
\SI{25}{Gb/s} of line rate and, after 8b/10b coding, \SI{20}{Gb/s} of
payload\roof{}. A 128-bit word in the processing clock carries
\SI{25.6}{Gb/s} at \SI{200}{MHz}, so the crossing is a width conversion into
the clock the pipeline already runs at rather than a reason to raise that
clock\roof{}. Full-frame buffering is not possible on chip. A
$2048\times2048$ sixteen-bit frame is $6.7\times10^{7}$ bits against
\SI{38.9}{\mega\nothing}b of block RAM and \SI{23.0}{\mega\nothing}b of
UltraRAM on the device, so the frame does not fit on chip even if the pipeline
used none of it\roof{}. This capacity limit motivates the streaming
design of \cref{tab:decisions}. The site windows do fit: at
\num{2500} sites of $8\times8$ sixteen-bit pixels they are
\SI{2.56}{\mega\nothing}b, \SI{6.6}{\percent} of the device's block
RAM\roof{}. The crossing therefore delivers pixels, region-of-interest
extraction discards the \SI{96}{\percent} of the frame that lies between sites,
and only the windows are retained.

The required crop-buffer capacity depends on the detector service rate and
the camera delivery pattern. An $8\times8$ window of sixteen-bit pixels
contains \num{1024} bits, equivalent to \SI{0.0512}{\micro\second} of
payload at \SI{20}{Gb/s}. The matched filter accepts one window per
\SI{200}{MHz} cycle; the serial CNN invocation takes about
\SI{9.0}{\micro\second} per site (\cref{app:impl:rf}).
The standard-scan readout of a 2048-row ROI takes about \SI{7.4}{ms},
giving a frame-average interval of \SI{2.96}{\micro\second} per site.
Buffer sizing must also account for bursts within a site row.
All \num{2500} crops occupy \SI{2.56}{Mb}, so crop buffering is distinct
from retaining a full raw frame. The line buffer and its delivery pattern
remain to be implemented. The loop budget charges kernel time after crop
acquisition and does not credit overlap with the proposed streaming ingest.

\subsection{Cycle accounting}\label{app:rtl:cycles}

\begin{figure}[htbp]
  \centering
  \includegraphics[width=\textwidth]{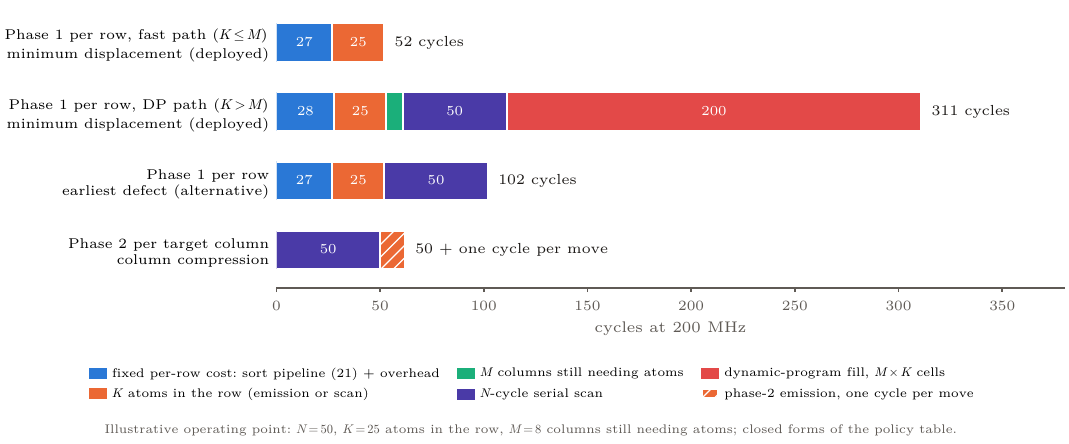}
  \caption{FSM state budgets, from the state tables of the RTL. Phase~1 costs
  the sort pipeline once per row whichever policy is selected; on the deployed
  policy the dynamic program is entered only when the row holds more atoms than
  there are columns still needing them, and it then dominates. Phase~2 is
  dominated by the serial column scan. Bars are design budgets from the
  closed forms of \cref{tab:rtl:policy} at $N=50$, $K=25$, $M=8$, rather than a
  measurement of a particular run.}
  \label{fig:rtl:cycles}
\end{figure}

The engine is a state machine per row and per column, so its cost is a sum of
state residencies rather than a pipeline throughput. \Cref{fig:rtl:cycles}
gives the per-row and per-column budgets at an illustrative operating point,
and \cref{tab:rtl:policy} the closed forms.

Three properties follow. The per-row cost is dominated by fixed work, not by
the number of atoms moved: at $N=50$ and half fill, \num{26} of the \num{52}
cycles of a row are the sort pipeline and the fixed overhead, and the emission
of \num{25} moves costs \num{26}. The dynamic program is the only superlinear
term, $M \times K$ cycles, and it is entered only in the late part of a sweep,
when the queues have drained enough that the row holds more atoms than the
remaining columns need. And phase~2 costs a serial scan of $N$ cycles per
target column whether or not that column has anything to move, so its total
grows as $N^{2}$ over a sweep although it performs no search at all
(\cref{fig:rtl:sweeps}).

For the array of \cref{app:frontend:verification} (\num{1270} atoms at half
fill, a \num{961}-site target), the measured engine time is \num{29184}
cycles\meassim{}, about \SI{0.15}{ms} at \SI{200}{MHz}; this is the
reference engine run listed in \cref{tab:latency}.

\subsection{Functional verification}\label{app:frontend:verification}

Integration is verified end to end in RTL simulation before silicon: a cocotb
testbench drives the reference top (loader, write-port multiplexer, occupancy
BRAM, engine) under Verilator, with occupancy produced by the detector rather
than injected by the testbench. Two cases exercise the occupancy contract of
\cref{app:frontend:contract}, one on each side of the feasibility boundary.

\begin{itemize}[leftmargin=1.2em,itemsep=4pt,topsep=3pt]\sloppy
  \item \textbf{Feasible array: the engine completes the pattern.} On an array
        holding more atoms than the target pattern requires, the engine
        consumes the detector's occupancy and asserts \texttt{done\_o} after
        emitting \num{1584} moves for one of the testcases. Each of the \num{1584} is checked against
        the occupancy state at the moment it is issued, on three conditions:
        the source site holds an atom, the destination site belongs to the
        target pattern, and both lie within the array bounds. All \num{1584}
        pass. Similar test patterns were generated and verified to be functionally correct for each generated move.

  \item \textbf{Infeasible array: the engine declines cleanly.} On an array
        holding fewer atoms than the target pattern requires, the engine
        asserts \texttt{abandoned\_o} rather than \texttt{done\_o}. It neither
        hangs nor emits a partial plan, so the downstream controller is never
        handed a move schedule it cannot complete.
\end{itemize}

One further check bounds what the integration itself changed. In the feasible
case the move count is identical to that of the standalone engine driven from
the same occupancy by a software harness, so the engine cannot distinguish
occupancy that reached it through the shared memory from occupancy injected
directly by a testbench: the three blocks placed in front of it
(\cref{app:frontend:scope}) are transparent to it. The rearrangement algorithm is common to
both paths, and the standalone and integrated engines have byte-identical port
lists.

\begin{table}[htbp]
  \caption{Verification levels. All simulation is Verilator with cocotb; the
  software model is the reference implementation of the same two-phase
  algorithm.}
  \label{tab:rtl:verify}
  \centering
  \small
  \begin{tabular}{@{}L{3.1cm}L{5.0cm}L{5.9cm}@{}}
    \toprule
    Level & What it drives & What it establishes \\
    \midrule
    Directed unit tests & Eight cases on the engine: reset, empty grid, atoms with no targets, single horizontal move, atom already placed, vertical move, several atoms in one row, back pressure held throughout & Each state-machine path is entered and leaves the move interface in a defined state; back pressure stalls emission rather than dropping moves \\
    Move-for-move equivalence & Seeded random cases in two testbenches, up to a full $50\times50$ grid: occupancy injected through the bypass port, the target preloaded, the plan captured beat by beat & The multiset of emitted moves equals the software model's exactly, and applying either to the initial grid yields the same final grid \\
    Interface co-simulation & Occupancy through the loader, the write-port arbiter and the external memory into the engine, feasible and infeasible arrays & The arbitrated path reproduces the engine's standalone behaviour; \texttt{done\_o} on a feasible array, \texttt{abandoned\_o} with no partial plan on an infeasible one \\
    Detector in the loop & Simulated site images through the generated detector netlist into the assembler & The assembled occupancy matches ground truth on the sampled rows, \SI{98}{\percent} against a \SI{88}{\percent} gate\meassim{}; a logit-packing or adapter error collapses this to chance \\
    \bottomrule
  \end{tabular}
\end{table}

The move-equivalence tests use a cocotb harness with Verilator. Each test case
is specified by NumPy occupancy and target arrays, which are converted into the
row bitmasks and queue-preload entries required by the RTL. The same arrays are
passed to the reference Python implementation, \texttt{run\_tetris\_plan},
using the minimum-displacement policy. The hardware and reference model are
therefore evaluated on the same inputs.

The comparison checks the emitted moves as a multiset and verifies that applying
the two plans produces the same final occupancy grid. Move ordering is not
asserted. These tests cover seeded cases up to a full $50\times50$ grid; they
do not establish equivalence for all possible inputs. Further verification could
use bounded formal analysis of a row assignment or constrained-random testing
with coverage of the assignment policy.

In the full-grid equivalence case, the $50\times50$ array contains
\num{1270} atoms and the centred $31\times31$ target contains \num{961}
sites. The hardware and software plans contain identical move multisets, and
both produce a defect-free target\meassim{}. This test evaluates the
assignment policy in addition to individual move legality. Occupancy is injected
through the bypass port; the detector and row assembler are tested separately
at the fourth verification level. A single simulation of all \num{2500}
sites through both detection and rearrangement has not been performed.

\section{Detector Datapath and Detection Accuracy}\label{app:detector}

Two detectors sit behind the occupancy contract: a calibrated matched filter
and a small CNN. Both have measured cycle counts and integrated
place-and-route results at \SI{200}{MHz} (\cref{app:impl}). The
\SI{668}{MHz} standalone result applies to forty fixed-template
matched-filter lanes. The loop model uses the runtime-loadable matched filter
(\cref{app:loop:closed}); the comparisons below also report CNN performance.

\paragraph{Data.} All training and evaluation images are generated with an
open-source neutral-atom imaging
simulator~\cite{winklmann2023imagesim,capstum2024simulator}, configured for the
Hamamatsu ORCA-Quest qCMOS sensor. The simulator models Poissonian shot noise,
dark current, structured (row, column, flicker) readout noise, background
illumination, and a diffraction PSF with Zernike aberrations; its authors note
that it characterises aberrations once across the field, places atoms only at
discrete sites, and does not reproduce every CMOS effect~\cite{winklmann2023imagesim}.
No real camera frames are used (\cref{sec:limits}). \Cref{tab:simcfg} lists the
configuration; every camera parameter is set explicitly. The scattering rate
is close to a previously reported \SI{41}{kHz}~\cite{covey2019repeated}.

We correct one defect in the simulator's CMOS readout. It holds each pixel's
photoelectron count in an integer and adds the real-valued row, column and
flicker noise to it, so every addition truncates toward zero; at a few
photoelectrons per pixel this discards most of the signal (\SI{59}{\percent} of
an atom's detected photons at \SI{30}{ms} in our earlier configuration). We hold the
count in floating point until readout, a one-line change; the unmodified build
reproduces our earlier data (27.6 against 25.6 detected photons per atom at
\SI{30}{ms}), and the patched build delivers the photons the configuration
implies. Every detector result in this paper uses the patched build. Earlier
drafts used the unmodified build with an earlier configuration
(\SI{0.1}{\micro\metre\per px} sampling, the simulator's default structured
noise at about eight times the sensor datasheet), which together left about a
fifth of the detected photons per atom at a given exposure; their accuracy
figures are superseded.

\begin{table}[htbp]
  \caption{Imaging-simulator configuration for all detector results.}
  \label{tab:simcfg}
  \centering
  \small
\begin{tabular}{@{}L{5.8cm}L{10.2cm}@{}}
    \toprule
    Parameter & Value \\
    \midrule
    Atom, imaging line & $^{88}$Sr, \SI{461}{nm} \\
    Scattering rate $\Gamma$ & \SI{40000}{photons\per s} \\
    Objective NA; collection $\eta=(1-\sqrt{1-\mathrm{NA}^2})/2$ & 0.65; 0.120 \\
    Quantum efficiency & 0.86 \\
    Detected photon rate per atom & \SI{4.13}{\per\milli\second} \\
    Sensor & ORCA-Quest qCMOS, \SI{4.6}{\micro\metre} pixels \\
    Object-plane sampling & \SI{0.4}{\micro\metre\per px} (magnification 11.5) \\
    Read noise & \SI{0.43}{e^-} rms (Standard scan) \\
    Row, column, flicker noise & \SI{0.02}{e^-}, \SI{0.02}{e^-}, \SI{0.01}{e^-} \\
    Gain, bias & \SI{0.11}{e^-\per ADU}, 200 ADU \\
    PSF & diffraction-limited with device Zernike terms, \SI{0.31}{\micro\metre} atom blur \\
    Background & \SI{3}{e^-\per px\per s}; \SI{100}{e^-\per px\per s} (high) \\
    Arrays & $14\times14$ at \SI{4}{\micro\metre}; $28\times28$ at \SI{2}{\micro\metre}; $36\times36$ at \SI{1.5}{\micro\metre} \\
    Fill, survival & 0.5, 1 (clean labels) \\
    Exposure & \SIrange{1}{50}{ms} \\
    Window & $8\times8$ px (\SI{3.2}{\micro\metre}) \\
    Sites per point & \num{43120} train, \num{43120} test (sparse array) \\
    \bottomrule
  \end{tabular}
\end{table}

\paragraph{Crop size.} The detector operates on a fixed-size pixel window
centred on each calibrated site position. The window size is a compile-time
parameter set by the imaging geometry: it must hold the atom's image with
margin and should exclude the neighbouring sites. At the
\SI{0.4}{\micro\metre\per px} object-plane sampling of \cref{tab:simcfg}
(\SI{4.6}{\micro\metre} camera pixels at $11.5\times$ magnification) we use
$8\times8$ pixels, \SI{3.2}{\micro\metre}. The window extends at least 3~px
(\SI{1.2}{\micro\metre}) from the site in every direction, about four times
the \SI{0.31}{\micro\metre} rms width of the simulated atom image, and it
excludes the neighbouring sites at \SI{2}{\micro\metre} pitch and above.

Earlier drafts used $12\times12$ pixels, chosen when the simulated sampling
was \SI{0.1}{\micro\metre\per px}, where that window spanned
\SI{1.2}{\micro\metre}. At the present sampling a $12\times12$ window spans
\SI{4.8}{\micro\metre}: at \SI{2}{\micro\metre} pitch it contains the
neighbouring sites, and at every pitch it adds background and read-noise
pixels that carry little signal. We ran every configuration of this appendix
at both sizes\simu{}. The $8\times8$ window holds about \SI{95}{\percent} of
the photons the $12\times12$ one does. The matched filter is insensitive to
the change, because its template falls to near zero away from the atom: its
fidelity moves by at most \SI{0.4}{\percent} between the two windows at
\SI{3}{ms} and above. The CNN is not: the smaller window raises the
\texttt{1-8-16-16} network from \SI{94.05}{\percent} to \SI{95.40}{\percent} at
\SI{2}{ms} on the sparse array and from \SI{95.13}{\percent} to
\SI{98.29}{\percent} at \SI{1.5}{\micro\metre} pitch and \SI{5}{ms}, because a
network whose global average pool discards where in the window the light falls
cannot separate a neighbour inside the window from the atom at its centre.

For the matched filter the optimum is broad. \Cref{tab:window} adds a
$7\times7$ window to the two above: from \SI{3}{ms} on, the three sizes are
within \SI{0.4}{\percent} of one another in every configuration, and no size
is best in all of them, because a template estimated from labelled windows
gives pixels without signal a weight near zero whatever the window holds.
Accuracy therefore does not fix the size, and three other considerations do.
The window must leave margin for an atom image broader than the simulated one
(real optics add aberration and defocus) and for a site centre that drifts by
a fraction of a pixel between calibrations; $8\times8$ leaves at least 3~px
on every side, and windows below $7\times7$ begin to cut into the image
itself. It should not reach the neighbouring site, which at
\SI{2}{\micro\metre} pitch is 5~px away: windows up to $9\times9$ keep it out,
and $10\times10$ or more brings it in. And it should suit the
learned lane, which shares the region-of-interest extraction and gains
markedly from a smaller window (above). $8\times8$ meets all three, and its
64-term dot product is a six-level adder tree; the DSP cost, 61 slices per
lane as built (\cref{tab:mfpnr}) against at most 49 for $7\times7$, is
immaterial because one lane keeps pace with the camera
(\cref{app:detector:law}). An even window cannot be centred on
a pixel: the site pixel sits at index~4 of 0 to 7, three pixels from one edge
and four from the other, and the region-of-interest extraction must use the
convention the template was calibrated with.

Misregistration does not separate the candidates either
(\cref{tab:misreg})\simu{}. Windows of $5\times5$ to $10\times10$ were cut from
the $12\times12$ crops, the template and threshold calibrated on centred
windows, and the test windows shifted. A half-pixel shift costs
\SIrange{0.1}{1.3}{\percent}, a one-pixel shift \SIrange{0.8}{5.2}{\percent}
and a two-pixel shift \SIrange{16}{28}{\percent}, nearly the same at every window
size, because the loss comes from the atom's image moving off the template's
peak, which a larger window does not help. What the matched filter needs is
site registration to a fraction of a pixel, which places the requirement on
the calibration of site positions rather than on the window.

\begin{table}[htbp]
  \caption{Matched-filter balanced fidelity (\%) under site misregistration,
  by window\simu{}. Template and threshold calibrated on centred windows;
  ``$\pm0.5$'' and ``$\pm1$'' are the worst of the four shifts along the axes
  by that many pixels. Half-pixel shifts are emulated by averaging the two
  neighbouring integer-shifted windows, which also averages pixel noise and
  slightly flatters those columns. Windows above $10\times10$ leave no room
  for a shift inside the $12\times12$ crop.}
  \label{tab:misreg}
  \centering
  \small
  \begin{tabular}{@{}lrrrrrrrrr@{}}
    \toprule
    & \multicolumn{3}{c}{Sparse, \SI{5}{ms}} & \multicolumn{3}{c}{\SI{1.5}{\micro\metre}, \SI{5}{ms}} & \multicolumn{3}{c}{High bg, \SI{5}{ms}} \\
    \cmidrule(lr){2-4}\cmidrule(lr){5-7}\cmidrule(l){8-10}
    Window & centred & $\pm0.5$ & $\pm1$ & centred & $\pm0.5$ & $\pm1$ & centred & $\pm0.5$ & $\pm1$ \\
    \midrule
    $5\times5$ & 99.74 & 99.61 & 98.95 & 98.63 & 97.66 & 94.14 & 98.06 & 97.47 & 94.22 \\
    $6\times6$ & 99.77 & 99.61 & 98.95 & 98.52 & 97.26 & 93.36 & 98.06 & 97.42 & 94.14 \\
    $7\times7$ & 99.79 & 99.62 & 98.93 & 98.39 & 97.30 & 93.70 & 98.11 & 97.53 & 94.65 \\
    $8\times8$ & 99.78 & 99.64 & 98.85 & 98.35 & 97.23 & 93.58 & 98.14 & 97.54 & 94.62 \\
    $9\times9$ & 99.79 & 99.67 & 98.99 & 98.23 & 97.08 & 93.53 & 98.13 & 97.54 & 94.65 \\
    $10\times10$ & 99.79 & 99.64 & 99.00 & 98.23 & 97.05 & 93.53 & 98.12 & 97.58 & 94.73 \\
    \bottomrule
  \end{tabular}
\end{table}

\begin{table}[htbp]
  \caption{Matched-filter fidelity (\%) against window size\simu{}.
  Configurations as in \cref{tab:xtalk}; each window is generated, calibrated
  and tested separately.}
  \label{tab:window}
  \centering
  \small
  \begin{tabular}{@{}lrrr@{}}
    \toprule
    Configuration & $7\times7$ & $8\times8$ & $12\times12$ \\
    \midrule
    \SI{4}{\micro\metre}, \SI{3}{ms}            & 97.68 & \textbf{97.69} & 97.32 \\
    \SI{4}{\micro\metre}, \SI{5}{ms}            & 99.65 & 99.77 & \textbf{99.79} \\
    \SI{4}{\micro\metre}, high bg., \SI{5}{ms}  & 98.06 & 97.86 & \textbf{98.13} \\
    \SI{2}{\micro\metre}, \SI{5}{ms}            & \textbf{99.65} & 99.57 & 99.60 \\
    \SI{1.5}{\micro\metre}, \SI{5}{ms}          & \textbf{98.44} & 98.22 & 98.19 \\
    \bottomrule
  \end{tabular}
\end{table}

Published per-site readout networks use comparable
windows: $10\times10$ pixels~\cite{phuttitarn2024ml} and
$28\times28$ crops with a boundary parameter swept from 2 to
14~px~\cite{kent2025matchedfilter}. Changing the window size changes the
multiply--accumulate count quadratically: at $8\times8$ the matched filter is
64 multiply--accumulates per site and the CNN \num{226944}, \SI{45}{\percent}
of the $12\times12$ count. The implemented RTL of both detectors
(\cref{app:impl}) is built for the $8\times8$ window.

\paragraph{Positioning.}
Matched filters can reach CNN-level readout error with two orders of
magnitude fewer parameters and four orders of magnitude fewer
multiplications~\cite{kent2025matchedfilter}\lit{}.
Classical estimators have also been compared against a Cram\'er--Rao
bound using the same simulator~\cite{winklmann2024detection}\lit{}.
This appendix reports the corresponding accuracy comparison for our
configuration; \cref{app:impl} gives the implemented cost of each detector,
and \cref{app:loop} converts the difference into atoms per assembly.
A global non-linear least-squares estimator fits site photon counts
through the optical point-spread function using weighted Gauss--Newton
iterations. The study reports its largest advantage at long exposures
and discusses its computational cost~\cite{winklmann2024detection}\lit{}.
We use the matched filter as the hardware baseline in this work.
The global estimator has not been implemented or benchmarked in the
present FPGA flow.

\subsection{Matched-filter lane}\label{app:detector:mf}

One site's decision is a dot product of the site window against a calibrated
template and a threshold, $\mathrm{occ} = (\sum_i \mathrm{pix}_i \cdot
\mathrm{tmpl}_i) > \Theta$; at the $8\times8$ window this is 64
multiply-accumulates and one comparison. The template is the
difference between mean occupied and mean empty crops and the threshold the
one that minimises $P_{\mathrm{FP}} + P_{\mathrm{FN}}$ on labelled windows; both
are calibrated in the outer loop and stored as \texttt{int8} coefficients and
one integer. Recalibration for a new optical setup replaces 64 coefficients and
one threshold. The lane is built in two forms. With the constants fixed at
synthesis, a ROM and a parameter, the tool folds the constant multiplies, but
recalibration is a rebuild. With \texttt{LOADABLE} set, each lane holds its
coefficients and threshold in registers written through a configuration
port, 65 register writes between frames and no rebuild, at the cost of 64
general multipliers per lane (\cref{tab:mfpnr}). The loop needs the second
form, since the outer loop refits the template as the device drifts, and the
integrated pipeline uses it (\cref{app:impl:pnr}). The calibration used here, on the sparse
\SI{5}{ms} data of \cref{app:detector}, scales the template to \texttt{int8}
and picks the best integer threshold on the training split; the integer
filter scores \SI{99.770}{\percent} on the test split against
\SI{99.773}{\percent} in floating point\meassim{}. The template is the simplest learned model the QDPU hosts: it is
fit from labelled measurement data and refit as the device drifts, exactly as
the CNN's weights would be, with no gradient training.

The array wrapper instantiates $P$ lanes and accepts $P$ crops per cycle.
Fixed-template lanes trade area for throughput at approximately constant
$F_{\max}$. In the loadable variants, the shared configuration-control path
also affects timing as $P$ grows (\cref{tab:mfpnr}). Each lane is fully pipelined: an input register, a product
register, one register per level of the adder tree and a registered compare,
nine cycles from window to bit at $8\times8$ (six adder levels) and eleven at
$12\times12$. The multipliers map to DSP slices and the
adder tree to fabric; letting the tool pack the adders into DSP slices as well
fails timing at $P=20$ (\cref{app:impl:pnr}). The input bus grows as
$P \times \mathit{pixels} \times 16$ bits, \num{40960} bits per cycle at
$P = 40$ for the $8\times8$ build, and must be driven from registers, which
the ROI line buffer does by construction; a line buffer delivering forty
windows per cycle is not yet built.

\subsection{Throughput law and bit-exactness}\label{app:detector:law}

A full $n$-site sweep costs $\lceil n/P \rceil + L$ cycles, where $L$ is the
lane's pipeline depth, nine cycles for the $8\times8$ lane. The law holds at
every point of \cref{tab:cycles} under Verilator with a cocotb harness (twelve
configurations, $n \in \{2500, 10000\}$ and $P$ from 1 to 40) with the
calibrated $8\times8$ template and threshold\meassim{}; the earlier
single-register lane ($L = 1$) and the pipelined $12\times12$ lane ($L = 11$)
obeyed it in the same harness. Every lane decision matches a software golden
model evaluating $(\mathrm{dot}(\mathrm{crop}, \mathrm{tmpl}) > \Theta)$ in
integer arithmetic, with zero mismatches; the test crops are concentrated
around the decision boundary, where an arithmetic error would show. The
runtime-loadable lane (\cref{app:detector:mf}) passes the same test at
$P \in \{1, 20, 40\}$ with the template written through its configuration
port, and again after a second, unrelated template and threshold are written
in its place\meassim{}. Bit-exactness is the relevant
criterion because the output bit is treated as ground truth downstream.

The sweep is never the bottleneck. At \SI{500}{MHz} a $100\times100$ array
is classified in $\lceil 10^4/20 \rceil + 9 = 509$ cycles,
\SI{1.02}{\micro\second}, at $P = 20$ and in 259 cycles,
\SI{0.52}{\micro\second}, at $P = 40$; a $2048\times2048\times16$-bit region
of interest requires at least \SI{3.4}{ms} of payload transfer at
\SI{20}{Gb/s}, before packet overhead. The sensor-readout budget is longer
(\cref{tab:if:cxp:budget}), so a single lane ($P = 1$,
\SI{5.0}{\micro\second} for 2500 sites) already keeps pace with the camera.

The two array sizes in \cref{tab:cycles} bracket current practice. Arrays of
1225 and 2000 sites are loaded and rearranged in published
systems~\cite{norcia2024iterative,pichard2024rearrangement2000}, a 6100-site
tweezer array has been demonstrated~\cite{manetsch2025tweezer6100}, and
$10^4$ sites is the scale at which \cref{app:platform:scaling} locates the
limits of the present data structures and of the optics. Sweep cost is linear
in $n$ at fixed $P$, for the matched filter here and for the CNN
(\cref{app:platform:scaling}), so the $n = 10\,000$ rows give the cost of the
same hardware one array generation out.

\begin{table}[htbp]
  \caption{Measured cycle counts for the pipelined $8\times8$ matched-filter
  lane with the calibrated template ($L = 9$).
  $\lceil n/P \rceil + 9$ holds exactly with zero lane mismatches against the
  software golden\meassim{}.
  Wall-clock is quoted at the \SI{200}{MHz} of the integrated design
  (\cref{app:impl})\roof{}; the lane routes at \SI{668}{MHz} to
  \SI{674}{MHz} (\cref{tab:mfpnr}). Sweep time is the time to classify all $n$
  sites at once.}
  \label{tab:cycles}
  \centering
  \small
  \begin{tabular}{@{}rrrrrr@{}}
    \toprule
    Array & Sites $n$ & Lanes $P$ & Cycles & Mismatches & Sweep time \\
    \midrule
    $50\times50$   & 2\,500  & 1  & 2\,509  & 0 & \SI{12.5}{\micro\second} \\
    $50\times50$   & 2\,500  & 2  & 1\,259  & 0 & \SI{6.30}{\micro\second} \\
    $50\times50$   & 2\,500  & 4  & 634     & 0 & \SI{3.17}{\micro\second} \\
    $50\times50$   & 2\,500  & 8  & 322     & 0 & \SI{1.61}{\micro\second} \\
    $50\times50$   & 2\,500  & 10 & 259     & 0 & \SI{1.30}{\micro\second} \\
    $50\times50$   & 2\,500  & 20 & 134     & 0 & \SI{0.67}{\micro\second} \\
    $50\times50$   & 2\,500  & 40 & 72      & 0 & \SI{0.36}{\micro\second} \\
    \midrule
    $100\times100$ & 10\,000 & 1  & 10\,009 & 0 & \SI{50.0}{\micro\second} \\
    $100\times100$ & 10\,000 & 4  & 2\,509  & 0 & \SI{12.5}{\micro\second} \\
    $100\times100$ & 10\,000 & 10 & 1\,009  & 0 & \SI{5.05}{\micro\second} \\
    $100\times100$ & 10\,000 & 20 & 509     & 0 & \SI{2.55}{\micro\second} \\
    $100\times100$ & 10\,000 & 40 & 259     & 0 & \SI{1.30}{\micro\second} \\
    \bottomrule
  \end{tabular}
\end{table}

\subsection{Learned lane: CNN}\label{app:detector:learned}

Sub-pixel position jitter and neighbour crosstalk can reduce the accuracy of a
fixed linear template (\cref{app:detector:xtalk}). We therefore evaluate a
small CNN, \texttt{Conv(1-8-16-16)} with global average pooling and a
two-layer dense head, as the learned readout model
(\cref{sec:unit:definition}). It uses \num{226944} MACs per site;
\cref{tab:cnn} gives its layer structure. The network takes one
$8\times8$ crop and returns an occupancy decision. All tensors use 12-bit
per-tensor quantisation (\cref{app:quant}).

\begin{table}[htbp]
  \caption{Layer structure of the CNN for an $8\times8$ input
  window, from the model definition (\texttt{atom\_readout/model/deploy/qat\_train.py},
  \texttt{QAT\_CONV=8,16,16}, \texttt{QAT\_CROP=8}, first-layer stride 1). Every
  convolution is followed by batch normalisation and ReLU. MACs are per site and
  count the convolution and dense multiply--accumulates; batch normalisation is
  a per-channel scale and shift and is not counted. Parameters do not depend on
  the window and include the four batch-normalisation values per channel
  (scale, shift, running mean and variance), 160 in all, of which 80 are
  trained. The same topology at \texttt{1-32-64-64} gives \num{3561600} MACs,
  $15.7\times$ more; at the earlier $12\times12$ window the two are
  \num{509184} and \num{8008320}.}
  \label{tab:cnn}
  \centering
  \small
  \begin{tabular}{@{}llrr@{}}
    \toprule
    Layer & Output shape & Parameters & MACs \\
    \midrule
    Input crop                                & $8\times8\times1$  & --     & --       \\
    Conv $3\times3$, 8 ch, same pad, BN, ReLU  & $8\times8\times8$  & 80 $+$ 32    & 4\,608   \\
    Conv $3\times3$, 16 ch, same pad, BN, ReLU & $8\times8\times16$ & 1\,168 $+$ 64 & 73\,728  \\
    Conv $3\times3$, 16 ch, same pad, BN, ReLU & $8\times8\times16$ & 2\,320 $+$ 64 & 147\,456 \\
    Global average pool ($8\times8$)          & 16                 & --     & --       \\
    Dense, 64 outputs, ReLU                   & 64                 & 1\,088 & 1\,024   \\
    Dense, 2 outputs                          & 2                  & 130    & 128      \\
    \midrule
    Total                                     &                    & 4\,946 & 226\,944 \\
    \bottomrule
  \end{tabular}
\end{table}

Channel width has little influence on accuracy. The \texttt{1-32-64-64}
network, $15.7\times$ the MACs, is within \SI{0.1}{\percent} of the
\texttt{1-8-16-16} network on every sparse configuration from \SI{3}{ms} up,
in full precision and under 12-bit quantisation-aware training
(\cref{tab:floor,tab:qat})\simu{}. It leads only below \SI{3}{ms}, by up to
\SI{1.4}{\percent} at \SI{1}{ms}, and by \SI{0.4}{\percent} at
\SI{1.5}{\micro\metre} pitch (\cref{tab:xtalk}). Accuracy is exposure-limited
rather than capacity-limited.

The network must also be trained for the exposure it will see. One model
trained on every configuration and exposure from \SI{1}{} to \SI{50}{ms}
together reaches \SI{96.7}{\percent} overall, and its accuracy on the sparse
array falls as exposure grows (\SI{98.7}{\percent} at \SI{10}{ms},
\SI{96.7}{\percent} at \SI{50}{ms}), every error a false positive: at
\SI{50}{ms} an empty window collects about eight electrons of stray light,
more than a \SI{1}{ms} atom delivers, so a decision rule sensitive enough for
short exposures calls long-exposure empties occupied. Restricting training to
the \SIrange{3}{7}{ms} training band closes most of the gap, and one model
per configuration closes it entirely (\cref{app:quant:qat}); a model trained
in one band fails outside it (\SI{52}{\percent} on the sparse array at
\SI{50}{ms} for the band model), so a CNN deployment requires one parameter set,
with its input normalisation, per exposure and configuration. The matched
filter has the same requirement in the form of one template and threshold per
configuration, at 64 coefficients rather than \num{4946} parameters.

Reducing the MAC count primarily reduces DSP usage;
at fixed reuse factor, it has little effect on the per-crop latency of the
streaming implementation (\cref{app:impl:rf}). The resulting resource margin
can instead support lower reuse factors or replicated detector cores. These
choices are evaluated within the shared budget for detection, rearrangement,
and decoding (\cref{app:impl:coresident}).

\subsection{Accuracy against the photon limit}\label{app:detector:floor}

We report accuracy as per-site imaging fidelity, the probability of
identifying a site's occupancy correctly, balanced over the two classes:
$F = 1 - (P_{\mathrm{FP}} + P_{\mathrm{FN}})/2$. At the fill fraction of 0.5
used throughout it equals the plain fraction of sites classified correctly.

The reference is an ideal observer for an independent-pixel model. In pixel
$i$, the reading in electrons is $x_i=k_i+n_i$, with
$k_i\sim\mathrm{Poisson}(\mu_i^{c})$ and
$n_i\sim\mathcal{N}(0,\sigma^2)$, where $\sigma=\SI{0.43}{e^-}$ and
$\mu_i^{c}$ is the mean over training sites of class $c$.
Thresholding $\sum_i\log[p(x_i\mid\mu_i^{1})/p(x_i\mid\mu_i^{0})]$
is Bayes-optimal under this specified model. Its reference fidelity is
estimated by Monte Carlo with $10^5$ samples per class.

The detector accuracies are evaluated on the imaging simulator, which also
contains structured noise and other dependencies absent from this
independent-pixel reference. The reference is therefore not a strict upper
bound on the tabulated detector fidelities; the CNN can exceed it, as in
the \SI{1}{ms} row of \cref{tab:floor}. In dense arrays, neighbour occupancy
also enters the reference only through its mean. Earlier drafts used an
independent-Gaussian reference; that value was not a bound on the correlated
simulator data and is not used here.

\Cref{tab:floor} compares the detectors with the ideal observer as a function
of exposure\simu{}. Below about twelve detected photons per atom
(\SI{3}{ms}) every detector is photon-limited and the CNN stays closer to the
ideal observer than the matched filter: at \SI{2}{ms}, \SI{95.4}{\percent}
(\texttt{1-8-16-16}) and \SI{96.0}{\percent} (\texttt{1-32-64-64}) against the
filter's \SI{91.6}{\percent} and the ideal observer's \SI{96.5}{\percent}. At
\SI{5}{ms}, nineteen photons per atom, the CNN reaches \SI{99.89}{\percent}
and the filter \SI{99.77}{\percent}, against \SI{99.93}{\percent} for the ideal
observer; both detectors pass \SI{99.9}{\percent} by \SI{7}{ms} and
\SI{99.99}{\percent} by \SI{10}{ms}. The CNN's advantage is therefore confined
to the photon-starved end, where no detector is accurate enough to assemble an
array (\cref{app:loop:closed}), and amounts to about \SI{1}{ms} of exposure at
the fidelities the loop needs. Imaging loss grows as
$1 - \exp(-t_{\mathrm{exp}}/\tau_{\mathrm{img}})$ per image
(\cref{app:loop:model}), so the exposure a detector saves is paid back in
atoms; \cref{app:loop} converts the difference, and the inference time spent
to obtain it, into atoms.

\begin{table}[htbp]
  \caption{Per-site imaging fidelity against exposure on a sparse array
  (\SI{4}{\micro\metre} pitch) with the $8\times8$ window\simu{}.
  Fidelity is $1-(\mathrm{FP}+\mathrm{FN})/2$ on \num{43120} test sites per row.
  The ideal observer is the Bayes-optimal detector of the independent-pixel
  Poisson--Gaussian model. An entry ``$>$x'' had no errors and gives the
  \SI{95}{\percent} bound. Photons are detected photoelectrons per atom inside
  the window. CNN columns are full-precision models trained per row;
  \cref{tab:qat} gives the 12-bit models.}
  \label{tab:floor}
  \centering
  \small
  \begin{tabular}{@{}lrrrrr@{}}
    \toprule
    & Photons & Ideal & Matched & CNN & CNN \\
    Exposure & per atom & observer & filter & 1-8-16-16 & 1-32-64-64 \\
    \midrule
    \SI{1}{ms} & 4 & 86.09 & 81.99 & 85.73 & \textbf{87.11} \\
    \SI{2}{ms} & 8 & 96.53 & 91.63 & 95.40 & \textbf{95.97} \\
    \SI{3}{ms} & 12 & 99.06 & 97.69 & 98.74 & \textbf{98.81} \\
    \SI{5}{ms} & 19 & 99.93 & 99.77 & \textbf{99.89} & 99.88 \\
    \SI{7}{ms} & 27 & 99.997 & 99.96 & 99.98 & \textbf{99.98} \\
    \SI{10}{ms} & 39 & $>$99.99 & $>$99.99 & $>$99.99 & $>$99.99 \\
    \SI{20}{ms} & 77 & $>$99.99 & $>$99.99 & $>$99.99 & $>$99.99 \\
    \SI{50}{ms} & 192 & $>$99.99 & $>$99.99 & $>$99.99 & $>$99.99 \\
    \bottomrule
  \end{tabular}
\end{table}

\subsection{Crosstalk and background}\label{app:detector:xtalk}

\Cref{tab:xtalk} varies the site pitch from \SI{4}{\micro\metre} to
\SI{1.5}{\micro\metre} and raises the background thirty-fold, at the two
exposures where detectors still differ\simu{}. With the $8\times8$ window
neither moves the detectors far apart. At \SI{2}{\micro\metre} crosstalk
costs every detector about \SI{0.2}{\percent} at \SI{5}{ms} and almost
nothing at \SI{10}{ms}. At \SI{1.5}{\micro\metre}, where the neighbours fall
inside the window, the \texttt{1-32-64-64} network leads the matched filter by
\SI{0.5}{\percent} at \SI{5}{ms} (\SI{98.70}{\percent} against
\SI{98.22}{\percent}) and the \texttt{1-8-16-16} network by \SI{0.1}{\percent};
at \SI{10}{ms} every detector is above \SI{99.8}{\percent}. Under high
background the matched filter leads both networks by \SI{0.3}{\percent} at
\SI{5}{ms}. The 12-bit models trained per configuration
(\cref{tab:qat}) widen the dense-array margin to \SIrange{0.6}{0.8}{\percent}
at \SI{1.5}{\micro\metre} and close the background gap. Crosstalk is the one
regime in which the learned lane holds a consistent, if small, accuracy
advantage; whether it is worth its inference time there is a question for
\cref{app:loop}, which models only the sparse array. With the earlier
$12\times12$ window the order was reversed at \SI{1.5}{\micro\metre}, with the
\texttt{1-8-16-16} network \SI{3}{\percent} behind the filter
(\cref{app:detector}, crop size).

\begin{table}[htbp]
  \caption{Per-site imaging fidelity by array pitch and background with the
  $8\times8$ window\simu{}; columns and conventions as in
  \cref{tab:floor}. High background is \SI{100}{e^-\per px\per s}, the
  baseline \SI{3}{e^-\per px\per s}. At \SI{2}{} and \SI{1.5}{\micro\metre}
  the ideal observer treats neighbour light as its mean and is optimistic.}
  \label{tab:xtalk}
  \centering
  \small
  \begin{tabular}{@{}lrrrrr@{}}
    \toprule
    & Photons & Ideal & Matched & CNN & CNN \\
    Configuration & per atom & observer & filter & 1-8-16-16 & 1-32-64-64 \\
    \midrule
    \SI{4}{\micro\metre}, \SI{5}{ms} & 19 & 99.93 & 99.77 & \textbf{99.89} & 99.88 \\
    \SI{4}{\micro\metre}, high bg., \SI{5}{ms} & 19 & 98.25 & \textbf{97.86} & 97.58 & 97.61 \\
    \SI{2}{\micro\metre}, \SI{5}{ms} & 19 & 99.85 & 99.57 & 99.67 & \textbf{99.69} \\
    \SI{1.5}{\micro\metre}, \SI{5}{ms} & 19 & 99.34 & 98.22 & 98.29 & \textbf{98.70} \\
    \midrule
    \SI{4}{\micro\metre}, \SI{10}{ms} & 39 & $>$99.99 & $>$99.99 & $>$99.99 & $>$99.99 \\
    \SI{4}{\micro\metre}, high bg., \SI{10}{ms} & 38 & 99.90 & \textbf{99.87} & 99.84 & 99.83 \\
    \SI{2}{\micro\metre}, \SI{10}{ms} & 38 & 99.999 & \textbf{99.998} & 99.995 & 99.995 \\
    \SI{1.5}{\micro\metre}, \SI{10}{ms} & 39 & 99.99 & 99.81 & 99.85 & \textbf{99.93} \\
    \bottomrule
  \end{tabular}
\end{table}

\section{Quantisation and Toolflow}\label{app:quant}

As machine learning (ML) models grow in parameter count, their on-chip
resource usage and latency grow with them, which makes some models infeasible
for latency-bound tasks. Quantisation reduces the precision of the arithmetic
(weights, biases and activations) with little loss in
accuracy~\cite{duarte2018hls4ml,aarrestad2021convhls4ml}. The CNN, the readout
QDM, is therefore trained with High Granularity Quantization~2 (HGQ2, built on
Keras~3), a differentiable fixed-point quantisation scheme that learns bit
widths against the training loss~\cite{sun2026hgq}. The resulting model is
exported to Vitis High-Level Synthesis through hls4ml, a hardware/software
co-design tool for implementing machine-learning models on
FPGAs~\cite{duarte2018hls4ml,aarrestad2021convhls4ml,coelho2021qkeras}.
This appendix summarises the workflow and the findings from these tools.

\subsection{Per-tensor quantisation-aware training (QAT)}\label{app:quant:qat}

The network is specified in \cref{tab:cnn}. Weights and activations use one
\texttt{ap\_fixed<W,I>} format per tensor, homogeneous across every axis,
at 12 bits.

\Cref{tab:qat} reports the selected 12-bit models on the corrected data.
The QAT runs train for up to 40 epochs with early stopping on the evaluation
split, whereas the full-precision runs use a fixed 30 epochs. That split
therefore participates in QAT model selection and is treated here as
validation data. The comparison describes the selected configurations;
it does not establish a quantisation-induced accuracy gain or an unbiased
estimate of held-out performance. The input normalisation is fitted from
each model's training data and stored with its weights.

\begin{table}[htbp]
  \caption{Model-selection accuracy (\%) for the $8\times8$ readout
  configurations\simu{}. The QAT models use this evaluation split for early
  stopping, so these are validation diagnostics rather than independent
  test-performance estimates. Per configuration means one model per row;
  band means one model for all six rows.}
  \label{tab:qat}
  \centering
  \small
  \begin{tabular}{@{}lrrrrrr@{}}
    \toprule
    & Matched & \multicolumn{3}{c}{\texttt{1-8-16-16}} & \multicolumn{2}{c}{\texttt{1-32-64-64}} \\
    Configuration & filter & FP32 & 12-bit & 12-bit band & FP32 & 12-bit \\
    \midrule
    \SI{4}{\micro\metre}, \SI{3}{ms}            & 97.69 & 98.74 & 98.93 & 98.53 & 98.81 & \textbf{98.98} \\
    \SI{4}{\micro\metre}, \SI{5}{ms}            & 99.77 & 99.89 & 99.92 & 99.79 & 99.88 & \textbf{99.94} \\
    \SI{4}{\micro\metre}, \SI{7}{ms}            & 99.96 & 99.98 & 99.98 & 99.75 & 99.98 & 99.98 \\
    \SI{4}{\micro\metre}, high bg., \SI{5}{ms}  & \textbf{97.86} & 97.58 & 97.84 & 96.91 & 97.61 & 97.83 \\
    \SI{2}{\micro\metre}, \SI{5}{ms}            & 99.57 & 99.67 & \textbf{99.73} & 98.82 & 99.69 & 99.70 \\
    \SI{1.5}{\micro\metre}, \SI{5}{ms}          & 98.22 & 98.29 & 98.86 & 98.50 & 98.70 & \textbf{99.00} \\
    \bottomrule
  \end{tabular}
\end{table}

Per-tensor quantisation matches the format organisation of the generated
RTL. Numerical agreement between a selected model and its implementation
must still be checked through the toolflow; format organisation alone does
not establish bit-exact equivalence.

The per-configuration and band-trained scores document the dependence on
exposure and background. A CNN deployment requires the corresponding
parameter set and normalisation constants. The current CNN weights are
compiled into the bitstream; runtime updates require the changes described
in \cref{app:if:additions}.

\subsection{Precision floor}\label{app:quant:8bit}

The tested toolflow and quantisation configurations did not provide an
eight-bit implementation with accuracy comparable to the higher-precision
baseline. Two issues were identified.

\paragraph{Finding 1: \texttt{ap\_fixed<W,I>} with $I \le 0$ is mishandled in csim.}
Observed with the tool versions of \cref{app:quant:toolflow}. Reproduction:
hold fractional resolution fixed and vary only the sign of $I$.
The positive-$I$ configuration scores \SI{87}{\percent}; the non-positive-$I$
configuration scores \SI{48.95}{\percent}, i.e.\ chance\simu{}. The failure is
silent. In \texttt{ap\_fixed<W,I>}, $W$ is the total number of bits and $I$
the number of them to the left of the binary point, so $I$ sets the range a
tensor can represent and $W - I$ its resolution. A tensor whose largest value
is $0.19$ never uses an integer bit; with $I \le 0$ every bit, and for $I < 0$
some implied leading zeros as well, goes to resolution. Small-magnitude layers
can require additional fractional precision. In the first convolution,
$|w|_{\max}=0.19$, so a positive-$I$ format allocates bits to an unused integer
range and reduces fractional precision at a fixed total width.

\paragraph{Finding 2: accuracy depends on the chosen precision and range.}
Independently of Finding~1, a positive-$I$ global weight sweep on the earlier,
photon-starved data (\cref{app:detector}, Data) gives
\SI{90.8}{\percent} at 12 bits, \SI{87.3}{\percent} at 10, \SI{76.0}{\percent}
at 9, and \SI{56.3}{\percent} at 8\simu{}. The cause is a dynamic-range conflict
that one per-tensor format cannot resolve: accumulators and batch-norm scales
reach $\approx 31$ and need six integer bits, while kernels lie within $\pm
0.45$ and need fractional bits. The deployed configuration is weights
\texttt{ap\_fixed<12,2>} and data \texttt{ap\_fixed<12,6>}, at
\SI{90.30}{\percent}, indistinguishable from the 18-bit baseline
(\SI{90.2}{\percent}) on that data; the corrected-data model-selection comparison is reported separately
(\cref{tab:qat}). On the corrected selection split the same format sensitivity is observed
(\cref{tab:wsweep})\simu{}: deploying the per-configuration $8\times8$
networks at weights \texttt{ap\_fixed<W,2>} and data \texttt{ap\_fixed<W,6>},
12 bits stays within \SI{0.2}{\percent} of the 18-bit result in every
configuration, 11 bits loses up to \SI{0.7}{\percent}, 10 bits up to
\SI{2.1}{\percent}, 9 bits up to \SI{7.3}{\percent}, and 8 bits fails, most
of all where the margin is thinnest (high background, \SI{1.5}{\micro\metre}
pitch). Twelve bits remains the deployed width.

\begin{table}[htbp]
  \caption{Balanced fidelity (\%) of the $8\times8$ \texttt{1-8-16-16}
  networks of \cref{tab:qat} deployed at $W$ bits (weights
  \texttt{ap\_fixed<W,2>}, data \texttt{ap\_fixed<W,6>}, accumulators
  \texttt{ap\_fixed<32,16>}), hls4ml C simulation on the model-selection split,
  \numrange{43120}{44064} windows per configuration\simu{}.}
  \label{tab:wsweep}
  \centering
  \small
  \begin{tabular}{@{}lrrrrrrr@{}}
    \toprule
    Configuration & 8 & 9 & 10 & 11 & 12 & 14 & 18 \\
    \midrule
    Sparse, \SI{3}{ms}                         & 83.49 & 97.23 & 98.33 & 98.75 & 98.79 & 98.82 & 98.84 \\
    Sparse, \SI{5}{ms}                         & 93.21 & 99.18 & 99.82 & 99.90 & 99.90 & 99.92 & 99.92 \\
    \SI{1.5}{\micro\metre}, \SI{5}{ms}         & 52.58 & 92.70 & 98.21 & 98.56 & 98.81 & 98.85 & 98.84 \\
    High background, \SI{5}{ms}                & 50.00 & 90.54 & 95.67 & 97.12 & 97.63 & 97.72 & 97.79 \\
    \bottomrule
  \end{tabular}
\end{table}

Hardware-aware quantisation has enabled a 10-bit implementation for a
different network and application~\cite{govorkova2026radhard}\lit{}.
The precision required in a particular design depends on the model,
arithmetic formats, and acceptable accuracy loss.

\subsection{Resource cost of precision}\label{app:quant:win}

\Cref{tab:csynth} gives the HLS-synthesis resource change from 18-bit to
per-tensor 12-bit on the reference network (\texttt{1-32-64-64}, RF$=$64,
\texttt{Resource} strategy) used for the precision study\csynth{}. Accuracy is
held; BRAM falls \SI{30}{\percent}, LUT \SI{22}{\percent}, FF
\SI{20}{\percent}, DSP \SI{11}{\percent}.

The ordering follows from what each resource holds: BRAM is weight ROM and
scales with operand width; LUT and FF scale with datapath width; the DSP count
is set by the multipliers the reuse factor leaves unrolled, and a 12-bit
operand still occupies a slice. Lowering precision therefore saves memory and
logic; DSP savings come from the reuse factor (\cref{app:impl:rf}).

\begin{table}[htbp]
  \caption{HLS synthesis on \texttt{xczu49dr} for the reference network
  \texttt{1-32-64-64} at RF$=$64, \texttt{Resource} strategy, 18-bit versus
  per-tensor 12-bit\csynth{}. Percentages are device utilisation.}
  \label{tab:csynth}
  \centering
  \small
  \begin{tabular}{@{}lrrr@{}}
    \toprule
    Resource & 18-bit & 12-bit per-tensor & $\Delta$ \\
    \midrule
    BRAM\_18K & 1\,113 (51\%) & 775 (35\%)    & $-30\%$ \\
    LUT       & 112\,519 (26\%) & 87\,641 (20\%) & $-22\%$ \\
    FF        & 117\,890 (13\%) & 93\,981 (11\%) & $-20\%$ \\
    DSP       & 1\,158 (27\%) & 1\,030 (24\%) & $-11\%$ \\
    \midrule
    Accuracy (earlier data) & 90.2\% & 90.3\% & --- \\
    \bottomrule
  \end{tabular}
\end{table}

\Cref{tab:csynth_vs_pnr} compares HLS estimates with place-and-route results
for the earlier $12\times12$ CNN. HLS predicts DSP usage exactly and
overestimates LUT and FF usage.

\begin{table}[htbp]
  \caption{HLS estimate versus place-and-route for the earlier $12\times12$ CNN
  (\texttt{1-8-16-16}, RF$=$8, 12-bit).}
  \label{tab:csynth_vs_pnr}
  \centering
  \small
  \begin{tabular}{@{}lrrl@{}}
    \toprule
    Resource & csynth\csynth{} & place-and-route\meashw{} & Note \\
    \midrule
    DSP       & 648 & 648 & exact \\
    LUT       & 50\,624 & 29\,284 & logic optimisation in synthesis and implementation \\
    FF        & 47\,348 & 33\,532 & logic optimisation in synthesis and implementation \\
    \bottomrule
  \end{tabular}
\end{table}

\subsection{Toolflow constraints}\label{app:quant:toolflow}

The following constraints were observed with HGQ2, hls4ml, Vitis HLS and
Vivado 2026.1 on Keras~3 and shaped the deployed configuration. They document
the path taken with those versions: later releases may behave differently, and
none of the four was pursued beyond the workaround noted. 

\begin{itemize}[leftmargin=1.2em,itemsep=2pt,topsep=2pt]\sloppy
  \item \texttt{QGlobalAveragePooling2D} raised an \texttt{ndim} assertion in
        the \texttt{bit\_exact} pass; \texttt{QAveragePooling2D(pool\_size)}
        followed by \texttt{Flatten} exported bit-exactly and is used instead.
        Whether the assertion reflects a limitation of the pass or a usage
        error on our side was not investigated.
  \item Under \texttt{io\_stream}, hls4ml accepted only homogeneously quantised
        activations; heterogeneous activation quantisation was accepted only
        under \texttt{io\_parallel}. This constraint, rather than any accuracy
        result, set the fully per-tensor configuration.
  \item C-synthesis of the fully unrolled \texttt{io\_parallel} convolution for
        the deployed network did not complete in the time allowed and was not
        pursued further; \texttt{io\_stream} synthesis of the same network
        completed. Whether \texttt{io\_parallel} would complete given more time
        or a different strategy was not tested. Under \texttt{io\_stream} the
        pixel-streaming loop rather than the MAC count sets latency
        (\cref{app:impl:rf}).
  \item With \texttt{WRAP} quantisers, exports made without
        \texttt{trace\_minmax} calibration carried incorrect ranges and raised
        no error; calibration is applied throughout.
\end{itemize}

In this evaluation, global normalisation preserves the absolute brightness
information used for occupancy classification. On the earlier data, at the
compared exposure, per-crop normalisation gave \SI{67}{\percent} accuracy,
compared with \SI{100}{\percent} for global normalisation\simu{}. We therefore
use global normalisation for this detector. The global mean and scale are fit
from each model's training data and stored with it; normalisation constants
from one data set applied to another are a silent source of error.
\section{Implementation Results}\label{app:impl}

\subsection{Place-and-route}\label{app:impl:pnr}

\Cref{tab:impl} gives post-place-and-route results for the integrated
pipeline on the \texttt{xczu49dr} (ZCU216 board), Vivado 2026.1,
out-of-context\meashw{}, with each of the two detectors at the $8\times8$
window and, for reference, with the earlier $12\times12$ CNN. The block rows
are the hierarchy of the matched-filter run and sum to its total. Both
$8\times8$ builds meet \SI{200}{MHz} without a post-route optimisation pass:
WNS \SI{+0.067}{ns} and WHS \SI{+0.042}{ns} with the matched filter, 0 of
\num{55993} endpoints failing, its worst path inside the rearrangement engine;
WNS \SI{+0.158}{ns} and WHS \SI{+0.028}{ns} with the CNN, 0 of \num{153233}.
The $12\times12$ CNN build needed \texttt{phys\_opt\_design} to close
(\cref{app:impl:timing}). The only constraint is the \SI{5.000}{ns} clock
period; no additional clock uncertainty or input jitter is specified, so
Vivado's default uncertainty applies. LUTs are dominated by the rearrangement engine and
all DSPs belong to the detector. The engine's LUT count varies between
implementation runs of the same RTL, from \num{69404} to \num{85766} across
the three builds here, so differences in the totals of that order are tool
variation rather than design; the detector rows are stable.

\begin{table}[htbp]
  \caption{Post-place-and-route results for the integrated pipeline on
  the \texttt{xczu49dr} (ZCU216 board), Vivado 2026.1, out-of-context,
  \SI{200}{MHz}\meashw{}. Block rows are the matched-filter build. Percentages
  are device utilisation; BRAM in 18\,K units.}
  \label{tab:impl}
  \centering
  \small
  \begin{tabular}{@{}lrrrr@{}}
    \toprule
    Block & LUT & FF & BRAM & DSP \\
    \midrule
    Matched filter ($8\times8$, one lane, loadable) & 1\,845 & 1\,866 & 0 & 64 \\
    Rearrangement engine (\texttt{tetris})  & 85\,766 & 47\,729 & 0  & 0 \\
    Site row assembler                      & 223     & 62      & 0  & 0 \\
    Crop register / loader / atom-grid BRAM & 10      & 42      & 2  & 0 \\
    \midrule
    \textbf{Integrated pipeline, matched filter} & \textbf{87\,844} & \textbf{49\,699} & \textbf{2} & \textbf{64} \\
    \phantom{xx}device utilisation  & 20.7\%  & 5.8\%   & 0.1\%  & 1.5\% \\
    \midrule
    Integrated pipeline, CNN $8\times8$ (\texttt{1-8-16-16}, RF=8) & 98\,798 & 80\,898 & 107 & 584 \\
    \phantom{xx}of which CNN                & 29\,168 & 33\,193 & 105 & 584 \\
    \phantom{xx}device utilisation  & 23.2\%  & 9.5\%   & 5.0\%  & 13.7\% \\
    \midrule
    Integrated pipeline, CNN $12\times12$, after \texttt{phys\_opt} & 112\,404 & 81\,300 & 114 & 648 \\
    \phantom{xx}of which CNN                & 29\,284 & 33\,532 & 112 & 648 \\
    \phantom{xx}device utilisation  & 26.4\%  & 9.6\%   & 5.3\%  & 15.2\% \\
    \bottomrule
  \end{tabular}
\end{table}

The matched-filter pipeline takes one site window per cycle on a registered
crop input, which stands in for the output register of the region-of-interest
line buffer; the line buffer itself is not built. In RTL simulation the
integrated top, with the calibration written through its configuration port,
assembles all 2500 sites of a $50\times50$ array bit-exact against the
software golden, \num{2510} cycles from the first window to the last
occupancy row\meassim{}. The CNN pipeline is the same back end with the CNN of
\cref{app:impl:rf} in place of the lane.

The matched-filter datapath (\cref{app:detector:mf}) was also placed and
routed on its own, out of context, on the same part and tool version, at a
\SI{500}{MHz} target (\cref{tab:mfpnr})\meashw{}: first at the earlier
$12\times12$ window, 144 multipliers per lane with the template of the
earlier calibration, where the lane was pipelined, then at $8\times8$ with the
template and threshold recalibrated on the data of \cref{app:detector}, with
the constants fixed at synthesis and runtime-loadable.

\begin{table}[htbp]
  \caption{Out-of-context place-and-route of the matched-filter datapath,
  \texttt{xczu49dr-ffvf1760-2-e}, Vivado 2026.1, \SI{2.0}{ns} constraint,
  through route\meashw{}. The harness rows feed the pixel inputs from
  registers, as the ROI line buffer will, and report the detector's resources
  without the harness; the plain out-of-context rows drive the inputs from
  unconstrained ports. No harness build has a failing hold endpoint, and the
  constant-template harness builds none failing setup; the loadable $P=20$ and
  $P=40$ builds miss the \SI{2.0}{ns} target and are quoted at their
  $F_{\max}$. Cycles are $\lceil 10^4/P \rceil + L$ for a $100\times100$ array.
  Percentages are device utilisation.}
  \label{tab:mfpnr}
  \centering
  \footnotesize
  \setlength{\tabcolsep}{4pt}
  \begin{tabular}{@{}llrrrrr@{}}
    \toprule
    Build & Lane RTL & WNS (ns) & $F_{\max}$ & DSP & LUT & FF \\
    \midrule
    \multicolumn{7}{@{}l}{\textit{$12\times12$ window}} \\
    1 lane      & combinational (original)          & $-5.39$ & \SI{135}{MHz} & 138 & 2\,177 & 2 \\
    $P=20$      & combinational (original)          & $-5.77$ & \SI{129}{MHz} & 2\,760 & 43\,498 & 21 \\
    1 lane      & pipelined, adders in DSP          & $+0.09$ & \SI{523}{MHz} & 210 & 64 & 140 \\
    $P=20$      & pipelined, adders in DSP          & $-3.14$ & \SI{195}{MHz} & 4\,200 (98\%) & 1\,280 & 2\,847 \\
    1 lane      & pipelined, adders in fabric       & $+0.52$ & \SI{674}{MHz} & 138 & 4\,143 & 4\,413 \\
    $P=20$      & pipelined, adders in fabric       & $+0.43$ & \SI{636}{MHz} & 2\,760 (65\%) & 82\,860 (19\%) & 88\,051 (10\%) \\
    \midrule
    $P=20$, harness & pipelined, adders in fabric   & $+0.46$ & \SI{651}{MHz} & 2\,760 & 82\,940 & 86\,771 \\
    \midrule
    \multicolumn{7}{@{}l}{\textit{$8\times8$ window, recalibrated template, harness}} \\
    1 lane      & pipelined, adders in fabric       & $+0.52$ & \SI{674}{MHz} & 61 & 1\,851 & 1\,877 \\
    $P=20$      & pipelined, adders in fabric       & $+0.51$ & \SI{673}{MHz} & 1\,220 (29\%) & 37\,040 (9\%) & 38\,329 (5\%) \\
    $P=40$      & pipelined, adders in fabric       & $+0.50$ & \SI{668}{MHz} & 2\,440 (57\%) & 74\,080 (17\%) & 76\,649 (9\%) \\
    \midrule
    \multicolumn{7}{@{}l}{\textit{$8\times8$ window, runtime-loadable template, harness}} \\
    1 lane      & pipelined, adders in fabric       & $+0.45$ & \SI{646}{MHz} & 64 & 1\,846 & 1\,866 \\
    $P=20$      & pipelined, adders in fabric       & $-0.34$ & \SI{428}{MHz} & 1\,280 (30\%) & 36\,269 (9\%) & 37\,174 (4\%) \\
    $P=40$      & pipelined, adders in fabric       & $-1.11$ & \SI{322}{MHz} & 2\,560 (60\%) & 72\,474 (17\%) & 74\,289 (9\%) \\
    \bottomrule
  \end{tabular}
\end{table}

The original lane computed the 144-term dot product combinationally, 26 to 27
logic levels, and reached \SI{135}{MHz}. Pipelining it (\cref{app:detector:mf})
closes the lane at \SI{674}{MHz}, but only if the adder tree is kept out of the
DSP slices: with the adders packed into DSPs a lane takes 210 slices, twenty
lanes take \SI{98}{\percent} of the device, and DSP-to-DSP routing fails at
\SI{195}{MHz}. With the adders in fabric, twenty lanes route at \SI{636}{MHz}
with \SI{65}{\percent} of the DSP slices. The plain out-of-context build
reports a hold violation of about \SI{0.08}{ns} on some \num{2000} paths from
input ports straight into the first register, an artefact of driving the
inputs from ports with zero delay; in a harness that feeds the pixel inputs
from registers the $P=20$ build has no failing setup or hold endpoint among
\num{308183}, with a worst setup slack of \SI{0.46}{ns} and a worst hold slack
of \SI{0.028}{ns}. Integration must feed the lane from registers.

At the $8\times8$ window the same RTL, with 64 coefficients, needs 61 DSP
slices per lane, because the tool folds the three zero coefficients of the
calibrated template, and $F_{\max}$ is flat in $P$: \SI{674}{MHz} for one lane,
\SI{673}{MHz} for twenty and \SI{668}{MHz} for forty, the critical path being
the same multiply or first adder in every build. Twenty lanes take
\SI{44}{\percent} of the DSP slices and \SI{45}{\percent} of the LUTs of the
$12\times12$ build. Forty lanes, which do not fit at $12\times12$ (\num{5520}
DSP slices against \num{4272}), take \SI{57}{\percent} of the DSP slices and
classify $10^4$ sites in 259 cycles, \SI{0.52}{\micro\second} at
\SI{500}{MHz}, with \SI{0.5}{ns} of setup slack.

The runtime-loadable lane costs 64 DSP slices instead of 61, since no
coefficient can be folded, and about the same fabric. One lane routes at
\SI{646}{MHz}. Replicated, it slows: \SI{428}{MHz} for twenty lanes and
\SI{322}{MHz} for forty, the worst path running from the shared
configuration-address register to the clock enables of the multipliers'
coefficient registers in every lane. That path is a fan-out of a
quasi-static control signal and could be pipelined per lane; as built, forty
loadable lanes classify $10^4$ sites in \SI{0.80}{\micro\second} at
\SI{322}{MHz}. The loop needs none of this: the ROI payload alone requires
at least \SI{3.4}{ms} of transfer time, so one lane is enough, classifying
2500 sites in \SI{12.5}{\micro\second} at \SI{200}{MHz} for 64 DSP slices,
\SI{1.5}{\percent} of the device, and that is the lane integrated in
\cref{tab:impl}.

\subsection{Per-crop latency and reuse factor}\label{app:impl:rf}

Except where the $8\times8$ build is named, every figure in this subsection
was measured on the $12\times12$ build of the CNN. Under \texttt{io\_stream} the pixel-streaming loop sets the per-crop kernel
latency, the cycles from \texttt{ap\_start} to \texttt{ap\_done}. The kernel
is invoked once per site and invocations are serialised, so this latency is
also the effective initiation interval; in a continuously streaming deployment
the two would separate. Reducing the network $15.7\times$
(\cref{app:detector:learned}) leaves the latency essentially unchanged at
fixed reuse factor. Across three C-synthesis estimates,
\begin{equation}
  L(\mathrm{RF}) \approx 1626 + 220\,\mathrm{RF}\ \text{cycles per crop,}
  \label{eq:ii}
\end{equation}
reproducing the estimates to within \SI{2}{\percent}\csynth{}: \num{15682}
cycles at RF$=$64, \num{3334} at 8, \num{2550} at 4. The fitted constant term
is line-buffer fill plus datapath register depth. For the deployed RF$=$8
configuration, RTL simulation measures \num{3246} cycles per crop\meassim{},
\SI{3}{\percent} below the C-synthesis estimate of \num{3334} and
\SI{4}{\percent} below the fit's \num{3386}; the measured value is the one
quoted for the $12\times12$ build, and the loop uses the $8\times8$
measurement below.

The floor makes reuse factor an inefficient lever below RF$=$8: RF$=$4 improves
latency $1.31\times$ at $1.9\times$ the DSP. RF$=$8 is the deployed operating
point.

The window, not the network, sets the constant. The $8\times8$ build, the
\texttt{1-8-16-16} network trained for the sparse \SI{5}{ms} data
(\cref{tab:qat}) at the same precision and RF$=$8, reports \num{1602} cycles
per crop against \num{3334} at $12\times12$\csynth{}, close to the ratio of
pixel counts, $64/144$, plus a fixed overhead; its hls4ml C simulation agrees
with the floating-point model on all of 2000 test windows, and its RTL
co-simulation passes. In RTL simulation\meassim{}, with each crop started as
the previous one completes, the $8\times8$ network takes \num{1797} cycles per
crop; the same harness gives \num{3286} for the $12\times12$ build, within
\SI{1.2}{\percent} of the \num{3246} measured in the integrated pipeline, and
both builds' logits match the Vitis co-simulation. Over 2500 sites at
\SI{200}{MHz} this is \SI{22.5}{ms} per core, against \SI{40.6}{ms} for the
$12\times12$ build; the network itself takes 584 DSP slices against 648.

Crops need not wait for the previous one to finish: the dataflow design
accepts the next window while the last layers of the previous one are still
running. Driven back to back, the measured interval between results is
\num{1165} cycles at $8\times8$ and \num{2651} at $12\times12$\meassim{},
which would bring one $8\times8$ core to \SI{14.6}{ms} per array. The loop
figures below use the serialised \SI{22.5}{ms}, the convention of the
$12\times12$ measurements.

This initialisation overhead occurs \num{2500} times per frame. Taking the fitted constant of
  \cref{eq:ii} as the fill, $2500 \times 1626 \approx \SI{20.3}{ms}$\csynth{},
  about half of the $12\times12$ build's \SI{40.6}{ms} one-core detection time, is attributable to
  line-buffer refill; the constant is the fit's intercept at RF$\,=0$, not a
  measured fill time. 

\subsection{Board build of the matched-filter pipeline}\label{app:impl:board}

The out-of-context top of \cref{tab:impl} has about \num{5200} I/O bits: the
$50\times50$ target input, the $50\times50$ assigned-rows output and the
1024-bit crop bus, more than the package has pins. For a board bitstream it
is therefore wrapped in an on-chip self-test. The wrapper generates every
input on chip: the matched filter's calibration writes, the target preload
and \num{2500} synthetic crops per iteration. It folds the occupancy rows, the
moves and the assigned rows into a CRC-32 signature. The \SI{300}{MHz} board
oscillator (Si570) drives an MMCM that generates the \SI{200}{MHz} clock;
reset asserts when the MMCM loses lock and releases synchronously. Status is
shown on the board's LEDs. Iterations alternate between checked runs, which
reseed the generators and must reproduce a fixed signature, and free runs on
new stimulus; each kind runs with the move stream always accepted and with it
accepted on about one cycle in sixteen, which models a slow controller.

\paragraph{Simulation.} In eight iterations covering all four kinds
(Verilator and cocotb), the assembled occupancy is bit-exact against the
matched filter's threshold rule and the RTL moves equal the engine's golden
model in every iteration, \numrange{1874}{1906} moves each. The checked
signature is identical with and without backpressure. The engine takes
\num{24694} cycles unthrottled and \num{45865} throttled on the checked
stimulus\meassim{}.

\paragraph{Implementation.} The bitstream closes \SI{200}{MHz} in context,
with board pins and the MMCM clock tree (\cref{tab:impl:board}). The
worst setup path is inside the engine, from its row-assignment logic into the
grid memory's data input. The bitstream has not yet been loaded on the board.

\begin{table}[htbp]
  \caption{Matched-filter pipeline as a ZCU216 bitstream with on-chip
  self-test, Vivado 2026.1, \SI{200}{MHz}\meashw{}, against the
  out-of-context build of \cref{tab:impl}.}
  \label{tab:impl:board}
  \centering
  \small
  \begin{tabular}{@{}lrr@{}}
    \toprule
    & Out of context & Board bitstream \\
    \midrule
    Setup slack (WNS) & \SI{+0.067}{ns} & \SI{+0.095}{ns} \\
    Hold slack (WHS) & \SI{+0.042}{ns} & \SI{+0.016}{ns} \\
    Recovery / removal slack & not reported & \SI{+0.254}{ns} / \SI{+0.097}{ns} \\
    Timing endpoints, failing of total & 0 of \num{55993} & 0 of \num{92077} \\
    \midrule
    LUT: matched filter & \num{1845} & \num{1847} \\
    LUT: rearrangement engine & \num{85766} & \num{72293} \\
    LUT: site row assembler & \num{223} & \num{533} \\
    LUT: self-test wrapper & --- & about \num{1430} \\
    LUT, total (device share) & \num{87844} (20.7\%) & \num{76302} (17.9\%) \\
    FF, total (device share) & \num{49699} (5.8\%) & \num{52123} (6.1\%) \\
    BRAM\_18K / DSP & 2 / 64 & 2 / 64 \\
    Bonded I/O / MMCM & --- & 30 / 1 \\
    Power, Vivado estimate & --- & \SI{2.39}{W} \\
    \bottomrule
  \end{tabular}
\end{table}

The detector is unchanged between the two builds, and the extra flip-flops
are the wrapper's: its generators, the 2500-bit target register, the signature
and counters. The engine's LUT count falls by \num{13473}. That is within the
range of \numrange{69404}{85766} seen across earlier implementation runs of
the same RTL (\cref{app:impl:pnr}), but the two builds differ in two ways
that have not been separated: the board build includes the fix below, and
the engine's inputs come from on-chip logic instead of ports.

\paragraph{Engine fix found by the self-test.} The backpressure iterations
exposed a defect in the rearrangement engine. Both of its phases decide a
move-FIFO write in one cycle and register it in the next, but they tested the
FIFO's full flag without counting the write already in flight. With one slot
left, a second write could be decided and land on a full FIFO, which drops
it. With a slow consumer, 595 of \num{1906} moves were lost. Every earlier
test held the move stream always ready and never reached this state. Both
phases now stall when the FIFO holds $\text{depth}-1$ entries. All moves are
recovered, unthrottled runs are unchanged cycle for cycle, and the engine's
existing tests pass. The loop-model results of \cref{app:loop} use the
software planner and are not affected.

\subsection{Timing closure}\label{app:impl:timing}

\paragraph{Cause.} At a \SI{5.000}{ns} constraint the initial implementation
reached $\approx \SI{115}{MHz}$ with \num{13307} failing endpoints. The
21-level critical path ran from the source-column encoder's ripple
prefix-popcount and $N \times N$ scatter into the
dynamic-programming cell arithmetic and the take-matrix scatter; the encoder
output depends only on the latched input row and was being recomputed on every
one of the $M \times K$ cycles of the dynamic-program fill ($K$ atoms in the
row, $M$ columns still needing atoms, \cref{tab:rtl:policy}).

\paragraph{Fix.} Registering the encoder output splits the path.

\paragraph{Cost.} Latency: none. The downstream consumer first reads the
registered value 22 cycles after the start pulse, so the added stage lies
within existing slack; cycle counts and the golden move-plan match are
unchanged. Area: $+\num{13503}$ LUT for 300 registers. The register is an
optimisation barrier: the scatter can no longer be collapsed into the
downstream 50:1 dynamic mux, so all slots are materialised.

\paragraph{Result.} WNS improves from \SI{-3.708}{ns} to \SI{-0.100}{ns}; a
post-route \texttt{phys\_opt\_design} pass closes the remaining \SI{0.1}{ns}
for 14 LUTs (\cref{tab:closure}). $F_{\max}$ is reported as $\ge
\SI{200}{MHz}$, the constraint having been met and the clock never swept
above it. The remaining limiter is the
encoder's ripple prefix-popcount. A prefix-sum tree could reduce its depth
from approximately 50 serial additions to about six levels. This alternative
was not implemented because registering the encoder output already met the
\SI{200}{MHz} target without changing the simulated cycle count: the extra
stage is hidden behind the sort flush. The tree's benefit at higher clock
targets remains to be evaluated.

\begin{table}[htbp]
  \caption{Timing closure at \SI{5.000}{ns}\meashw{}.}
  \label{tab:closure}
  \centering
  \small
  \begin{tabular}{@{}lrrr@{}}
    \toprule
    & baseline & $+$register & $+$\texttt{phys\_opt} \\
    \midrule
    WNS (ns)            & $-3.708$ & $-0.100$ & \textbf{0.000} \\
    Failing endpoints   & 13\,307  & 89       & \textbf{0} \\
    $F_{\max}$ (MHz)    & $\approx$115 & $\approx$196 & $\ge$\textbf{200} \\
    CLB LUTs            & 98\,901  & 112\,390 & 112\,404 \\
    \bottomrule
  \end{tabular}
\end{table}

\subsection{Latency budget}\label{app:impl:latency}

\Cref{tab:latency} decomposes FPGA-side latency per assembly cycle for the
$50\times50$ array at \SI{200}{MHz}. Detection and rearrangement serialise. The row assembler signals
  completion (\texttt{occ\_done\_o}) after the last occupancy row is written,
  and the controller pulses the engine's \texttt{start\_i} only after it: on
  start the engine takes the occupancy memory's write port
  (\cref{app:frontend:contract}), so a row written after that point would be
  lost. The stages therefore add.

\begin{table}[htbp]
  \caption{Component timings for the $50\times50$ design at
  \SI{200}{MHz}; these are not an end-to-end feedback measurement.
  The engine row is the separate \num{961}-site target verification run.
  Exposure and mean transport refer to the corrected-planner loop model
  (\cref{app:loop:latency}); transport varies with each cycle's schedule.}
  \label{tab:latency}
  \centering
  \small
  \begin{tabular}{@{}L{4.4cm}rL{3.6cm}L{3.3cm}@{}}
    \toprule
    Stage & Cycles & Time & Provenance \\
    \midrule
    Camera exposure (matched-filter operating point) & --- & \SI{10.0}{ms} & phys \\
    Sensor readout $+$ CXP ingest, 2048-row ROI & --- & \SI{7.4}{ms} (standard scan) or \SI{36}{ms} (ultra quiet) & datasheet; \cref{tab:if:cxp:budget} \\
    Detection, matched filter, $P=1$ ($2500 + 9$) & 2\,509 & \SI{12.5}{\micro\second} & meas-sim \\
    Detection, CNN $8\times8$, 1 core ($2500 \times 1797$) & 4\,492\,500 & \SI{22.5}{ms} & meas-sim \\
    Detection, CNN $8\times8$, 4 cores & $\approx$1\,123\,125 & \SI{5.6}{ms} & derived \\
    Row assembler $+$ loader preload & $\approx 1\,000$ & $< \SI{0.01}{ms}$ & derived: one row or queue entry per cycle (\cref{app:frontend:contract}) \\
    Rearrangement engine & 29\,184 & \SI{0.15}{ms} & meas-sim \\
   Egress to controller: FPGA to server memory, OPNIC SDK stream, NVQLink to the OPX1000 & --- & about \SI{1}{\micro\second} per PCIe transfer inside the OPC, memory to OPNIC included (vendor figure); stream hand-off unmeasured; vendor round trip $< \SI{4}{\micro\second}$~\cite{qm_dgxq_blog2025} & spec, unmeasured \\
    Atom transport (modelled schedule) & --- & about \SI{10}{ms}, varying by cycle & phys \\
    \bottomrule
  \end{tabular}
\end{table}

\subsection{Co-residency with a real-time decoder}\label{app:impl:coresident}

\Cref{tab:coresident} combines the implemented array-assembly pipeline
(\cref{tab:impl}) with the separately evaluated four-round decoder specified
in \cref{app:decoder}. That decoder has a C-synthesis clock estimate of
\SI{3.270}{ns} and a standalone post-route period of \SI{3.514}{ns},
corresponding to a derived kernel latency of \SI{0.84}{ms}.
Its resource figures are the C-synthesis estimates in \cref{tab:dec_compare}.
The budget uses one loadable matched-filter lane for the loop and replicated
CNN cores as alternative hardware-sizing cases.

\begin{table}[htbp]
  \caption{Co-residency budget: implemented pipeline\meashw{} with the qLDPC
  decoder\csynth{}, all at the $8\times8$ window. The one-lane matched-filter
  and one-core CNN rows are the integrated builds of \cref{tab:impl}; the
  others replace the one lane by the separately routed loadable lanes of
  \cref{tab:mfpnr}, or replicate the routed CNN core, and are
  derived\roof{}. Percentages use \num{425280} LUT,
  \num{4272} DSP and \num{2160} BRAM\_18K on the target device
  \cite{amd_ug1390_zcu216}.}
  \label{tab:coresident}
  \centering
  \small
  \begin{tabular}{@{}lrrrr@{}}
    \toprule
    Configuration & Detect & LUT & DSP & BRAM \\
    \midrule
    Matched filter, $P=1$, no decoder & \SI{12.5}{\micro\second} & 20.7\% & 1.5\% & 0.1\% \\
    \textbf{Matched filter, $P=1$, $+$ decoder} & \SI{12.5}{\micro\second} & 61.0\% & \textbf{40.2\%} & 37.4\% \\
    Matched filter, $P=20$, $+$ decoder & \SI{0.67}{\micro\second} & 69.1\% & 68.7\% & 37.4\% \\
    Matched filter, $P=40$, $+$ decoder & \SI{0.36}{\micro\second} & 77.6\% & 98.7\% & 37.4\% \\
    \midrule
    CNN, 1 core, no decoder  & \SI{22.5}{ms} & 23.2\% & 13.7\% & 5.0\% \\
    CNN, 6 cores, no decoder & \SI{3.7}{ms}  & 57.5\% & 82.0\% & 29.3\% \\
    CNN, 1 core $+$ decoder  & \SI{22.5}{ms} & 63.6\% & 52.4\% & 42.2\% \\
    CNN, 2 cores $+$ decoder & \SI{11.2}{ms} & 70.4\% & 66.1\% & 47.1\% \\
    CNN, 4 cores $+$ decoder & \SI{5.6}{ms}  & 84.2\% & 93.4\% & 56.8\% \\
    CNN, 6 cores $+$ decoder & \SI{3.7}{ms}  & 97.9\% & 120.8\% & 66.5\% \\
    \bottomrule
  \end{tabular}
\end{table}

\begin{enumerate}[leftmargin=1.2em,itemsep=2pt,topsep=2pt]
  \item \textbf{The matched-filter configuration leaves component-level margin.}
        One loadable lane and the decoder total \SI{61.0}{\percent} LUT,
        \SI{40.2}{\percent} DSP and \SI{37.4}{\percent} BRAM.
        Twenty or forty lanes add throughput that the camera does not require;
        forty lanes already reach \SI{98.7}{\percent} DSP before interfaces.
  \item \textbf{CNN replication consumes the interface margin.}
        Each additional core uses \num{584} DSP slices. Four cores and the
        decoder total \SI{84.2}{\percent} LUT and \SI{93.4}{\percent} DSP,
        excluding ingress and egress. These totals do not establish that the
        complete design fits (\cref{app:if:additions}).
        The loop comparison uses one CNN core (\cref{tab:cores}); the
        replicated rows here are resource projections.
  \item \textbf{Synthesis estimates require implementation checks.}
        Decoder resources are planning inputs, not guaranteed upper bounds.
        Joint placement and routing may change resource use, timing and
        routing feasibility.
\end{enumerate}

The estimates exclude camera ingest, its GTY interface and the PCIe block.
The decoder has not been placed and routed together with the pipeline.

\subsection{Power}\label{app:impl:power}

Vivado's post-route power estimate for the matched-filter board build of
\cref{app:impl:board} at \SI{200}{MHz} is \SI{2.39}{W} on chip:
\SI{1.17}{W} device static and \SI{1.22}{W} dynamic\meashw{}
(\cref{tab:impl:power}). The rearrangement engine accounts for
\SI{0.94}{W}, \SI{77}{\percent} of the dynamic power, and the matched filter
for \SI{0.04}{W}. The estimate is vectorless: no switching activity from
simulation was supplied, so internal nodes run at the tool's default toggle
rates, and Vivado rates its confidence as low. It also assumes the engine
switches continuously, whereas in the loop it is active only while planning,
about \SI{0.12}{ms} per plan (\num{24694} cycles, \cref{app:impl:board}). An estimate from
simulated activity, and estimates for the CNN and the CNN-plus-decoder
configurations, remain open (\cref{app:open}).

\begin{table}[htbp]
  \caption{Post-route power estimate, matched-filter pipeline as a ZCU216
  bitstream with on-chip self-test, \SI{200}{MHz}, vectorless, Vivado
  confidence low\meashw{}.}
  \label{tab:impl:power}
  \centering
  \small
  \begin{tabular}{@{}lr@{\qquad}lr@{}}
    \toprule
    By block & W & By resource & W \\
    \midrule
    Rearrangement engine & 0.938 & Logic (LUT, FF, carry) & 0.429 \\
    Site row assembler & 0.087 & Signals (routing) & 0.347 \\
    Matched filter & 0.040 & Clocks & 0.313 \\
    Self-test wrapper & 0.036 & MMCM & 0.098 \\
    Clocking, I/O and top level & 0.113 & DSP, I/O, BRAM & 0.028 \\
    \midrule
    Dynamic & 1.215 & Dynamic & 1.215 \\
    Device static & 1.170 & & \\
    \textbf{Total on chip} & \textbf{2.385} & & \\
    \bottomrule
  \end{tabular}
\end{table}

\section{Atom-Loss Model and Closed-Loop Evaluation}\label{app:loop}

This appendix evaluates atom loss during array assembly. It defines the loss
model and operating points, decomposes the cycle time, and estimates the
contribution of each loss channel to array lifetime.

\subsection{Model and parameters}\label{app:loop:model}

An assembly cycle images the array, decides occupancy per site, and moves atoms
into the target sites; if the assembly is incomplete the cycle repeats. Four
channels remove atoms:
\begin{description}[leftmargin=1em,itemsep=1pt,topsep=2pt]
  \item[Imaging] per-readout survival $\exp(-t_{\mathrm{exp}}/\tau_{\mathrm{img}})$; reduced by shorter exposure.
  \item[Transport] a fixed loss probability per moved atom; reduced by fewer
        moves and fewer rearrangement cycles.
  \item[Collision] an atom moved onto an occupied trap loses both atoms. It
        arises when the detector misses an atom and the planner moves another
        atom onto its site; reduced by fewer missed atoms.
  \item[Dead time] survival $\exp(-t_{\mathrm{cycle}}/\tau_{\mathrm{vac}})$ over
        the whole cycle wall-clock, including any host round-trip; reduced by
        removing the host from the feedback path.
\end{description}
\Cref{tab:params} lists the parameters. The model is calibrated against
a published survival and exposure
baseline~\cite{muzifalconi2025yb}\lit{}. Every attempt is simulated on the
$50\times50$ array the hardware is built for, with a centred $35\times35$
target, and charged the transport time of its own move schedule. The
reference point is $\tau_{\mathrm{vac}} = \SI{60}{s}$ and \SI{0.1}{\percent}
loss per moved atom; lossless transport is the comparison case.

\paragraph{Metrics.} An attempt succeeds when the target is defect-free
within nine detect--rearrange cycles. The figures reported are the success
probability $P_{\mathrm{df}}$ and the wall-clock per successful assembly,
the mean time per attempt, failed attempts included, divided by
$P_{\mathrm{df}}$. Atoms lost per attempt, all channels, are reported
alongside. At \num{1225} target atoms a slow loop does not lose a few atoms
more per assembly; it fails to complete the target, because the atoms lost
during each cycle must be re-supplied by the next.

\paragraph{Geometries.} The results of this paper are produced at three
geometries, and each quantity below is labelled by the one it belongs to. The
detector accuracies of \cref{app:detector} come from the imaging simulator: a
$14\times14$ array at \SI{4}{\micro\metre} pitch sampled at
\SI{0.4}{\micro\metre\per px}, \SI{10}{px} per site, read through an
$8\times8$-pixel crop (\cref{tab:simcfg}). The closed-loop results of this
appendix, and every hardware figure of \cref{app:impl}, are for the
$50\times50$ array with a $35\times35$ target; only \cref{fig:fidreq}, a
study of how the required detector fidelity scales with target size, uses
smaller grids. The site pitch of the target apparatus is not yet fixed. The
\num{2048}-pixel region of interest of the FPGA design is a sizing
assumption: at the \SI{0.4}{\micro\metre\per px} of \cref{tab:simcfg} the
simulated \SI{4}{\micro\metre} pitch is \SI{10}{px} per site, and a
$50\times50$ array at that pitch spans about \num{500} pixels. The detector
is therefore evaluated at the simulated pitch only, and a re-simulation at the
apparatus's pitch is left until that pitch is known (\cref{sec:limits}).

\begin{table}[htbp]
  \caption{Parameters of the atom-loss model.}
  \label{tab:params}
  \centering
  \scriptsize
  \begin{tabular}{@{}L{2.6cm}L{2.9cm}L{3.4cm}L{6.3cm}@{}}
    \toprule
    Parameter & Value & Role & Source \\
    \midrule
    Grid & $50\times50$ & all closed-loop results & the array of the hardware design (\cref{app:impl}) \\
    Target & centred $35\times35$ (\num{1225} atoms) & all closed-loop results & the target of the rearrangement engine's schedule (\cref{app:rtl:algorithm}) \\
    Trap fill probability & 0.6 (about \num{1500} atoms loaded) & initial occupancy & simulation setting, as in the engine sweep of \cref{app:rtl:algorithm}; the RTL verification runs of \cref{app:frontend:verification} load \num{1270} of \num{2500} sites \\
    Refill & none & no reservoir reload between cycles & simulation setting; a real apparatus reloads between assemblies, so assemblies per load compares configurations rather than predicting throughput \\
    Attempts per configuration & \num{1000} (\num{2000} for the operating-point sweep) & statistics & simulation setting \\
    Maximum cycles per attempt & 9 & abandon threshold & simulation setting \\
    Planner & row sweep into the target columns, then column compression & move schedule & the engine's algorithm (\cref{app:rtl:algorithm}) with its column step corrected in the software reference to avoid moving onto an atom that stays; this correction is not in the RTL, whose original planner is compared in \cref{tab:fom} \\
    $\tau_{\mathrm{img}}$ & \SI{70}{s} & imaging-loss lifetime & calibrated to the imaging-survival baseline of~\cite{muzifalconi2025yb}\lit{}; equivalent to \SI{0.1}{\percent} loss per \SI{70}{ms} image \\
    $\tau_{\mathrm{vac}}$ & \SIrange{5}{1380}{s}, ref.\ \SI{60}{s} & vacuum-limited trap lifetime & \SIrange{5}{60}{s}: room-temperature glass cells; \SI{1380}{s}: the \SI{23}{min} of~\cite{manetsch2025tweezer6100}; cryogenic systems (\SI{6000}{s}~\cite{schymik2021cryo}) lie above the range\lit{} \\
    Transport loss & \SI{0.1}{\percent} per moved atom per path; lossless for comparison & transport channel & assumed; moves over \SI{610}{\micro\metre} at about \SI{99.95}{\percent} fidelity are reported~\cite{manetsch2025tweezer6100}\lit{}; at \SI{1}{\percent} no $35\times35$ assembly completes \\
    $t_{\mathrm{host}}$ & \SI{150}{ms} & host round-trip in the host-in-loop configuration & assumed: a PC in the loop with camera transfer and software analysis, swept over \SIrange{10}{150}{ms} (\cref{fig:breakeven}); dedicated systems are faster: about \SI{10}{ms} of host time~\cite{dadpour2025lowlatency}, \SI{282}{\micro\second} for 24 tweezers~\cite{hu2026lowlatency}, about \SI{0.5}{ms} per cycle~\cite{qm2026readout}\lit{} \\
    Exposure sweep & \SIrange{7}{12}{ms}; \SIrange{21}{33}{ms} at a third of the photon rate & operating-point search & simulation setting \\
    Imaging model & caps-tum simulator, ORCA-Quest configuration~\cite{winklmann2023imagesim,capstum2024simulator} & photon floor, detector accuracy & NA 0.65, \SI{40}{kHz}, \SIrange{1}{50}{ms}, \SI{0.43}{e^-}, \SI{0.4}{\micro\metre\per px}, $8\times8$ crop (\cref{tab:simcfg}) \\
    Detector error rates & empirical per-class FP and FN from the sparse-array evaluation, whose balanced fidelity is summarised in \cref{tab:floor}; CNN rates use the full-precision \texttt{1-8-16-16} models; linear interpolation between exposures & detection errors & \cref{app:detector} \\
    Detection time & matched filter \SI{12.5}{\micro\second}; CNN \SI{22.5}{ms} ($8\times8$, one core) & cycle wall-clock & separate RTL timing; the CNN cost is used as a proxy for the operating-point model, without streaming overlap (\cref{app:loop:latency}) \\
    Camera readout & \SI{7.4}{ms} per cycle & cycle wall-clock & 2048-row region of interest, standard scan (\cref{app:platform:camera}) \\
    Rearrangement & \SI{0.15}{ms} per cycle, plus the attempt's own minimum-jerk AOD transport & cycle wall-clock & \cref{app:loop:latency} \\
    \bottomrule
  \end{tabular}
\end{table}

All closed-loop results come from one model
(\texttt{atom\_readout/simcheck/loop50.py}), run at the operating points by
\texttt{loop50\_opt.py}; configurations that appear in more than one table
report the same values.

\subsection{Closed-loop operating points}\label{app:loop:closed}

\begin{figure}[htbp]
  \centering
  \includegraphics[width=0.7\textwidth]{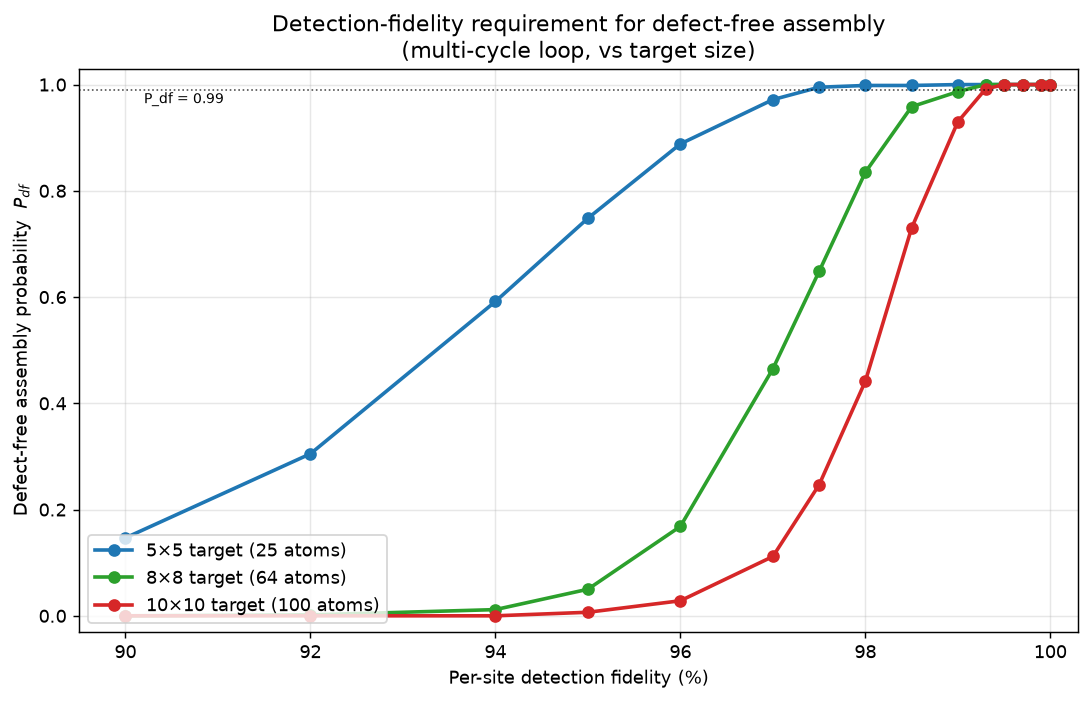}
  \caption{Multi-cycle defect-free assembly probability $P_{\mathrm{df}}$
  against per-site detection fidelity for $5\times5$, $8\times8$ and
  $10\times10$ targets\phys{}. Detection is modelled as a symmetric error,
  each site's occupancy read wrongly with probability $1-f$; moves are planned
  on the detected map with a parallel Hungarian planner and executed on the
  true array; an assembly succeeds when the target is defect-free within nine
  detect--rearrange cycles. Targets are centred in $11\times11$,
  $18\times18$ and $22\times22$ grids, with no atom loss and no refill;
  600 shots per point.}
  \label{fig:fidreq}
\end{figure}

Assembly can require multiple cycles because detection errors and atom loss leave defects after rearrangement. With every site read
correctly and no loss, the planner completes the $35\times35$ target from a
\SI{60}{\percent} load in a single cycle in every attempt. A missed atom is
moved onto, losing two atoms, and a site read as occupied when it is empty is
used as a source that delivers nothing; either leaves a defect for the next
cycle. The per-site detection fidelity required for a given multi-cycle
defect-free probability rises with target size (\cref{fig:fidreq}): at
$P_{\mathrm{df}} = 0.99$, from about \SI{97.5}{\percent} for 25 atoms to about
\SI{99.5}{\percent} for 100 atoms.

Each configuration runs at the exposure that minimises its wall-clock per
successful assembly at the reference point ($\tau_{\mathrm{vac}} =
\SI{60}{s}$, \SI{0.1}{\percent} loss per move), found by sweeping
\SIrange{7}{12}{ms}\phys{}\simu{}. Longer exposures cost imaging loss and
cycle time, and detection errors cost cycles, each of which makes about
\num{1300} lossy moves. For the matched filter on the FPGA the wall-clock per
assembly falls from \SI{0.34}{s} at \SI{7}{ms} to \SI{0.157}{s} at
\SI{10}{ms} and rises again to \SI{0.177}{s} at \SI{12}{ms}; the CNN on one
core has its minimum at the same \SI{10}{ms} (\SI{0.43}{s}), and the matched
filter behind the host at \SI{11}{ms}, where its success probability is flat
within the sampling error from \SIrange{10}{12}{ms}. Neither detector produced
errors in the evaluated simulated samples at exposures of \SI{10}{ms} or longer.
At these exposures, the loop model assigns zero detection error to both;
their assembly outcomes therefore differ only through inference latency.
An earlier rule, the least exposure reaching $P_{\mathrm{df}} \ge 0.99$
without vacuum or transport loss, put the operating points at \SI{7.5}{ms}
(matched filter) and \SI{7.0}{ms} (CNN). It favoured the CNN, whose lower
miss rate at those exposures saves cycles, but under transport loss neither
detector is at its best there: the matched filter at \SI{7.5}{ms} succeeds in
\SI{58}{\percent} of attempts at \SI{0.28}{s}, against \SI{85}{\percent} at
\SI{0.16}{s} at \SI{10}{ms}. Earlier drafts placed the operating points at
\SI{70}{} and \SI{120}{ms}, and later at \SI{4.0}{} and \SI{5.0}{ms} on an
$11\times11$ grid with a 25-atom target; the first came from the
photon-starved data that \cref{app:detector} supersedes, the second from a
geometry at which a single cycle was rarely enough.

\subsection{Loop wall-clock decomposition}\label{app:loop:latency}

\begin{figure}[htbp]
  \centering
  \includegraphics[width=0.75\textwidth]{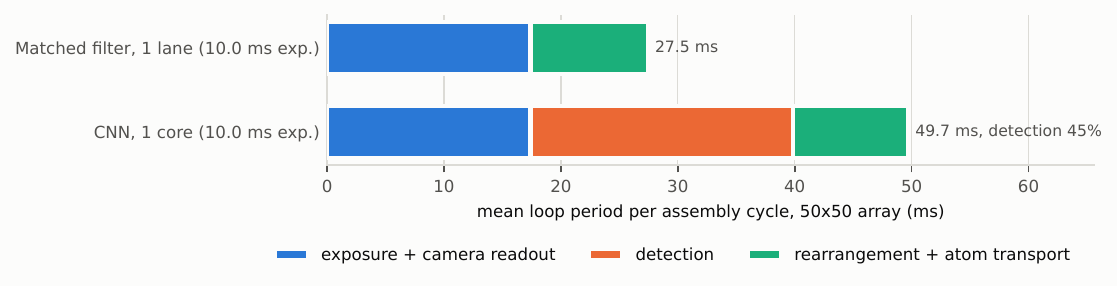}
  \caption{Mean loop period per assembly cycle, $50\times50$ array with a
  $35\times35$ target, with one matched-filter lane and with the $8\times8$
  CNN on one core, each at its operating exposure of \SI{10}{ms}
  (\cref{app:loop:closed})\phys{}. Camera readout is \SI{7.4}{ms}; detection
  is the CNN's serialised \SI{22.5}{ms}, and the matched filter's
  \SI{12.5}{\micro\second} is not visible at this scale; transport is the mean
  over cycles of each cycle's own schedule ($\tau_{\mathrm{vac}} =
  \SI{60}{s}$, \SI{0.1}{\percent} loss per move).}
  \label{fig:waterfall}
\end{figure}

The CNN evaluates one crop at a time, at \num{1797} cycles per crop for the
$8\times8$ build\meassim{} (\cref{app:impl:rf}), on one core: over \num{2500}
sites at \SI{200}{MHz} this is \SI{22.5}{ms}, against \SI{40.6}{ms} for the
$12\times12$ build. One matched-filter lane classifies the same array in
\SI{12.5}{\micro\second} at \SI{200}{MHz} (\cref{app:impl:pnr}). Every cycle
is charged its exposure, the \SI{7.4}{ms} camera readout, the detector's
datapath time once the crops are in memory, \SI{0.15}{ms} of rearrangement
(\num{29184} cycles) and the minimum-jerk AOD transport of that cycle's own
move schedule, about \SI{10}{ms}. At the reference point the mean loop period
is therefore \SI{27.5}{ms} with the matched filter and \SI{49.7}{ms} with the
CNN, \SI{45}{\percent} of it detection (\cref{fig:waterfall}). With the
matched filter the loop is bound by camera readout and atom transport, which
together take \SI{64}{\percent} of the period.

\Cref{tab:cores} charges the CNN's inference time against its accuracy. At
its operating exposure the CNN has no accuracy to offer: with its detection
time set to zero it reproduces the matched filter's outcome exactly. Its
\SI{22.5}{ms} per cycle is therefore pure dead time, and it lowers the
success probability from 0.85 to 0.69 and nearly triples the wall-clock per
assembly (\SI{0.43}{s} against \SI{0.16}{s}); with lossless transport both
complete almost every attempt, the CNN at \SI{0.14}{s} against
\SI{0.05}{s}.

\begin{table}[htbp]
  \caption{Cycle time and assembly outcome including inference latency
  ($\tau_{\mathrm{vac}} = \SI{60}{s}$, $35\times35$ target in the
  $50\times50$ array, detector on the FPGA, \SI{10}{ms} exposure)\phys{}.
  Cycle time is the mean loop period of \cref{fig:waterfall}. The CNN is the
  $8\times8$ build on one core; its error rates are per class
  (\cref{tab:params}). Wall-clock: per successful assembly. Lost: atoms per
  attempt, standard error \numrange{0.1}{0.3}.}
  \label{tab:cores}
  \centering
  \footnotesize
  \setlength{\tabcolsep}{4pt}
  \begin{tabular}{@{}lrrrrrrr@{}}
    \toprule
    & & & \multicolumn{3}{c}{\SI{0.1}{\percent} per move} & \multicolumn{2}{c}{lossless transport} \\
    \cmidrule(lr){4-6}\cmidrule(l){7-8}
    Detector & Detect & Cycle & Success & Wall-clock & Lost & Success & Wall-clock \\
    \midrule
    \textbf{Matched filter, 1 lane} & \SI{0.01}{ms} & \SI{27.5}{ms} & \textbf{0.846} & \textbf{\SI{0.16}{s}} & 11.0 & 1.000 & \SI{0.05}{s} \\
    CNN, 1 core                     & \SI{22.5}{ms} & \SI{49.7}{ms} & 0.686 & \SI{0.43}{s} & 16.3 & 0.983 & \SI{0.14}{s} \\
    CNN, no detection time          & ---           & \SI{27.4}{ms} & 0.846 & \SI{0.16}{s} & 11.0 & 1.000 & \SI{0.05}{s} \\
    \bottomrule
  \end{tabular}
\end{table}

\subsection{Loss channels and array lifetime}\label{app:loop:lifetime}

\paragraph{Imaging.} At $\tau_{\mathrm{img}} = \SI{70}{s}$ one \SI{10}{ms}
image of \num{1500} atoms costs about 0.21 atoms; over an attempt, failed
attempts included, imaging costs about one atom (\num{1.02} with the matched
filter, \num{1.25} with the CNN at the reference point)\phys{}. It is the
smallest channel at every trap lifetime.

\paragraph{Transport and collisions.} An attempt makes about \num{1300} moves
per cycle, so \SI{0.1}{\percent} per move costs \num{6.7} atoms per attempt
with the matched filter on the FPGA, three fifths of all loss at the
reference point, and \SI{1}{\percent} per move, about 13 atoms per cycle,
prevents any assembly from completing. With both detectors error-free at the
operating exposure, the corrected planner makes no collisions; the planner as
built makes about 45 per attempt (\cref{tab:fom}).

\begin{figure}[htbp]
  \centering
  \includegraphics[width=\textwidth]{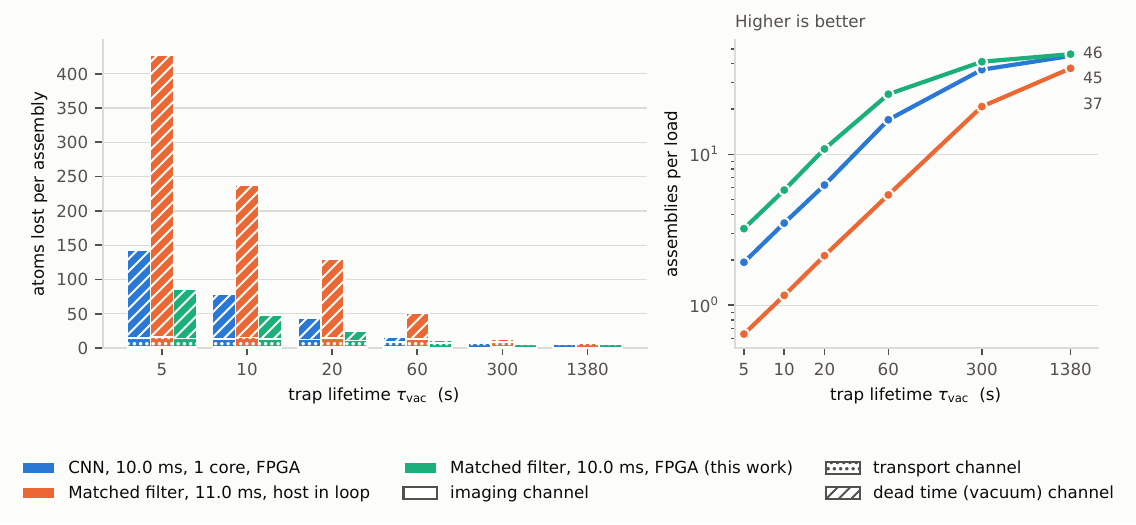}
  \caption{Atom loss per attempt by channel (left) and assemblies per
  stochastic load (right) against trap lifetime $\tau_{\mathrm{vac}}$\phys{}.
  Stacked from the bottom: imaging (solid), transport (dotted) and dead-time
  (hatched) loss; the corrected planner makes no collisions at these
  exposures. $50\times50$ array at \SI{60}{\percent} fill with a centred
  $35\times35$ target, at most nine detect--rearrange cycles per attempt, no
  refill, \SI{0.1}{\percent} loss per moved atom, \num{1000} attempts per
  point; per-class detector errors from the imaging simulator ($8\times8$
  window) at each configuration's operating exposure, the CNN on one core,
  $\tau_{\mathrm{img}} = \SI{70}{s}$. Assemblies per load is the number of
  attempts the \num{275} atoms loaded beyond the target would last: (loaded
  $-$ \num{1225}) / atoms lost per attempt.}
  \label{fig:lifetime}
\end{figure}

\paragraph{Dead time.} The dead-time channel is the one the readout path
controls. At $\tau_{\mathrm{vac}} = \SI{60}{s}$ the FPGA loop loses \num{3.3}
atoms per attempt to it with the matched filter and \num{7.2} with the CNN on
one core; with the \SI{150}{ms} host round-trip it loses \num{37.8}, and the
assembly succeeds in \SI{8.2}{\percent} of attempts instead of
\SI{84.6}{\percent}\phys{}. Each cycle with the host costs about \num{4.4}
atoms of the \num{1500} to vacuum, more than three of them in the target, and
they must be re-supplied in the next cycle, which costs as many again. Over
$\tau_{\mathrm{vac}} \in [5, 1380]\,\si{s}$ dead time is
\SIrange{2}{84}{\percent} of all loss with the matched filter on the FPGA and
\SIrange{11}{96}{\percent} with the host in the loop (\cref{fig:lifetime},
\cref{tab:lifetime}).

\begin{table}[htbp]
  \caption{Assembly success probability and wall-clock per successful
  assembly against trap lifetime, $35\times35$ target in the $50\times50$
  array, \SI{0.1}{\percent} loss per moved atom, no refill\phys{}. Each
  configuration runs at its own operating exposure (\cref{app:loop:closed});
  the CNN runs on one core. ---: no attempt of \num{1000} succeeded.}
  \label{tab:lifetime}
  \centering
  \small
  \begin{tabular}{@{}lcccccc@{}}
    \toprule
    $\tau_{\mathrm{vac}}$ & \SI{5}{s} & \SI{10}{s} & \SI{20}{s} & \SI{60}{s} & \SI{300}{s} & \SI{1380}{s} \\
    \midrule
    \multicolumn{7}{@{}l}{\emph{Success probability}} \\
    \textbf{Matched filter, \SI{10}{ms}, on-FPGA} & 0.003 & 0.096 & 0.447 & \textbf{0.846} & 0.945 & 0.963 \\
    CNN, \SI{10}{ms}, 1 core, on-FPGA             & 0.000 & 0.007 & 0.115 & 0.686 & 0.925 & 0.956 \\
    Matched filter, \SI{11}{ms}, host in loop     & 0.000 & 0.000 & 0.000 & 0.082 & 0.790 & 0.937 \\
    \midrule
    \multicolumn{7}{@{}l}{\emph{Wall-clock per successful assembly}} \\
    \textbf{Matched filter, \SI{10}{ms}, on-FPGA} & \SI{82}{s} & \SI{2.4}{s} & \SI{0.44}{s} & \textbf{\SI{0.16}{s}} & \SI{0.11}{s} & \SI{0.10}{s} \\
    CNN, \SI{10}{ms}, 1 core, on-FPGA             & ---        & \SI{64}{s}  & \SI{3.7}{s}  & \SI{0.43}{s} & \SI{0.22}{s} & \SI{0.19}{s} \\
    Matched filter, \SI{11}{ms}, host in loop     & ---        & ---         & ---          & \SI{19}{s}   & \SI{1.2}{s}  & \SI{0.77}{s} \\
    \bottomrule
  \end{tabular}
\end{table}

\paragraph{Net.} At the reference point the pipeline with the matched filter
on the FPGA completes the $35\times35$ target in \SI{85}{\percent} of
attempts at \SI{0.16}{s} per successful assembly; with the \SI{150}{ms} host
round-trip it completes \SI{8}{\percent} at \SI{19}{s}, about 120 times the
wall-clock (\cref{tab:fom}). The host loop catches up only as the trap
lifetime grows: at \SI{300}{s} it succeeds in \SI{79}{\percent} of attempts,
at $11\times$ the wall-clock, and at \SI{1380}{s} in \SI{94}{\percent}, at
$7.5\times$ (\cref{tab:lifetime}). With lossless transport the ordering is
the same: \SI{100}{\percent} against \SI{23}{\percent} at \SI{60}{s}. The
CNN on one core succeeds less often than the matched filter at every
lifetime and takes about twice its wall-clock from \SI{300}{s}, nearly three
times at \SI{60}{s}.

\begin{figure}[htbp]
  \centering
  \includegraphics[width=0.6\textwidth]{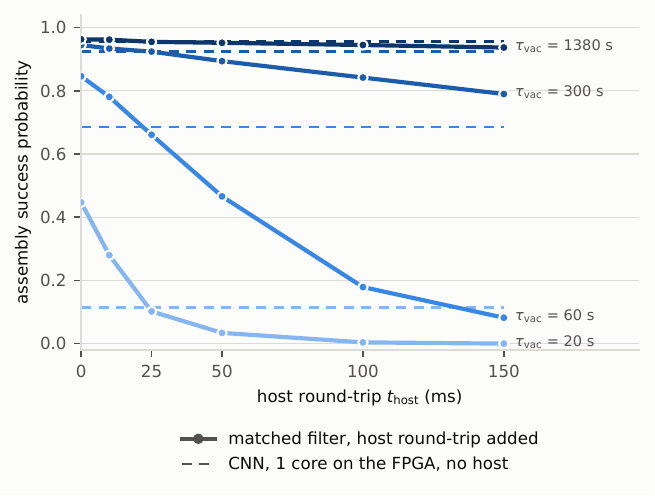}
  \caption{Assembly success against the host round-trip $t_{\mathrm{host}}$
  added to every cycle, at four trap lifetimes\phys{}: matched filter at
  \SI{10}{ms} (solid, markers at the simulated round-trips; the
  \SI{150}{ms} point at its own operating exposure of \SI{11}{ms}) and, for
  comparison, the CNN at \SI{10}{ms} on one FPGA core with no host (dashed,
  same shade). $50\times50$ array, centred $35\times35$ target,
  \SI{60}{\percent} fill, no refill, at most nine cycles per attempt,
  \SI{0.1}{\percent} loss per moved atom, \num{1000} attempts per point.
  $\tau_{\mathrm{vac}} = \SI{1380}{s}$ is the \SI{23}{min} lifetime
  of~\cite{manetsch2025tweezer6100}.}
  \label{fig:breakeven}
\end{figure}

\paragraph{Sensitivity.} The dead-time conclusion depends on the host
round-trip $t_{\mathrm{host}}$, which has not been measured on the target
apparatus; the \SI{150}{ms} used above is an assumption. At the reference
point every millisecond counts (\cref{fig:breakeven}): \SI{10}{ms} of host
round-trip lowers the success probability from 0.85 to 0.78, \SI{25}{ms} to
0.66, \SI{50}{ms} to 0.47 and \SI{100}{ms} to 0.18; at \SI{20}{s} the host
loop completes fewer than \SI{4}{\percent} of attempts from \SI{50}{ms}, and
only at \SI{1380}{s} is the penalty small at every round-trip studied. The
CNN's \SI{22.5}{ms} of inference acts as dead time of the same kind: at
\SI{20}{s} and \SI{60}{s} it costs about as much success as a \SI{25}{ms} host
round-trip. The conclusions also hold at a photon rate three times below
that of \cref{tab:simcfg}, closer to the exposures laboratories report, with
each configuration re-optimised over \SIrange{21}{33}{ms}: the matched filter
on the FPGA succeeds in \SI{70}{\percent} of attempts at \SI{0.40}{s}
(\SI{30}{ms}), the CNN on one core in \SI{51}{\percent} at \SI{0.91}{s}
(\SI{28.5}{ms}), and the host loop in \SI{6}{\percent} at \SI{30}{s}.
Measuring $t_{\mathrm{host}}$ on the existing host pipeline, from frame
arrival to trigger, fixes the size of the penalty.

\paragraph{Figures of merit.} \Cref{tab:fom} gives the figures of merit
usually reported for rearrangement at the reference point\phys{}. The FPGA
loop completes a $35\times35$ assembly in \SI{0.16}{s} of wall-clock with the
matched filter and \SI{0.43}{s} with the CNN on one core, against
\SI{19}{s} with the matched filter behind the host and \SI{29}{s} with the
CNN behind it. The planner as built, compared at the same \SI{10}{ms}, loses
45 atoms per attempt to collisions, all from its column step since detection
is error-free, against none with the corrected step. With lossless transport
it is the worse of the two (0.995 against 1.000 success, 3.4 cycles against
1.8, \SI{0.09}{s} against \SI{0.05}{s} per assembly); at \SI{0.1}{\percent}
per move it succeeds more often (0.943 against 0.846, \SI{0.14}{s} against
\SI{0.16}{s}). The difference lies in the repair cycles. Both planners
re-plan the whole array from the detected map every cycle, a full row sweep
and column compression, so after the first cycle they fill the two or so
remaining holes by shifting atoms already in place: about \num{1130} moves
per repair cycle with the corrected step and \numrange{670}{940} as built, nearly all
of atoms that start inside the target (\num{300} attempts per planner). At
\SI{0.1}{\percent} per move a repair cycle then loses about one atom with the
corrected step and 0.7 as built, each opening a new hole; the corrected
planner completes a repair cycle without loss about a third of the time, the
planner as built about half. The planner as built makes nearly all of its
collisions in the first cycle, where the \num{275} spare atoms absorb them.
With error-free detection, the failures under transport loss in
\cref{tab:fom} therefore come from this repair-cycle churn, not from
detection or dead time. Every configuration uses the same planner, so the
comparisons between configurations stand, but the absolute success under
transport loss is that of a full re-plan. A planner that fills holes directly
from the reservoir in repair cycles, in tens of moves rather than a thousand,
would raise it; this is left as future work for the engine
(\cref{app:rtl:algorithm}).

\begin{table}[htbp]
  \caption{Rearrangement figures of merit at $\tau_{\mathrm{vac}} =
  \SI{60}{s}$\phys{}: $50\times50$ array, centred $35\times35$ target,
  \SI{60}{\percent} fill, no refill, at most nine cycles, \num{1000} attempts
  per row, corrected planner unless stated, each configuration at its
  operating exposure, the CNN on one core. Success: target complete within
  nine cycles. Cycles: mean over successful attempts. Cycle time: mean over
  all cycles. Wall-clock per success: mean time per attempt, failed attempts
  included, divided by the success probability. Lost: atoms lost per attempt,
  all channels.}
  \label{tab:fom}
  \centering
  \footnotesize
  \setlength{\tabcolsep}{4pt}
  \begin{tabular}{@{}lccccc@{}}
    \toprule
    Configuration & Success & Cycles & Cycle time & Wall-clock / success & Lost \\
    \midrule
    \multicolumn{6}{@{}l}{\emph{\SI{0.1}{\percent} loss per moved atom}} \\
    \textbf{Matched filter, \SI{10}{ms}, on-FPGA} & 0.846 & 4.10 & \SI{27.5}{ms} & \textbf{\SI{0.16}{s}} & 11.0 \\
    CNN, \SI{10}{ms}, 1 core, on-FPGA          & 0.686 & 4.49 & \SI{49.7}{ms} & \SI{0.43}{s} & 16.3 \\
    Matched filter, \SI{11}{ms}, host in loop  & 0.082 & 5.24 & \SI{178}{ms}  & \SI{19}{s}   & 51.1 \\
    CNN, \SI{10}{ms}, 1 core, host in loop     & 0.061 & 5.26 & \SI{200}{ms}  & \SI{29}{s}   & 56.3 \\
    Matched filter, \SI{10}{ms}, on-FPGA, planner as built & 0.943 & 4.59 & \SI{26.4}{ms} & \SI{0.14}{s} & 54.7 \\
    \midrule
    \multicolumn{6}{@{}l}{\emph{Lossless transport}} \\
    Matched filter, \SI{10}{ms}, on-FPGA       & 1.000 & 1.80 & \SI{29.3}{ms} & \SI{0.05}{s} & 1.6 \\
    CNN, \SI{10}{ms}, 1 core, on-FPGA          & 0.983 & 2.63 & \SI{50.8}{ms} & \SI{0.14}{s} & 4.0 \\
    Matched filter, \SI{11}{ms}, host in loop  & 0.225 & 4.84 & \SI{178}{ms}  & \SI{6.4}{s}  & 37.1 \\
    Matched filter, \SI{10}{ms}, on-FPGA, planner as built & 0.995 & 3.41 & \SI{26.5}{ms} & \SI{0.09}{s} & 47.5 \\
    \bottomrule
  \end{tabular}
\end{table}

\section{Decoder Implementation on Two FPGA Targets}\label{app:decoder}

This appendix maps the 12-bit learned message-passing model specified in
\cref{app:decoder:model} to two FPGA targets: the \texttt{xczu49dr} RFSoC
as a programmable-logic (PL) kernel, and the Versal AI Edge \texttt{xcve2802},
with arithmetic split between the PL and the AI Engine (AIE) array.
The builds use Vitis HLS and, for the Versal partition, the AIE tools of
Vitis 2026.1 at a \SI{3.33}{ns} (\SI{300}{MHz}) target.
The trained checkpoint has a logical error rate of \SI{0.98}{\percent}
in software evaluation at $p = 0.01$ under code-capacity noise.
The hardware implementations reproduce the integer reference on the
validation vectors described in \cref{app:decoder:model}.
A separate quantised GraphMP ranking kernel with a min-sum stage runs at
\SI{250}{MHz} on the \texttt{xczu49dr}, receiving syndromes over NVQLink
(\cref{app:if:nvqlink}). Each benchmark shot processes one syndrome input;
the message-passing rounds below are internal algorithmic iterations.

\subsection{Model and bit-exact contract}\label{app:decoder:model}

The decoder targets the LP20 $[[1122,148,\le 20]]$ lifted-product code. Its
Tanner graph has $C = 990$ check nodes, $F = 2244$ bit nodes and $E = 7920$
edges. Each of $R = 4$ rounds gathers check-to-bit messages, updates every bit
node, gathers bit-to-check messages and updates every check node, in that order.
A node update is a residual MLP on the node's feature and its aggregate,
$\mathrm{LN}(x + W_2\,\mathrm{ReLU}(W_0[x,\mathrm{agg}]))$, with hidden width
$H = 32$, followed by a normalisation. The layer is weight-tied across rounds,
so one physical update unit per direction is iterated four times, and the model
has about \num{6.5}k parameters, all on chip. Weights and activations are
quantised to 12 bits (W12A12) after training. The reciprocal square root of the
normalisation is replaced by GateNorm, a lookup table indexed by the bit fields
of the variance, which removes every division and square root from the
datapath.

Numerical agreement is checked in stages: the trained model agrees with a NumPy
reference to within $10^{-3}$ in the logits; the reference produces the same decoded outputs as an
integer model; and the HLS kernel reproduces the integer model's logits exactly
on golden vectors (8 shots, \num{17952} logits). Every kernel change below was
required to pass those C-simulation checks, and the final designs were re-checked in RTL
co-simulation and, for the AIE, in the cycle-accurate simulator.

\subsection{Programmable-logic kernel on the \texttt{xczu49dr}}\label{app:decoder:zu49dr}

The first synthesisable kernel did not fit the device: \SI{287}{\percent} of
its LUTs and \SI{275}{\percent} of its BRAM\csynth{}, with a \SI{17.40}{ns}
critical path and a \SI{13}{h} C-synthesis. Five bit-exact changes, listed in
\cref{tab:dec_zu49dr_steps}, brought it inside the part at the target clock.

\begin{table}[htbp]
  \caption{Decoder on the \texttt{xczu49dr}: bit-exact changes and their
  effect\csynth{}. Latency at \SI{3.33}{ns}.}
  \label{tab:dec_zu49dr_steps}
  \centering
  \small
  \begin{tabular}{@{}L{6.2cm}rrl@{}}
    \toprule
    Change & Cycles/shot & Latency & Effect \\
    \midrule
    Variance accumulator from \texttt{double} to exact 64-bit integer &
      --- & --- & critical path \SI{17.40}{ns} $\to$ \SI{2.33}{ns} \\
    GateNorm index read from IEEE-754 bit fields &
      --- & --- & GateNorm \num{76} $\to$ \num{40} cycles \\
    Adjacency as CSR data walked by a rolled loop &
      \num{4823578} & \SI{15.77}{ms} & fits: LUT \SI{11.2}{\percent}, BRAM \SI{24.5}{\percent} \\
    Node updates batched as two GEMMs per round &
      \num{401466} & \SI{1.31}{ms} & \num{8.3} $\to$ \num{99} MAC/cycle \\
    Gather streamed as one flat CSR edge loop &
      \num{238126} & \SI{0.78}{ms} & gathers at II$=$1 over \num{7920} edges \\
    \bottomrule
  \end{tabular}
\end{table}

\begin{enumerate}[leftmargin=1.2em,itemsep=2pt,topsep=2pt]
  \item \textbf{Timing.} The \SI{17.40}{ns} path is a single
        \texttt{double} multiply--accumulate in the recurrence of an II$=$1
        pipeline. An exact integer accumulator bound to one DSP multiply
        removes this path. Vitis HLS does not strength-reduce the
        \texttt{double} multiplication by a power of two in this configuration,
        so the GateNorm index is computed from IEEE-754 bit fields.
  \item \textbf{Area.} Replicated gather-index ROMs account for
        \SI{97}{\percent} of the BRAM. The fully unrolled gather reads every
        entry in one cycle, requiring HLS to duplicate the ROMs for read ports
        and giving a storage efficiency of \SI{0.56}{\percent}. Storing
        adjacency as CSR data and traversing it with a rolled loop reduces BRAM
        \num{9.0}-fold, LUT \num{28.6}-fold, and C-synthesis time from
        \SI{13}{h} to \SI{33}{s}.
  \item \textbf{Latency.} Because the layer is weight-tied, a round is two
        matrix--matrix products, $(N \times 2H)(2H \times H)$ and
        $(N \times H)(H \times H)$, which hold \SI{99.8}{\percent} of the
        arithmetic. Node-by-node execution required pipeline fill and drain
        \num{12936} times per shot. Batching each phase into one flattened
        pipeline reduced latency by a factor of \num{12.0}. Streaming the
        gather as one edge loop, with a flag on each destination's final
        edge, provided a further latency reduction by a factor of \num{1.69}.
\end{enumerate}

The resulting design, with \texttt{MLP\_LANES}$\,=8$ output neurons per cycle
per update unit, is given in \cref{tab:dec_compare}. Doubling the lanes
reaches \SI{0.53}{ms} at \SI{82}{\percent} LUT and \SI{77}{\percent} DSP, and
32 lanes do not fit\csynth{}. Place-and-route of the standalone kernel misses
\SI{300}{MHz} at \SI{3.514}{ns}\meashw{}, so the achievable latency is
\SI{0.84}{ms} per shot. This is the design budgeted in
\cref{tab:coresident}; its \num{1655} DSPs are what bind detector replication
when the two share the part.

\subsection{PL and AI Engine partition on the \texttt{xcve2802}}\label{app:decoder:versal}

\paragraph{Why a partition.} Moving the PL-only kernel to an AIE-equipped
Versal part does not help by itself: those parts trade fabric DSPs for the AIE
array, and the kernel needs \SI{132}{\percent} of the \texttt{xcve2802}'s
\num{1312} DSPs\csynth{}. Moving only the node-update GEMMs to the AIE does not
help much either. A round on the \texttt{xczu49dr} spends \SI{44}{\percent} of
its cycles in the GEMMs, \SI{28}{\percent} in the floating-point
normalisation and \SI{27}{\percent} in the gathers\csynth{}, which caps the
speed-up from offloading the GEMMs with the stages still in sequence at about
\num{1.8}$\times$. The gain comes from streaming each half-round so that the
gather, the AIE hand-off and the normalisation overlap.

\paragraph{Architecture.} The AIE tiles run the two GEMMs of each node update,
which are \SI{99.8}{\percent} of the MACs; the PL keeps the CSR gathers and the
normalisation, with arithmetic unchanged from the PL-only kernel. Each
half-round is one dataflow region,
\[
  \text{gather} \to \text{dispatch} \Rightarrow \text{AIE tiles} \Rightarrow
  \text{collect} \to \text{stats} \to \text{LN},
\]
where $\Rightarrow$ marks a bank of AXI4-Stream ports. Nodes are dealt round-robin over 17 tiles for the bit direction
($2244 = 17 \times 132$) and 8 tiles for the check direction (990 padded to
992), one 128-bit PLIO pair per tile. Each node crosses as eight row-major
beats, and the tile's DMA scatters them into the $4\times4$ operand tiles of
\texttt{aie::mmul<4,4,4>}, so neither side transposes in logic. The AIE kernel
takes \num{715} cycles per four-node call, \SI{0.57}{\micro\second} at
\SI{1.25}{GHz}, which out-runs the gather by about \num{1.4}$\times$ in both
directions\meassim{}. Two further bit-exact changes fold the initialisation
and the readout into the first and last half-rounds, and drop the final check
half-round, whose output the readout never reads.

\paragraph{Timing closure.} The first dataflow build had a \SI{23.66}{ns}
estimated clock and \SI{80}{\percent} LUT\csynth{}. Every failing path was in
loop control rather than datapath: a 17-input adder counting finished ports in
one cycle, an OR tree of busy flags, and a 17:1 select feeding a counter.
Replacing per-port \num{4096}-bit buffers by a node round-robin with one row
buffer per port halved the LUTs, and running the dispatch loop at II$=$2, which
throughput allows since rows arrive at least three cycles apart, brought the
estimate to \SI{2.77}{ns}. Place-and-route closes at \SI{3.327}{ns}, WNS
\SI{+0.003}{ns}, WHS \SI{+0.010}{ns}\meashw{}.

\paragraph{Latency.} RTL co-simulation, with the AIE results pre-loaded into the
input streams, gives \num{7950} cycles per bit half-round and \num{8056} per
check half-round, within \SI{1.6}{\percent} of a bare gather: everything
downstream of the gather is hidden behind it\meassim{}. A shot is four bit and
three check half-rounds, \num{55968} PL cycles or \SI{186.6}{\micro\second},
plus an AIE tail of \SIrange{5}{15}{\micro\second}, so
\SIrange{192}{202}{\micro\second} per shot.

\begin{table}[htbp]
  \caption{The decoder on the two targets at a \SI{3.33}{ns} target. Utilisation
  as a share of each device.}
  \label{tab:dec_compare}
  \centering
  \small
  \begin{tabular}{@{}lrr@{}}
    \toprule
    & \texttt{xczu49dr}, PL only & \texttt{xcve2802}, PL $+$ AIE \\
    \midrule
    Clock, estimated & \SI{3.270}{ns}\csynth{} & \SI{2.77}{ns}\csynth{} \\
    Clock, post-route & \SI{3.514}{ns}\meashw{} & \SI{3.327}{ns}\meashw{} \\
    Latency per shot & \SI{0.84}{ms}\roof{} & \SIrange{192}{202}{\micro\second}\meassim{} \\
    LUT & \num{171635} (40.4\%)\csynth{} & \num{163021} (31.3\%)\meashw{} \\
    FF & \num{186815} (22.0\%)\csynth{} & \num{195258} (18.7\%)\meashw{} \\
    DSP & \num{1655} (38.7\%)\csynth{} & \num{704} (53.7\%)\meashw{} \\
    BRAM\_18K & \num{805} (37.3\%)\csynth{} & \num{558} (46.5\%)\meashw{} \\
    AIE tiles & --- & 25 (17 bit, 8 check) \\
    \bottomrule
  \end{tabular}
\end{table}

\subsection{Findings}\label{app:decoder:findings}

\begin{enumerate}[leftmargin=1.2em,itemsep=2pt,topsep=2pt]
  \item \textbf{The partition is \num{4.2}$\times$ faster per shot} than the
        PL-only kernel at its post-route clock, at \num{2.4}$\times$ fewer fabric DSPs.
        It comes from overlapping the stages of a half-round rather than
        from AIE throughput: once the GEMMs leave the PL, the gathers, at one edge per
        cycle, are \SI{88}{\percent} of the latency, and AIE compute is off the
        critical path.
  \item \textbf{Resource and timing attribution.} Analysis of the HLS
        scheduler and per-array memory reports identified the recurrence
        arithmetic responsible for the timing failure and ROM replication
        responsible for the BRAM overflow on the \texttt{xczu49dr}.
  \item \textbf{Post-route timing differs from HLS estimates.} The estimated
        periods of \SI{3.270}{ns} and \SI{2.77}{ns} became
        \SI{3.514}{ns} and \SI{3.327}{ns} after place-and-route on the
        \texttt{xczu49dr} and \texttt{xcve2802}, respectively. The
        \SIrange{7}{20}{\percent} increase illustrates the need to confirm
        synthesis estimates through implementation, as also observed for
        detector resources (\cref{tab:csynth_vs_pnr}).
  \item \textbf{The AIE tools have simulator-only failure modes.} A register
        split of a 32-lane \texttt{int16} vector
        (\texttt{aie::filter\_even}/\texttt{filter\_odd}) is correct in the
        functional x86 simulator but halts the AIE-ML core in the
        cycle-accurate simulator; it was found by bisection and avoided by
        moving all tiling into the DMA. The dense \texttt{int16} GEMM shapes
        also differ between AIE generations, so the kernel as built targets
        AIE-ML only.
\end{enumerate}

\section{Platform Boundary}\label{app:platform}

\begin{figure}[htbp]
  \centering
  \includegraphics[width=\textwidth]{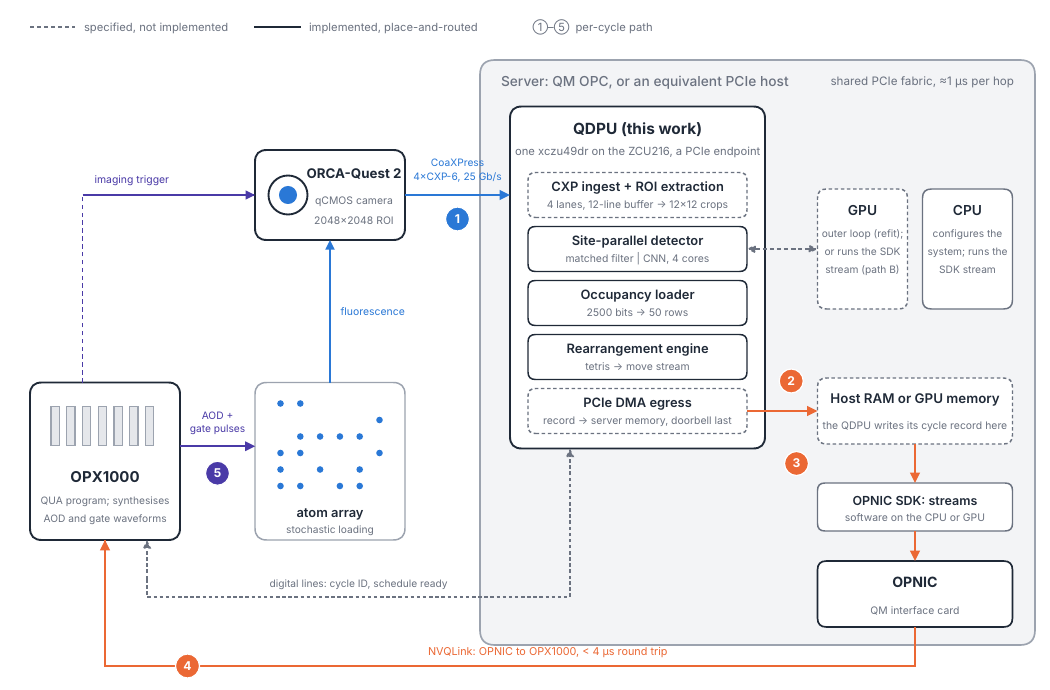}
  \caption{QDPU boundary. Fluorescence is digitised by the camera and
  enters the QDPU over CoaXPress. The QDPU, the OPNIC, the CPU and the GPU
  share the PCIe fabric of one server, Quantum Machines' OPC or an equivalent
  host. Once per cycle the QDPU writes occupancy and the move stream into
  host or GPU memory; a stream defined with the vendor's OPNIC SDK, running
  on the CPU or GPU, carries the record to the OPNIC, which forwards it to the
  OPX1000 over NVQLink; the OPX1000 synthesises the AOD and gate waveforms.
  Dashed blocks are specified but not implemented.}
  \label{fig:opx}
\end{figure}

The QDPU sits between two commercial instruments: a Hamamatsu ORCA-Quest~2
camera upstream and a Quantum Machines OPX1000 controller downstream
(\cref{fig:opx}). In a conventional deployment a frame grabber, a server and a
feedback card occupy this position, and every frame crosses the server's
software on its way to the controller; the QDPU is specified to take all three
roles on one device. Physically the device is a PCIe endpoint in the server
that also hosts the controller's interface card, the OPNIC, and the outer
loop's GPU: Quantum Machines' OPC (OP-Central), the vendor's validated host, or any PCIe
server of the same topology. The server's CPU configures the
system and allocates the memory the devices exchange data through, and runs the vendor's OPNIC software that streams the QDPU's output to the controller (\cref{app:platform:controller}). This appendix
specifies each boundary and its implementation status.

\begin{table}[htbp]
  \caption{Summary of the three links. Rates are line rates; payload rates
  after encoding are in the per-link subsections. Status as in
  \cref{app:platform:ifaces}.}
  \label{tab:if:summary}
  \centering
  \small
  \begin{tabular}{@{}L{1.8cm}L{2.6cm}L{3.3cm}L{2.5cm}L{2.3cm}l@{}}
    \toprule
    Link & Peer & PHY / protocol & Line rate & FPGA resources & Status \\
    \midrule
    Ingress & ORCA-Quest~2 (C15550-22UP) & CoaXPress~1.1, 4$\times$CXP-6, 75\,$\Omega$ coax; GenICam control & $4\times\SI{6.25}{Gb/s}$ & 1 GTY quad; CXP host IP; DDR4 frame buffer (optional) & \stSpec \\
    Egress & OPX1000 via OPNIC & PCIe Gen3 (x4--x8) DMA into server memory; OPNIC SDK stream; NVQLink to the controller & up to \SI{63}{Gb/s} (x8) & PCIE4 hard block; 1--2 GTY quads & \stSpec \\
    Outer loop & NVIDIA GPU in the same server (variant: GPU cluster via ConnectX NIC) & PCIe, GPUDirect RDMA into GPU memory (variant: NVQLink, RoCEv2 over 10 or \SI{25}{GbE}; \SI{10}{GbE} measured, \cref{app:if:nvqlink}) & shared with egress (variant: 10 or \SI{25}{Gb/s} per SFP28 cage) & shared PCIE4 block (variant: Ethernet PCS/PMA and MAC, RoCE engine, 1--4 SFP28 lanes) & \stOpen \\
    \bottomrule
  \end{tabular}
\end{table}

\subsection{Interface specifications}\label{app:platform:ifaces}

\Cref{tab:if_camera,tab:if_outer,tab:if_ctrl} specify the three external
interfaces of the QDPU. Status is one of \stImpl{} (in the place-and-routed
design), \stSpec{} (defined, not implemented) or \stOpen{} (undefined).

\begin{table}[htbp]
  \caption{Camera interface (upstream).}
  \label{tab:if_camera}
  \centering
  \small
  \begin{tabular}{@{}L{2.6cm}lL{4.2cm}L{2.6cm}L{2.2cm}l@{}}
    \toprule
    Channel & Dir. & Payload & Width / rate & Latency budget & Status \\
    \midrule
    Video downlink & in & $2048\times2048\times16$-bit ROI frames (sensor $4096\times2304$) & quad CXP-6, $4 \times \SI{6.25}{Gb/s}$ (\SI{20}{Gb/s} payload) & readout-bound: \SI{7.4}{ms} standard / \SI{36}{ms} ultra quiet per ROI & \stSpec \\
    ROI extraction & int. & $8\times8\times16$-bit crops, 2500 per frame (\SI{2.6}{Mb}) & line-buffered, 8 lines; no frame buffer & 8 lines behind the stream: \SIrange{0.03}{0.14}{ms} by scan mode & \stSpec \\
    Exposure trigger & out & trigger pulse & OPX1000 digital output to the camera's external trigger input (default); CXP low-speed uplink as alternative & camera trigger latency, to be measured at bring-up & \stSpec \\
    Transceivers & --- & --- & 4 GTY lanes (one quad) & --- & \stSpec \\
    \bottomrule
  \end{tabular}
\end{table}

\begin{table}[htbp]
  \caption{Outer-loop interface (GPU cluster).}
  \label{tab:if_outer}
  \centering
  \small
  \begin{tabular}{@{}L{2.6cm}lL{4.2cm}L{2.6cm}L{2.2cm}l@{}}
    \toprule
    Channel & Dir. & Payload & Width / rate & Latency budget & Status \\
    \midrule
    Data collection & out & crops or occupancy with labels; syndromes & $\le \SI{2.6}{Mb}$ per frame (all crops), \num{2500} b per frame (occupancy) & min to hr & \stOpen \\
    Policy update & in & detector template and threshold, or CNN weights; decoder parameters & matched filter: 64 \texttt{int8} coefficients plus one integer threshold; CNN: \num{4786} weights and biases at 12 bits, \SI{7179}{B} when packed (about \SI{7.2}{kB}), plus normalisation constants; decoder: about \SI{9.8}{kB} & min to hr & \stOpen \\
    Physical link & --- & --- & NVQLink/RoCEv2 at \SI{10}{GbE}, measured on the board (\cref{app:if:nvqlink}); alternative paths in \cref{app:if:gpu} & --- & \stImpl \\
    \bottomrule
  \end{tabular}
\end{table}

\begin{table}[htbp]
  \caption{Controller interface (downstream, OPX1000 $+$ OPNIC).}
  \label{tab:if_ctrl}
  \centering
  \small
  \begin{tabular}{@{}L{2.6cm}lL{4.2cm}L{2.6cm}L{2.2cm}l@{}}
    \toprule
    Channel & Dir. & Payload & Width / rate & Latency budget & Status \\
    \midrule
    Occupancy & out & 2500 bits per frame & PCIe Gen3 x4, \SI{316}{B} per cycle (\cref{tab:if:opx:payload}) & $< \SI{2}{\micro\second}$ link time; $< \SI{4}{\micro\second}$ vendor round trip to the controller~\cite{qm_dgxq_blog2025} & \stSpec \\
    Move stream & out & from/to (row, col) per move, ready/valid & 24-bit coordinates packed into a 32-bit word per move; $4D$ bytes for $D$ move descriptors (\cref{tab:if:opx:payload}) & within the same transfer & \stSpec \\
    Shuttle API & out & undefined in this iteration (\cref{app:open}) & --- & --- & \stOpen \\
    Control API & out & corrections and measurement selection from the decoder, real-time regime (\cref{sec:workloads:decoding}); undefined in this iteration & --- & --- & \stOpen \\
    Server memory and SDK stream & --- & the cycle record of \cref{tab:if:opx:payload} & host memory (path A) or GPU memory (path B), written by FPGA DMA; a stream defined with the OPNIC SDK carries it to the OPNIC (\cref{app:platform:controller}) & about \SI{1}{\micro\second} per PCIe hop; stream hand-off unmeasured & \stSpec \\
    OPNIC to OPX1000 & --- & the same record & vendor fibre link (NVQLink); no FPGA resource & $< \SI{4}{\micro\second}$ round trip & \stSpec \\
    \bottomrule
  \end{tabular}
\end{table}

\subsection{Camera ingest}\label{app:platform:camera}

The imaging camera is a Hamamatsu ORCA-Quest~2 qCMOS (C15550-22UP):
$4096\times2304$ effective pixels, \SI{4.6}{\micro\metre} pixel pitch,
\SI{0.30}{e^-} readout noise and \SI{25}{fps} at full frame in ultra quiet
scan~\cite{hamamatsu2024orcaquest2}. The pipeline is evaluated on a
$2048\times2048$ sub-array readout of this sensor; the simulator used for all
detector results models this camera family~\cite{winklmann2023imagesim}. The
camera streams over CoaXPress as a quad CXP-6 device: four lanes at
\SI{6.25}{Gb/s}, \SI{25}{Gb/s} raw and \SI{20}{Gb/s} of payload after 8b/10b
coding~\cite{hamamatsu2024orcaquest2}. The two scan modes trade read noise for
readout time: \SI{0.30}{e^-} rms at \SI{25}{fps} full frame in ultra quiet
scan, \SI{0.43}{e^-} rms at \SI{120}{fps} in standard scan, which for the
2048-row ROI is about \SI{36}{ms} against \SI{7.4}{ms} of readout
(\cref{tab:if:cxp:budget}). The imaging simulator's read noise of
\SI{0.43}{e^-} (\cref{tab:simcfg}) corresponds to the standard scan, so the
\SI{7.4}{ms} figure is the one the detector results assume; \cref{tab:latency}
carries both. The link requires multi-gigabit transceivers, budgeted at one
GTY quad of the sixteen on the target device, with per-lane clock recovery,
lane deskew and bonding, and a CXP link layer including the low-speed uplink
that carries the exposure trigger. The uplink is what allows the loop to close
on-chip. The ingest datapath is specified and not implemented; closing the loop on the device depends on it.

\paragraph{Protocol stack.} CoaXPress is asymmetric. The \emph{downlink}
(camera to host) carries stream packets at the lane rate; the \emph{uplink}
(host to camera) is a low-speed channel, \SI{20.83}{Mb/s} in CXP~1.1, that
carries register reads and writes (the GenICam control plane) and trigger
packets~\cite{jiia2021coaxpress}. The FPGA-side stack is:
\begin{enumerate}[leftmargin=1.6em,itemsep=2pt,topsep=3pt]
  \item \textbf{Analog front-end.} A CoaXPress equaliser/driver per lane
        between the coax and the GTY, with the low-speed uplink superimposed
        on the same conductor, and, if required, PoCXP \SI{24}{V} injection.
        The ZCU216 has no CoaXPress connector, so this is an FMC card
        (commercial CXP-host FMC, or a custom one).
        
  \item \textbf{Transceivers.} Four GTY receivers at \SI{6.25}{Gb/s} with
        8b/10b decode and comma alignment, 32-bit user width, recovered user
        clock $6.25\,\mathrm{GHz}/40 = \SI{156.25}{MHz}$ per lane.
  \item \textbf{CXP link layer.} Packet deframing, CRC-32 check, lane deskew
        and bonding into one stream, stream-packet reassembly into image
        lines, and the uplink encoder. Licensed rather than written
        (\cref{app:if:bringup}); the core presents an AXI4-Stream of pixels
        with start-of-frame / end-of-line sideband, plus an AXI4-Lite port for
        the control plane.
  \item \textbf{Control plane.} GenICam register access (ROI, scan mode,
        exposure, trigger source) is driven by the RFSoC's Arm cores through
        the CXP core's AXI4-Lite port. It is configured once per run and is
        not in the per-cycle path.
  \item \textbf{Pixel path.} The AXI4-Stream crosses from the
        \SI{156.25}{MHz} recovered-clock domain into the \SI{200}{MHz}
        processing domain through an asynchronous FIFO
        (\cref{app:if:clock}) and feeds the ROI extractor of
        \cref{app:frontend:contract} directly, line by line.
\end{enumerate}

\paragraph{Streaming, not buffering.} The ROI extractor needs only the
$8$ image rows that span one row of sites before it can emit that row's
$8\times8$ crops, so the pixel path requires a line buffer of
$8\times2048\times16$ bits $=\SI{262}{kb}$ (about 8 of the device's block
RAMs) rather than a frame buffer. Detection can then start roughly one
site-row's worth of lines after the first pixel instead of after the last.
A DDR4 frame buffer in the PL memory is still useful, but off the
critical path: it holds raw frames for the outer loop's data collection
(\cref{app:if:gpu}) and for debugging.

\paragraph{Bandwidth and readout time.} \Cref{tab:if:cxp:budget} gives the
arithmetic. Four CXP-6 lanes carry at most $4\times6.25\times0.8 =
\SI{20}{Gb/s}$ of payload before packet overhead. A $2048\times2048$ ROI at
16 bits is $6.7\times10^{7}$ bits and needs \SI{3.4}{ms} of link time; the full
sensor needs \SI{7.5}{ms}. In either case the link is faster than the sensor
in its low-noise mode, so \emph{sensor readout, not the CoaXPress link, sets
the ingest time}: the ultra-quiet scan at \SI{25}{fps} full frame reads a
2048-row ROI in about $2048/2304\times\SI{40}{ms}\approx\SI{36}{ms}$, while the
standard scan at \SI{120}{fps} reads it in about \SI{7.4}{ms}\roof{}. This is
the ingest row of \cref{tab:latency}, and which of the two values applies
depends on the scan mode the imaging simulator assumes.

\begin{table}[htbp]
  \caption{Ingress bandwidth and time per frame, 16-bit pixels, four CXP-6
  lanes (\SI{20}{Gb/s} payload after 8b/10b, before packet overhead). The
  sensor readout times are scaled from the datasheet frame rates by rows
  read\roof{}.}
  \label{tab:if:cxp:budget}
  \centering
  \small
  \begin{tabular}{@{}lrrrr@{}}
    \toprule
    Readout & Bits / frame & Link time & Sensor, standard & Sensor, ultra quiet \\
    \midrule
    $2048\times2048$ ROI & $6.7\times10^{7}$ & \SI{3.4}{ms} & \SI{7.4}{ms} & \SI{36}{ms} \\
    $4096\times2304$ full & $1.5\times10^{8}$ & \SI{7.5}{ms} & \SI{8.3}{ms} & \SI{40}{ms} \\
    \bottomrule
  \end{tabular}
\end{table}

\paragraph{Trigger.} The exposure trigger can reach the camera in two ways:
over the CXP uplink as a trigger packet, which keeps the whole loop on one
cable but carries a protocol-dependent latency and jitter; or as a TTL edge
on the camera's external trigger input, driven by a PL GPIO or, more
naturally, by the OPX1000 digital output that already sequences the imaging
light (\cref{app:if:sequence}). The second keeps the OPX1000 as the single
time reference for the experiment and is the specified default.

\subsection{Controller link}\label{app:platform:controller}

The OPX1000 is a modular hybrid controller: each of its front-end modules
carries a pulse processor that synthesises analogue waveforms, DC to
\SI{750}{MHz} on the low-frequency module and up to \SI{10.5}{GHz} on the
microwave module, under a program written in QUA, the vendor's pulse-level
language, which the pulse processor executes with deterministic
timing~\cite{qm_opx1000}. It is the instrument that physically moves and gates
the atoms. The link specified here carries decisions rather than waveforms:
once per cycle, occupancy and the move stream leave the FPGA over PCIe into
server memory and reach the OPX1000 through the OPNIC's NVQLink hop; the
OPX1000 owns AOD waveform playback and gate pulses and turns the schedule into
waveforms itself. 

\paragraph{Controller.} The OPX1000 owns everything that must be
nanosecond-deterministic: AOD and Rydberg waveform playback, gate pulses,
and the imaging and trigger timing~\cite{qm_opx1000}. Its pulse processor
can branch on classical values delivered to the running QUA program. The
QDPU therefore does not synthesise waveforms (\cref{app:platform:aod}); it
delivers two things per cycle, the occupancy vector and the move schedule,
and the OPX1000 turns the schedule into multi-tone chirps.

\paragraph{Physical path.} Quantum Machines' route for low-latency data into a
running QUA program is the OP-Network Interface Card (OPNIC): a PCIe card
installed in the server, carrying its own FPGA, that links to the OPX1000 over a
dedicated fibre connection and exchanges data with the server's CPU and GPU
memory by DMA~\cite{qm_opnic}. The vendor describes the OPNIC's data network as
NVQLink-compatible~\cite{qm_opnic}, and the OPNIC to OPX1000 hop is the
NVQLink hop of this system; its round trip is under
\SI{4}{\micro\second}~\cite{qm_dgxq_blog2025}. The OPNIC, the CPU, the GPU and
the QDPU share the PCIe fabric of one server, about \SI{1}{\micro\second}
between any two of them; that server is the vendor's OPC, its validated host,
or any PCIe server of the same topology.

The vendor-supported integration transfers data through server memory.
The OPNIC exchanges data with host memory and, through GPUDirect
RDMA~\cite{nvidia_gpudirect_rdma}, GPU memory. The supported solution
does not include a peer-to-peer path from a third-party endpoint to the
OPNIC. The vendor gives a planning estimate of approximately
\SI{1}{\micro\second} for a PCIe transfer inside the OPC, including
the transfer from server memory to the OPNIC. The OPNIC behaves as a
network interface: its operation and its data streams are defined in software
through the vendor's SDK~\cite{qm_opnic}, and that software, on the CPU or the
GPU, is what carries data between server memory and the card. The FPGA
therefore never addresses the OPNIC. It writes the cycle record of
\cref{tab:if:opx:payload} into a buffer in server memory that the host
allocates once at start-up, and a stream defined with the SDK carries the
record from there to the OPNIC:
\begin{itemize}[leftmargin=1.2em,itemsep=2pt,topsep=3pt]
  \item \textbf{Path A, host memory (primary).} The FPGA's DMA engine writes
        the record into a pinned buffer in host memory, and a stream run on
        the CPU hands it to the OPNIC. Two PCIe hops, no GPU in the path, and
        the standard use of the AMD DMA subsystem, so this is the path the
        bring-up of \cref{app:if:bringup} takes.
  \item \textbf{Path B, GPU memory (variant).} The FPGA writes the record into
        a GPU buffer through GPUDirect RDMA, and a stream run on the GPU hands
        it to the OPNIC. The hop count is the same; it becomes the natural
        choice once a GPU-resident step (\cref{app:if:gpu}) sits between
        detection and the controller.
  \item \textbf{FPGA side.} The \texttt{xczu49dr}'s PCIE4 hard block as
        endpoint, Gen3 x4 or x8, with the AMD DMA/Bridge subsystem (XDMA or
        QDMA) turning the egress AXI4-Stream of \cref{app:frontend:contract}
        into \SI{256}{B} posted writes at the buffer's bus address, programmed
        once by the host. User clock \SI{250}{MHz}.
\end{itemize}
The latency of the stream hand-off, from the record landing in memory to its
departure through the OPNIC, is set by the vendor's software. The vendor's
planning figure for the memory-to-OPNIC transfer is about
\SI{1}{\micro\second}; the hand-off has not been measured (\cref{app:open}).

\paragraph{Payload.} \Cref{tab:if:opx:payload} gives the record layout for
one cycle of the $50\times50$ array. The occupancy map contains \num{2500}
bits, padded to \SI{316}{B}, and each move descriptor occupies one 32-bit
word. Including the header and doorbell, the record contains $336 + 4D$
bytes for $D$ descriptors emitted across both rearrangement phases.
At the modelled operating point of about \num{1300} moves per cycle
(\cref{app:loop:lifetime}), this is approximately \SI{5.5}{kB}, or
\SI{0.20}{MB/s} at the \SI{27.5}{ms} mean cycle period
(\cref{app:loop:latency}). At this operating point, low hand-off latency
is the principal egress requirement; sustained bandwidth demand is modest.

\begin{table}[htbp]
  \caption{Egress payload per cycle, $50\times50$ array, and its layout in server memory. Word size 32 bits.}
  \label{tab:if:opx:payload}
  \centering
  \small
  \begin{tabular}{@{}L{2.6cm}L{5.4cm}rl@{}}
    \toprule
    Region & Contents & Size & Status \\
    \midrule
    Header & cycle ID, move count, CRC-32, detector version & \SI{16}{B} & \stSpec \\
    Occupancy & \num{2500} bits, row-major, padded to 79 words & \SI{316}{B} & \stSpec \\
    Move list & per move: $(r_\text{from}, c_\text{from}, r_\text{to}, c_\text{to})$ 6 b each, group tag, valid & $4D$ bytes & \stSpec \\
    Doorbell & cycle ID written last; checked by whichever agent forwards the record, and again by the QUA program & \SI{4}{B} & \stSpec \\
    \bottomrule
  \end{tabular}
\end{table}

\paragraph{Ordering.} PCIe posted writes from one requester arrive in memory
in order, so writing the doorbell last is sufficient: any reader that sees the
doorbell match the expected cycle ID sees a complete record. The stream that
hands the record to the OPNIC reads the doorbell before it sends, and the QUA
program checks it again on arrival. The CRC
guards against a stale or partially overwritten record if the OPX1000 falls
behind; the QUA program rejects a mismatched CRC and leaves the array
unmoved for that cycle. The \emph{group tag} marks moves that may run
concurrently, and its maximum group size is the AOD tone count the
controller can sustain (\cref{app:platform:aod}).

\paragraph{Hard-wired side channel.} Two OPX1000 digital lines complement the
PCIe path: one OPX1000 output tells the FPGA the frame it is about to receive
belongs to cycle $k$ (so both ends agree on the cycle ID without a
round trip), and one FPGA output to an OPX1000 digital input raises a
``schedule ready'' edge for programs that wait on an input rather than
polling.

\subsection{Outer loop: GPU in the server, or over NVQLink}\label{app:if:gpu}

\paragraph{NVLink versus NVQLink.}\label{app:if:gpu:nvlink} NVLink is NVIDIA's
GPU-to-GPU and GPU-to-CPU fabric. It is not exposed on AMD FPGAs: the only
route for third-party silicon onto NVLink is NVLink Fusion, licensed to a
small set of ASIC and CPU partners~\cite{nvidia_nvlink_fusion}, and the
\texttt{xczu49dr} has no NVLink PHY. The FPGA reaches a GPU in one of two
ways. A GPU in the same server is reached over the PCIe fabric of
\cref{app:platform:controller}, with GPUDirect RDMA placing the FPGA's writes
in GPU memory~\cite{nvidia_gpudirect_rdma}; this is the configuration of the
vendor's OPC and the primary one here. A GPU in another chassis is reached
over \emph{NVQLink}, NVIDIA's open architecture for coupling quantum
controllers to GPUs~\cite{nvidia_nvqlink,caldwell2025nvqlink}, whose
FPGA-facing side is Ethernet with remote DMA into GPU memory. NVLink still
appears in the system, one level up: it connects the GPUs to each other, so
that the outer loop's training and any GPU-resident decoding scale across a
node without passing through host memory (\cref{tab:if:gpu:traffic}).

\paragraph{Physical path.} Inside the server the path is
\begin{center}
  FPGA (PCIE4 block, DMA engine) $\to$ PCIe fabric $\to$ GPUDirect RDMA
  $\to$ GPU memory,
\end{center}
the same endpoint and DMA engine as the egress of
\cref{app:platform:controller} with a second buffer, in GPU memory, as the
target; the outer loop then costs no additional transceivers. Across chassis
the NVQLink path is
\begin{center}
  FPGA (RoCE engine, 10 or \SI{25}{GbE} MAC) $\to$ SFP28 $\to$ ConnectX NIC $\to$
  GPUDirect RDMA $\to$ GPU memory,
\end{center}
with the NIC and GPU on the same PCIe switch so that RDMA writes land in
GPU memory without a host-memory copy~\cite{nvidia_gpudirect_rdma}. On the
FPGA the transport is RoCEv2: UDP/IP over an Ethernet MAC, with an RDMA
engine that issues RDMA WRITE operations into buffers registered in advance.
NVIDIA's Holoscan Sensor Bridge FPGA IP implements this sensor-to-GPU
path~\cite{nvidia_holoscan_hsb}; it is the configuration built and measured
in \cref{app:if:nvqlink}. The ZCU216 carries four SFP28 cages and no
QSFP28~\cite{amd_ug1390_zcu216}, two cages to a quad on quads 128 and 129.
A 100\,GbE CAUI-4 port, whose four lanes must share one quad, cannot be
formed from them, so this variant runs one to four independent 10 or
\SI{25}{GbE} links through a breakout to the NIC's QSFP28. 100\,GbE would
need a QSFP28 card on the FMC+ slot, whose lanes the camera and PCIe links
already take (\cref{tab:if:gty}).

\paragraph{What crosses the link.} The outer loop has no per-cycle deadline
(\cref{sec:unit:interfaces}), but the microsecond round trip of either path
makes a second, faster use possible, so the link is specified for two traffic
classes (\cref{tab:if:gpu:traffic}).
\begin{itemize}[leftmargin=1.2em,itemsep=2pt,topsep=3pt]
  \item \textbf{Data collection and policy update} (the outer loop proper).
        Out: $8\times8\times16$-bit crops, \SI{1024}{b} per site,
        $2.6\times10^{6}$ bits per $50\times50$ frame, with the detector's
        output and, when available, a label from a second image; optionally
        whole raw frames. In: QDM weights for the detector
        (\texttt{1-8-16-16}, 12-bit, \cref{tab:cnn}) and the decoder. Even
        raw full frames at \SI{10}{Hz} require about \SI{1.5}{Gb/s},
        approximately \SI{15}{\percent} of the measured \SI{10}{GbE} line rate
        before protocol overhead.
        Weight updates are written to a shadow copy of the weight memories
        and swapped at a cycle boundary, so the detector never runs on a
        half-written model.
  \item \textbf{GPU-in-the-loop inference} (optional). For a workload too
        large for the FPGA, the FPGA streams its inputs to the GPU and waits
        for the result inside the cycle. At a few microseconds round trip
        this fits the millisecond budgets of \cref{app:loop}; its candidates
        are the \num{10000}-site rearrangement of
        \cref{app:platform:scaling} and decoders larger than the resident
        one. The result returns by a write into an FPGA-side buffer, GPU DMA
        over PCIe or RDMA WRITE over NVQLink, and joins the egress path of
        \cref{app:platform:controller}.
\end{itemize}

\begin{table}[htbp]
  \caption{Outer-loop traffic, $50\times50$ array. Rates at a \SI{10}{Hz}
  cycle for scale\roof{}.}
  \label{tab:if:gpu:traffic}
  \centering
  \small
  \begin{tabular}{@{}L{2.8cm}lL{3.8cm}L{2.2cm}L{2.0cm}l@{}}
    \toprule
    Class & Dir. & Payload & Rate & Deadline & Status \\
    \midrule
    Crops $+$ labels & out & $2500\times8\times8\times16$ b & \SI{26}{Mb/s} & none & \stOpen \\
    Raw frames & out & $4096\times2304\times16$ b & \SI{1.5}{Gb/s} & none & \stOpen \\
    QDM weights & in & detector $+$ decoder parameters & per refit & min to hr & \stOpen \\
    GPU inference & both & workload-specific & workload-specific & $\mu$s & \stOpen \\
    GPU--GPU (NVLink) & --- & gradients, decoder state & node-internal & none & \stOpen \\
    \bottomrule
  \end{tabular}
\end{table}

\subsection{NVQLink link measured on the ZCU216}\label{app:if:nvqlink}

The NVQLink variant has been built and measured on a ZCU216 board linked to an
NVIDIA DGX Spark (GB10: Grace CPU and integrated GPU, ConnectX-7 NIC). The
design is a link test, separate from the pipeline of \cref{tab:impl}: a test
source and, in a second build, a GraphMP decoding kernel stand in for the QDPU's
traffic. It ports the RFSoC~4x2 reference design of NVIDIA's Holoscan Sensor
Bridge (HSB) FPGA IP, release 3.0.0~\cite{nvidia_holoscan_hsb}, to the
\texttt{xczu49dr}.

\paragraph{Link.} The reference design uses a 100\,GbE hard MAC, which the
ZCU216's SFP28 cages cannot carry (\cref{app:if:gpu}). The measured link is
10GBASE-R on one SFP28 cage, through a QSFP28-to-SFP28 breakout into a
ConnectX-7 port. The PHY is AMD's \texttt{xxv\_ethernet} in its PCS/PMA-only
mode, which needs no licence; AMD's 10G/25G MAC does, so the MAC is
\texttt{eth\_mac\_10g} from the open-source \texttt{verilog-ethernet} library.
The HSB datapath is 64 bits wide at \SI{156.25}{MHz} with one sensor channel,
and the FPGA's clock locks to a PTP master on the host. The build closes
timing at WNS \SI{+1.69}{ns}\meashw{}. HSB's software reset must not reach
the PCS: resetting the PCS stops the \SI{156.25}{MHz} clock that HSB needs to
release its own reset, which deadlocked the first build.

\paragraph{Round trip.} A RAM player in the FPGA sends 128-byte messages
stamped with the PTP-locked FPGA time; the host returns each one, and a
timestamp logic analyser records its arrival at \SI{10}{ns} resolution. The
host forwards either on one pinned Grace core (HSB's
\texttt{cpu\_roce\_loopback}) or on the GPU through DOCA GPUNetIO
(\texttt{gpu\_roce\_loopback}). Packet logic analysers on both sides of the
MAC split the round trip into its FPGA and off-FPGA parts
(\cref{tab:nvqlink:rt}). Over full captures of \num{16384} round trips, the
CPU path ranges from \SIrange{2.54}{7.19}{\micro\second} and the GPU path
from \SI{6.13}{\micro\second} to about \SI{650}{\micro\second}\measbd{}.

\begin{table}[htbp]
  \caption{FPGA-to-host-to-FPGA round trip over NVQLink at \SI{10}{GbE},
  128-byte messages, ZCU216 to DGX Spark, about \num{8200} joined round trips
  per path at \SI{20.7}{\micro\second} spacing\measbd{}. Values in
  \si{\micro\second}.}
  \label{tab:nvqlink:rt}
  \centering
  \small
  \begin{tabular}{@{}L{7.0cm}rrrr@{}}
    \toprule
    & \multicolumn{2}{c}{CPU forwarding} & \multicolumn{2}{c}{GPU forwarding} \\
    Stage & median & p99 & median & p99 \\
    \midrule
    FPGA transmit: timestamp to MAC frame start & 0.24 & 0.25 & 0.24 & 0.25 \\
    Between MAC-side measurement points: transceivers, cable, NIC, host & 2.21 & 2.51 & 5.87 & 244 \\
    FPGA receive: MAC frame start to timestamp & 0.28 & 0.29 & 0.28 & 0.29 \\
    \midrule
    Round trip & 2.74 & 3.03 & 6.40 & 245 \\
    \bottomrule
  \end{tabular}
\end{table}

\begin{enumerate}[leftmargin=1.2em,itemsep=2pt,topsep=2pt]
  \item \textbf{FPGA packet-processing logic contributes \SI{0.52}{\micro\second}.}
        This covers the timestamp-to-MAC and MAC-to-timestamp intervals;
        transceivers are included in the intermediate segment of
        \cref{tab:nvqlink:rt}.
  \item \textbf{CPU forwarding gives lower latency and a tighter tail.}
        The CPU and GPU medians are \SI{2.74}{\micro\second} and
        \SI{6.40}{\micro\second}, respectively, with p99 values of
        \SI{3.03}{\micro\second} and \SI{245}{\micro\second}.
  \item \textbf{The GPU latency tail persists at both tested message rates.}
        Similar tails were observed at message spacings of
        \SI{10}{\micro\second} and \SI{20.7}{\micro\second}.
        These measured distributions favour CPU forwarding for
        latency-sensitive feedback messages on this DGX Spark configuration.
  \item \textbf{The HSB 3.0.0 host software needed fixes for the GPU path.}
        Its GPU RoCE operator was built but not registered for Python, was
        not compiled position-independent, and passed a device function
        where a kernel was expected; NVIDIA's GPU loopback benchmark also
        needed a smaller frame and a default clock to run.
\end{enumerate}

\paragraph{A GraphMP kernel over the link.} A second build places an LP20
decoding kernel behind HSB: GraphMP in a quantised ranking configuration with
a min-sum stage, built as an HLS kernel alongside the configurations of
\cref{app:decoder}. It runs at \SI{250}{MHz} behind asynchronous FIFOs and
width converters, and the build closes at WNS \SI{+0.24}{ns}\meashw{}. For
each shot the host sends the 990-bit syndrome as one 128-byte RDMA SEND, and
the FPGA returns a 4800-byte record, holding a score for each of the 2244 bit
nodes, the min-sum hard decision and a convergence flag, as one RDMA WRITE of
four packets. BP+OSD post-processing stays on the host. The host client
drives HSB's CPU transceiver from one pinned Grace core without a polling
thread. Every result was compared byte for byte with the PyTorch golden
record: all \num{2100} test shots, five passes at each of five pipeline
depths, \num{52500} results, all bit-exact (\cref{tab:nvqlink:ranker}).

\begin{table}[htbp]
  \caption{GraphMP ranking kernel (LP20) over NVQLink at \SI{10}{GbE}:
  latency from host submission to receipt of the FPGA result, excluding
  host-side BP+OSD post-processing. CPU forwarding; \num{10500} results
  per row (five passes over \num{2100} test shots), all bit-exact\measbd{}.}
  \label{tab:nvqlink:ranker}
  \centering
  \small
  \begin{tabular}{@{}rrrr@{}}
    \toprule
    Shots in flight & Median (\si{\micro\second}) & p99 (\si{\micro\second}) & Throughput (shots/s) \\
    \midrule
    1 & 50.5 & 69.7 & \num{19220} \\
    2 & 87.5 & 126.1 & \num{22176} \\
    4 & 175.6 & 252.5 & \num{22176} \\
    8 & 351.8 & 505.4 & \num{22176} \\
    16 & 704.3 & 1011.2 & \num{22175} \\
    \bottomrule
  \end{tabular}
\end{table}

Single-shot latency is the kernel's compute, \SIrange{44}{65}{\micro\second}
depending on the number of min-sum iterations; the link adds about
\SIrange{4}{5}{\micro\second}, of which \SI{3.9}{\micro\second} is
serialising the 4800-byte result at \SI{10}{Gb/s}. Throughput saturates at
about \num{22200} shots/s, the kernel's compute bound of about
\SI{45}{\micro\second} per shot; the results use about \SI{106}{MB/s} of the
link's \SI{1.2}{GB/s}, so deeper pipelines add only queueing delay. The same
kernel fed through the board's processing-system Ethernet over TCP took about
\SI{1.5}{ms} per shot and reached \num{8700} shots/s. For shots that terminate
at the min-sum stage, returning only the hard decision and status would
avoid transferring the score vector.

\paragraph{Not yet measured.} The link test does not carry the traffic of
\cref{tab:if:gpu:traffic}: crops, labels and weights need the
\texttt{crop\_tap} and \texttt{weight\_loader} blocks of
\cref{tab:if:blocks2}, and the pipeline has not been joined to the link. The
decoding kernel's results land in host memory rather than GPU memory, and
\SI{25}{GbE} has not been built.

\subsection{Clock domains and transceiver budget}\label{app:if:clock}

The implemented pipeline uses a single \SI{200}{MHz} clock domain. The
proposed external interfaces use separate clocks and communicate with the
processing domain through asynchronous FIFOs
(\cref{tab:if:clocks}). These crossings isolate link timing from the
processing clock. Their integration and the timing of the complete design
require verification after the interface blocks are added.

\begin{table}[htbp]
  \caption{Clock domains. In the pipeline only the \SI{200}{MHz} domain is
  implemented; the Ethernet domain is implemented in the separate NVQLink
  link test (\cref{app:if:nvqlink}).}
  \label{tab:if:clocks}
  \centering
  \small
  \begin{tabular}{@{}lL{3.2cm}rL{4.4cm}l@{}}
    \toprule
    Domain & Source & Frequency & Crossing into processing & Status \\
    \midrule
    \texttt{clk} & board oscillator / MMCM & \SI{200}{MHz} & --- & \stImpl \\
    \texttt{cxp\_rx} & GTY recovered clock, one lane, others deskewed to it & \SI{156.25}{MHz} & async FIFO, pixel stream, $\ge 12$-line depth upstream & \stSpec \\
    \texttt{pcie\_user} & PCIE4 block & \SI{250}{MHz} & async FIFO, egress record; crops out / weights in when the GPU shares the server & \stSpec \\
    \texttt{eth\_usr} & Ethernet PCS TX clock, \SI{10}{GbE}; built in the link test of \cref{app:if:nvqlink} & \SI{156.25}{MHz} & async FIFOs to the 250 MHz ranking kernel (link-test configuration) & \stImpl \\
    \texttt{axil} & PS & \SI{100}{MHz} & register slices, control plane only & \stSpec \\
    \bottomrule
  \end{tabular}
\end{table}

\Cref{tab:if:gty} budgets the transceivers against what the board exposes.
The \texttt{xczu49dr} has 16 GTY lanes in four quads, but the ZCU216 routes
twelve of them to connectors: eight to its single FMC+ slot (quads 130 and
131) and four to its SFP28 cages (two each on quads 128 and
129)~\cite{amd_ug1390_zcu216}. The board has no PCIe edge connector, so the
PCIe link also leaves through the FMC+ slot. With the GPU in the server the
design needs eight lanes, CoaXPress x4 and PCIe x4, which the FMC+ slot
provides; the NVQLink variant adds one to four lanes on the SFP28 cages. The egress
payload allows PCIe x4 by a wide margin.

\begin{table}[htbp]
  \caption{GTY budget, \texttt{xczu49dr} on the ZCU216 (16 GTY lanes in 4
  quads; board routing from the user guide~\cite{amd_ug1390_zcu216}).}
  \label{tab:if:gty}
  \centering
  \small
  \begin{tabular}{@{}L{3.4cm}rL{2.6cm}L{6.6cm}@{}}
    \toprule
    Link & Lanes & Board route & Note \\
    \midrule
    CoaXPress, 4$\times$CXP-6 & 4 & FMC+ DP0--3 & RX only at line rate; uplink via FMC front-end \\
    PCIe Gen3 x4 to the server & 4 & FMC+ DP4--7 & cable adapter to a server slot; x8 would take all eight FMC+ lanes \\
    10 or \SI{25}{GbE} to ConnectX (variant) & 1--4 & SFP28 cages & one independent link per cage, \num{10.3125} or \SI{25.78}{Gb/s}; CAUI-4 cannot span quads 128 and 129; one \SI{10}{GbE} link measured (\cref{app:if:nvqlink}) \\
    Not routed & 4 & quads 128/129, ch2--3 & one to the clock device, one to a test connector, two unconnected \\
    \midrule
    Total & 16 & & \\
    \bottomrule
  \end{tabular}
\end{table}

\subsection{Per-cycle handshake}\label{app:if:sequence}

The proposed assembly cycle uses the OPX1000 as its timing reference. Time zero
is the imaging trigger, and the following steps coordinate acquisition,
processing, and transport.
\begin{enumerate}[leftmargin=1.6em,itemsep=2pt,topsep=3pt]
  \item \textbf{Arm.} The OPX1000 raises the cycle-ID line to the FPGA and
        fires the camera's external trigger together with the imaging light.
  \item \textbf{Expose and read out.} The camera exposes (\SI{10}{ms} at the
        matched-filter operating point, \cref{tab:latency}) and streams the ROI over
        CoaXPress as it reads out (\cref{tab:if:cxp:budget}).
  \item \textbf{Detect.} The ROI extractor emits crops as soon as each
        site-row's 8 lines have arrived; the detector classifies them;
        the row assembler builds the occupancy of \cref{app:frontend:contract}.
        In parallel, crops are copied to the outer-loop transmit FIFO for data
        collection; that copy is lossy by design and never back-pressures
        detection.
  \item \textbf{Plan.} After the last row, the rearrangement engine reads the
        occupancy and emits the move stream.
  \item \textbf{Hand off.} The DMA engine writes header, occupancy and
        moves into server memory, then the doorbell, and raises the
        ``schedule ready'' line. The OPNIC SDK stream carries the record to
        the OPNIC, which forwards it to the OPX1000 over NVQLink.
  \item \textbf{Move.} The QUA program receives the record, checks cycle ID
        and CRC, and plays the moves group by group. If the cycle ID is stale
        or the CRC fails, it skips the move and re-images.
  \item \textbf{Refit (asynchronous).} The GPU accumulates crops and labels,
        refits the QDM, and at some later cycle writes new weights over PCIe
        into the FPGA's shadow memories and sets a swap flag honoured at the
        next cycle boundary.
\end{enumerate}
Steps 1--6 are the fast loop and involve the camera, the FPGA, server
memory, the vendor's OPNIC stream and the OPX1000; step 7 is the
slow loop, and nothing in steps 1--6 waits on it.

\paragraph{Watchdog.} If step 5 has not completed by a deadline set from
\cref{tab:latency}, the FPGA writes an empty schedule with an error code, so
the OPX1000 is never left waiting on a lost frame. A missed CXP packet
(CRC error) poisons the frame, and the same path reports it.

\subsection{Bring-up order}\label{app:if:bringup}

The proposed bring-up sequence validates each interface and measures its
contribution to the latency budget in \cref{tab:latency}.
\begin{enumerate}[leftmargin=1.6em,itemsep=2pt,topsep=3pt]
  \item \textbf{CXP interface.} Stream the actual ORCA-Quest~2 into a
        PL capture buffer through a commercial CXP-host FMC and licensed CXP
        IP. This step validates the analogue front-end and measures
        trigger-to-first-pixel and readout time in the chosen scan mode.
  \item \textbf{Pixel path into the pipeline.} Connect the stream to the ROI
        extractor through the CDC FIFO and check bit-exactness against the
        simulator crops for recorded frames. Yields the ``CXP ingest $+$ ROI
        extraction'' row.
  \item \textbf{PCIe into server memory.} Bring up the PCIE4 endpoint and
        the DMA engine, through the FMC+ cable adapter, against a pinned
        buffer in host memory. This verifies PCIe enumeration and DMA for path A.
  \item \textbf{OPNIC stream.} Define the stream with the OPNIC SDK and
        close the loop with a QUA program that echoes the doorbell on a
        digital output. Yields the ``egress to controller'' row as the
        FPGA-write-to-OPX-edge time, measured on a scope, including the
        stream hand-off of \cref{app:open}.
  \item \textbf{GPU path.} Retarget the data-collection DMA to a GPU buffer
        through GPUDirect RDMA in the same server, first for data collection,
        then weight swap. For a GPU in another chassis, bring up the Ethernet
        link and RoCE engine against a ConnectX NIC instead; this has been
        done at \SI{10}{GbE} with test traffic (\cref{app:if:nvqlink}).
\end{enumerate}
Steps 1--4 close the fast loop; step 5 is needed only for
the outer loop and can proceed in parallel after step 2.

\subsection{Required FPGA-side additions}\label{app:if:additions}

The place-and-routed design of \cref{tab:impl} ends at two internal
boundaries: it takes $8\times8$ crops on an input port and delivers
occupancy and moves on an output port (\cref{app:frontend:contract}).
Everything between those ports and the three cables is new. This subsection
lists it: the new blocks (\cref{tab:if:blocks,tab:if:blocks2}), the changes it forces on
existing blocks, the resource cost (\cref{tab:if:resources}), and the board,
software, constraint and verification work that comes with it. Each new
block is one of three kinds: \emph{licensed} IP from a third party,
\emph{AMD} IP wrapping a hard block or shipped with Vivado, or \emph{RTL} to
be written in this repository in the style of \cref{tab:rtl:modules}.

\begin{table}[htbp]
  \caption{New FPGA blocks, by link: ingress and egress. Names in
  \texttt{typewriter} are proposed module names for blocks to be written.
  Domain as in \cref{tab:if:clocks}. All \stSpec{} except where marked
  \stOpen.}
  \label{tab:if:blocks}
  \centering
  \small
  \begin{tabular}{@{}L{3.7cm}lL{6.6cm}l@{}}
    \toprule
    Block & Kind & Role & Domain \\
    \midrule
    \multicolumn{4}{@{}l}{\emph{Ingress (camera)}} \\
    GTY transceiver wizard, 4 lanes & AMD & \SI{6.25}{Gb/s} RX, 8b/10b, comma align, 32-bit user width & \texttt{cxp\_rx} \\
    CXP host core & licensed & deframe, CRC, lane bonding, stream reassembly, uplink encoder, AXI4-Lite control port & \texttt{cxp\_rx} \\
    \texttt{async\_fifo} & RTL & Gray-code dual-clock FIFO, AXI4-Stream wrapper; the library \texttt{FIFO} is single-clock & any pair \\
    \texttt{roi\_line\_buffer} & RTL & 8-line circular buffer, $8\times2048\times16$ b in BRAM; emits an $8$-row column slice per pixel clock & \texttt{clk} \\
    \texttt{roi\_crop\_gen} & RTL & site-coordinate table ($x_0, y_0$ per site row and column, loaded from PS) $\to$ $8\times8$ crops into the detector input port, in site order & \texttt{clk} \\
    \texttt{frame\_tap} $+$ AXI DMA & RTL $+$ AMD & optional raw-frame copy to PL DDR4 for data collection and debug; never back-pressures the pixel path & \texttt{clk} \\
    \texttt{cxp\_frame\_monitor} & RTL & frame/line counters, CRC-error and lane-loss flags, frame-poison signal to the watchdog & \texttt{clk} \\
    \midrule
    \multicolumn{4}{@{}l}{\emph{Egress (OPX1000)}} \\
    PCIE4 block $+$ DMA/Bridge Subsystem (XDMA or QDMA) & AMD & Gen3 x4 endpoint through the FMC+ adapter; AXI4 slave for host access to the CSRs and shadow weight memories, AXI4 master for FPGA-initiated DMA into server memory & \texttt{pcie\_user} \\
    \texttt{egress\_packer} & RTL & occupancy rows and move stream $\to$ 32-bit words of \cref{tab:if:opx:payload}; running CRC-32; move count and group tags & \texttt{clk} \\
    \texttt{egress\_dma} & RTL & words $\to$ AXI4 burst writes into the record buffer at its host-programmed bus address, in host or GPU memory; moves streamed as produced, header written after the last move, doorbell last & \texttt{pcie\_user} \\
    \texttt{cycle\_sync} & RTL & two-flop synchronisers on the OPX1000 digital lines; latches cycle ID at arm; drives ``schedule ready'' & \texttt{clk} \\
    \texttt{loop\_watchdog} & RTL & per-cycle deadline counter; on timeout or frame poison, forces an empty record with an error code & \texttt{clk} \\
    \bottomrule
  \end{tabular}
\end{table}

\begin{table}[htbp]
  \caption{New FPGA blocks, continued: outer loop and common blocks.
  Conventions as in \cref{tab:if:blocks}.}
  \label{tab:if:blocks2}
  \centering
  \small
  \begin{tabular}{@{}L{3.7cm}lL{6.6cm}l@{}}
    \toprule
    Block & Kind & Role & Domain \\
    \midrule
    \multicolumn{4}{@{}l}{\emph{Outer loop (GPU)} \quad \stOpen} \\
    Second DMA channel of \texttt{egress\_dma} & RTL & crops and decisions into a GPU buffer on the same server (GPUDirect RDMA); shares the PCIE4 block with the egress & \texttt{pcie\_user} \\
    Ethernet PCS/PMA $+$ MAC & AMD $+$ open source & NVQLink variant: \texttt{xxv\_ethernet} in its licence-free PCS/PMA mode and the \texttt{verilog-ethernet} 10G MAC, 64-bit AXI4-Stream; built and measured at \SI{10}{GbE} (\cref{app:if:nvqlink}) & \texttt{eth\_usr} \\
    RoCEv2 engine & licensed or AMD & NVQLink variant: UDP/IP, queue pairs, RDMA WRITE initiator and target: Holoscan Sensor Bridge IP~\cite{nvidia_holoscan_hsb}, built and measured (\cref{app:if:nvqlink}), or AMD ERNIC & \texttt{eth\_usr} \\
    \texttt{crop\_tap} & RTL & lossy tee of detector-input crops and decisions into the TX FIFO, with a drop counter & \texttt{clk} \\
    \texttt{weight\_loader} & RTL & PCIe or RDMA target buffer $\to$ shadow weight memories; checksum; swap request at a cycle boundary & \texttt{clk} \\
    \midrule
    \multicolumn{4}{@{}l}{\emph{Common}} \\
    Clocking and reset & AMD $+$ RTL & MMCM for \texttt{clk}; GTY reference clocks; per-domain reset synchronisers and ordered release after link-up & all \\
    \texttt{qdpu\_csr} & RTL & AXI4-Lite register file: configuration, site table, record buffer address, deadlines, status and error counters, timestamp counter & \texttt{axil} \\
    Integrated logic analysers & AMD & bring-up only; removed from the timing-closure build & per link \\
    \bottomrule
  \end{tabular}
\end{table}

\paragraph{Changes to existing blocks.} Three of the implemented blocks
change.
\begin{itemize}[leftmargin=1.2em,itemsep=2pt,topsep=3pt]
  \item \textbf{CNN detector: runtime weights.} The hls4ml flow of
        \cref{app:quant:toolflow} compiles the detector's weights into the
        bitstream as constants, which the synthesis tool folds into LUTs and
        DSP coefficients. A refit from the GPU then requires a new bitstream,
        and \cref{sec:unit:interfaces}'s ``parameters in'' is not a runtime
        operation. Runtime weights mean re-generating the detector with its
        weights in writable memories (double-buffered, so the swap is
        atomic at a cycle boundary), which removes constant folding. That
        costs LUTs and possibly DSPs, and it changes the per-crop latency
        of \cref{app:impl:rf}, so it has to go back through
        place-and-route. Until then, the outer loop can collect data and
        deliver a refit as a partial or full reconfiguration between runs.
  \item \textbf{Detector input: from a port to a stream.} The system top
        (\texttt{atom\_rearrangement\_cnn\_top}) currently reads crops from a
        testbench-driven port. It gains back-pressure from the detector
        cores to \texttt{roi\_crop\_gen}, and \texttt{roi\_line\_buffer}
        must never overflow while the cores stall. A stall-free design sizes
        the line buffer for the worst case: the time to classify one site
        row against the time the camera takes to deliver the next 8 lines.
        The matched-filter lane classifies a row of 50 sites in 59 cycles,
        \SI{0.3}{\micro\second}, and keeps pace with the camera. One CNN
        core takes $50\times1797/\SI{200}{MHz} = \SI{0.45}{ms}$ per row,
        while 8 lines of the ultra-quiet scan arrive in about
        \SI{0.14}{ms}, so the CNN falls behind the camera and either the
        buffer grows to most of a frame or the detector is allowed to
        finish after readout\roof{}.
  \item \textbf{Engine output: record framing.} The move FIFO of
        \texttt{tetris} feeds \texttt{egress\_packer} instead of a testbench
        sink. The engine itself is unchanged; the end of the move
        stream must be signalled so the packer knows when to write the
        header and doorbell. \texttt{tetris} already exposes
        \texttt{done\_o} and \texttt{abandoned\_o}: the first triggers the
        header and doorbell, and the second is written into the header as
        an error code, so the QUA program can tell an infeasible target
        from a lost frame.
\end{itemize}

\paragraph{Resource cost.} \Cref{tab:if:resources} gives order-of-magnitude
planning estimates for the proposed interfaces. Actual resources depend on
the configured IP and optional blocks and require implementation checks.
The listed interface blocks are budgeted without DSP slices, but their LUT
and BRAM requirements still affect feasibility.

One matched-filter lane and the decoder occupy about \SI{61.0}{\percent}
of LUTs and \SI{37.4}{\percent} of BRAM before interfaces
(\cref{tab:coresident}). Adding the table's \SIrange{14}{29}{\percent} LUT
range leaves component-level margin for that configuration. Four CNN cores
and the decoder already use about \SI{84.2}{\percent} LUT; the same range
would bring the total to approximately \SIrange{98}{113}{\percent}.
The complete four-core configuration is therefore not established as
feasible. Transceiver routing is a further constraint
(\cref{tab:if:gty}), and joint place-and-route is required for each topology.

\begin{table}[htbp]
  \caption{Estimated resource cost of the additions on \texttt{xczu49dr}
  (\num{425280} LUT, \num{2160} BRAM\_18K and \num{4272} DSP
  \cite{amd_ug1390_zcu216})\roof{}. Ranges are planning estimates, not
  measurements. Totals include optional frame capture and the remote-GPU
  Ethernet variant; the selected topology determines which blocks are present.}
  \label{tab:if:resources}
  \centering
  \small
  \begin{tabular}{@{}lrrrl@{}}
    \toprule
    Block group & LUT & BRAM\_18K & DSP & Hard blocks \\
    \midrule
    CXP host core $+$ GTY wrapper & 10--20k & 10--20 & 0 & 4 GTY \\
    ROI line buffer $+$ crop generator & 2--5k & 22--30 & 0 & --- \\
    Frame tap $+$ AXI DMA $+$ PL DDR4 controller & 10--15k & 10--20 & 0 & PL DDR4 I/O \\
    PCIe DMA/bridge $+$ write engine $+$ packer & 15--30k & 20--40 & 0 & PCIE4, 4 GTY \\
    Ethernet PCS/MAC $+$ RoCEv2 engine (variant) & 20--50k & 40--100 & 0 & 1--4 GTY \\
    Crop tap, weight loader, CSR, sync, watchdog & 2--5k & 5--20 & 0 & --- \\
    \midrule
    Total, additions & 60--125k & 110--230 & 0 & \\
    \phantom{xx}share of device & 14--29\% & 5--11\% & 0\% & \\
    Matched-filter pipeline (\cref{tab:impl}) & 87\,844 & 2 & 64 & \\
    CNN $8\times8$ pipeline (\cref{tab:impl}) & 98\,798 & 107 & 584 & \\
    \bottomrule
  \end{tabular}
\end{table}

\paragraph{Board-level hardware.} Two additions are physical, not logic.
\begin{itemize}[leftmargin=1.2em,itemsep=2pt,topsep=3pt]
  \item \textbf{CoaXPress FMC.} Four CXP equaliser/driver channels, uplink
        injection, DIN~1.0/2.3 or micro-BNC connectors matching the camera,
        and PoCXP if the camera draws power over the coax. Its FMC lanes
        must route to one GTY quad. A commercial CXP-host FMC is the first
        choice for bring-up (\cref{app:if:bringup}); a card that shares the
        slot with the PCIe adapter (below) is the deployment step.
  \item \textbf{PCIe path into the server.} The ZCU216 is a bench board
          with no PCIe edge connector~\cite{amd_ug1390_zcu216}; it reaches the
          server's PCIe fabric through a cable adapter in any slot of the
          server that hosts the OPNIC, CPU and GPU
          (\cref{app:platform:controller}). The board has one FMC+ slot, so the
          CoaXPress front-end and the PCIe link share it on a single carrier
          with four GTY lanes each: four CXP-6 receivers 
          (\cref{app:platform:camera}) and a PCIe Gen3~x4 link to the server. Four
          lanes are ample for egress (\SI{\approx 31}{Gb/s} against a move
          list of tens of kilobits per shot), and the camera keeps its native
          coax link with no media converter in the latency path. Moving the
          camera onto the SFP28 cages over CoaXPress-over-Fibre is the
          fallback if no such carrier is available; it frees the FMC+ for
          PCIe but adds a coax-to-fibre converter and a different host IP
          core.
\end{itemize}
The SFP28 cages serve the NVQLink variant of the outer loop, and the OPX1000
digital lines need only level-compatible I/O on a board header.

\paragraph{Processor-side software.}
Detection and rearrangement scheduling execute in FPGA logic. Configuration
and run control use the RFSoC's Arm cores, running bare-metal software or Linux:
\begin{itemize}[leftmargin=1.2em,itemsep=2pt,topsep=3pt]
  \item GenICam camera configuration through the CXP core: ROI, scan mode,
        exposure, trigger source, pixel format;
  \item site-coordinate table for \texttt{roi\_crop\_gen}, written after each
        optical calibration;
  \item for the NVQLink variant, RoCE queue-pair setup and memory-key
        exchange with the GPU host, over the PS Ethernet port;
  \item run control through \texttt{qdpu\_csr}: arm, deadlines, error
        counters, and a status readout for the experiment's sequencer.
\end{itemize}
Three host-side pieces run on the server: a start-up utility that enumerates
the FPGA, allocates and pins the record buffer in host or GPU memory and
programs its bus address into \texttt{qdpu\_csr}; the vendor's OPNIC SDK
stream, on the CPU or GPU, that carries each record to the OPNIC
(\cref{app:platform:controller}); and, on the GPU, the receive buffers, the
refit job and the weight push.

\paragraph{Constraints and timing closure.}
The crossings in \cref{tab:if:clocks} use asynchronous FIFOs or synchronisers,
with constraints for the asynchronous clock groups and crossing paths. The
Ethernet user domain, \SI{156.25}{MHz} at \SI{10}{GbE} and
\SI{390.625}{MHz} at \SI{25}{GbE}, sets the timing target for the RoCE engine
and its FIFOs; the \SI{10}{GbE} link test closes at WNS \SI{+1.69}{ns}
(\cref{app:if:nvqlink}). Transceiver, PCIe, and Ethernet locations follow
the board routing (\cref{tab:if:gty}). The complete design requires
in-context implementation because the results in \cref{tab:impl} exclude
board-level I/O and clock-tree effects.

\paragraph{Verification.} Each link gets a simulation model in the existing
Verilator and cocotb framework before any hardware:
\begin{itemize}[leftmargin=1.2em,itemsep=2pt,topsep=3pt]
  \item a CXP stream model that replays simulator frames as the CXP core's
        AXI4-Stream output, so \texttt{roi\_line\_buffer} and
        \texttt{roi\_crop\_gen} are checked crop-for-crop against the
        simulator's own crops;
  \item an AXI4 slave model of the record buffer, checking record layout,
        CRC, write ordering and the doorbell-last rule, including watchdog
        and poisoned-frame cases;
  \item a loopback model of the RoCE engine for \texttt{crop\_tap} and
        \texttt{weight\_loader}, checking that a weight swap takes effect
        only at a cycle boundary;
  \item formal proofs for \texttt{async\_fifo} alongside the library's
        existing FIFO proofs.
\end{itemize}
The vendor IP (CXP core, PCIe, Ethernet MAC, RoCE) is not simulated at the protocol
level here; it is replaced by these models in simulation and exercised on
hardware in the bring-up order of \cref{app:if:bringup}.

\subsection{Scaling to \num{10000} sites}\label{app:platform:scaling}

Throughout this subsection $n$ is the number of sites and $N$ the number of
sites along one side of the square array. Thus $n = N^2$, as in
\cref{app:frontend,app:detector:law}: the implemented design is $N = 50$, $n = 2500$, and the next array generation is
taken as $n = 10^4$ (\cref{app:detector:law}). Three quantities scale
differently with $n$, and only one of them is a property of the pipeline as
built.

\begin{itemize}[leftmargin=1.2em,itemsep=4pt,topsep=3pt]\sloppy
  \item \textbf{Detection time scales linearly in $n$ and stays negligible
        with the matched filter.} One lane classifies $n = 10^4$ sites in
        $10^4 + 9$ cycles, \SI{50}{\micro\second} at \SI{200}{MHz}, against
        at least \SI{3.4}{ms} to transfer the $2048\times2048$ ROI payload;
        four fixed-template lanes (244 DSP slices) reduce detection to about
        \SI{12.5}{\micro\second}. The CNN's
        \SI{22.5}{ms} for $n = 2500$ at one core would become
        $\approx \SI{90}{ms}$, more than half the \SI{150}{ms} host
        round-trip whose removal is the largest gain in
        \cref{app:loop:lifetime}, and four cores with the decoder resident
        already take \SI{93}{\percent} of the DSP slices
        (\cref{tab:coresident}). A learned detector at $10^4$ sites therefore
        needs a larger DSP budget, a cheaper network per site, or a shorter
        per-crop latency than the \num{1797} cycles of \cref{app:impl:rf}.

  \item \textbf{Rearrangement operates on row- and column-sized data structures.}
        For $n=N^2$ sites, each row-assignment problem has at most $N$
        sources and $N$ target columns. Its $M\times K$ take-bit matrix
        therefore contains at most $N^2=n$ bits. At $n=10^4$ sites, this
        matrix occupies at most \SI{1.25}{kB}. The sorter ranks $N$ column
        priorities; padding $N=100$ to 128 inputs gives a 28-stage bitonic
        network with \num{1792} compare--exchange pairs. These structural
        counts follow from the row-wise architecture of
        \cref{app:rtl:blocks}\algo{}.

  \item \textbf{The sensor fixes the product of sites across and pixels per
        site pitch; where to sit on that curve is a choice.} At a site pitch of
        $S$ sensor pixels, the $4096\times2304$ sensor images at most $\lfloor
        4096/S \rfloor \times \lfloor 2304/S \rfloor$ sites. The evaluated
        configuration (\cref{tab:simcfg}) samples the atom plane at
        \SI{0.4}{\micro\metre\per px}, so the \SI{4}{\micro\metre} pitch of the
        sparse array is $S = 10$ px and the sensor holds up to $409\times230$
        sites; a $100\times100$ array fits in a $1000\times1000$-pixel region at
        that pitch, and a $2048\times2048$ region holds $204\times204$ sites.
        Denser pitches trade sites across for crosstalk, whose cost
        \cref{tab:xtalk} measures with the $8\times8$ window. This is an instrument trade; the pipeline itself is
        unaffected.
\end{itemize}

\subsection{Waveform synthesis, on the controller side}\label{app:platform:aod}

Moves leave the QDPU as (from, to) site pairs. Their conversion to AOD drive,
multi-tone chirped DDS, is the controller's task and lies outside the QDPU. It
enters this specification as a constraint on the egress: the tone count the
controller can sustain bounds the number of atoms moved concurrently, and
therefore the parallelism the rearrangement schedule may assume.

\section{Open Items}\label{app:open}

The following are open at the time of writing. They are listed so that the
provenance tags of the appendices can be read against them.

\begin{enumerate}[leftmargin=1.2em,itemsep=2pt,topsep=2pt]
  \item \textbf{Hardware status.} The matched-filter pipeline closes
        \SI{200}{MHz} as a ZCU216 bitstream with an on-chip self-test, with
        recovery and removal met for its synchronous-release reset
        (\cref{app:impl:board}). The CNN pipeline is place-and-routed out of context only
        (\cref{app:frontend:scope}).
  \item \textbf{Region-of-interest line buffer.} Both $8\times8$ pipelines
        close \SI{200}{MHz} with a registered crop input standing in for it
        (\cref{tab:impl}); build the line buffer that cuts the windows from
        the camera stream and feeds the detector from registers. Decide
        whether the CNN is driven back to back (\num{1165} rather than
        \num{1797} cycles per crop, \cref{app:impl:rf}).
  \item \textbf{Ingest and egress latency.} The sensor-readout and egress rows
        of \cref{tab:latency} are derived from the camera datasheet and the
        link specification and await measurement after integration
        (\cref{app:platform}); the scan mode the imaging simulator assumes
        decides which readout time applies.
  \item \textbf{Measurements not yet made.} The $F_{\max}$ ceiling of the
        pipeline above the \SI{200}{MHz} constraint (\cref{app:impl:timing});
        per-lane pipelining of the matched filter's configuration port, which
        limits loadable lanes at $P \ge 20$ (\cref{tab:mfpnr}); csynth LUT/FF
        for the $8\times8$ CNN; a power
        estimate from simulated switching activity, and power estimates for
        the CNN and CNN-plus-decoder configurations. The vectorless estimate
        for the matched-filter pipeline is \SI{2.39}{W}
        (\cref{app:impl:power}).
  \item \textbf{Host round trip.} The \SI{150}{ms} host-in-loop round trip of
        \cref{tab:params} is an assumption. With the detector charged its
        inference time there is no break-even round trip, and the conclusion
        holds for any host round trip above a few milliseconds
        (\cref{app:loop:lifetime}); the round trip has not been measured on
        the target apparatus.
  \item \textbf{Simulator-conditional accuracy.} No detection-accuracy figure
        was measured on a real camera frame; fixed-pattern noise is the leading
        sim-to-real risk (\cref{sec:limits}).
  \item \textbf{Interfaces.} The Shuttle and Control APIs of \cref{tab:if_ctrl}
        are undefined in this iteration. The latency of the hand-off from
        server memory through the vendor's OPNIC stream to the controller has
        not been measured; the vendor's planning figure is about
        \SI{1}{\micro\second} for any PCIe transfer inside the OPC, so it does
        not change the millisecond-scale conclusions of \cref{app:loop}
        (\cref{app:platform:controller}).
  \item \textbf{Board route.} The ZCU216 has one FMC+ slot and no PCIe edge
        connector, so the camera link and the PCIe link share one FMC+ carrier
        with four GTY lanes each; that carrier has yet to be sourced or built,
        with CoaXPress-over-Fibre into the SFP28 cages as the fallback
        (\cref{app:if:additions}), which would then share the cages with
        NVQLink.
  \item \textbf{NVQLink.} The outer-loop transport is measured at
        \SI{10}{GbE} with test traffic and a GraphMP ranking kernel
        (\cref{app:if:nvqlink}). Open: \SI{25}{GbE}; the GPU path's
        tail latency on the DGX Spark and the proxy-free path of a discrete
        GPU; and the crop, label and weight traffic of
        \cref{tab:if:gpu:traffic}.
\end{enumerate}

\section{Glossary of Special Terms and Acronyms}\label{app:glossary}

\begin{itemize}
    \item AIE = AI Engines.
    \item AMD = FPGA manufacturer.
    \item AOD = Acousto-Optic Deflectors.
    \item ap\_fixed<W,I> = "ap\_fixed with $W$ bitwidth and $I$ integer bits" fixed-point datatype. 
    \item ASIC = Application-Specific Integrated Circuit.
    \item AXI-stream = FPGA on-chip communication protocol.
    \item BP = Belief Propagation.
    \item BRAM = "Block RAM" FPGA resource.
    \item BRAM\_SDP = Simple dual-port block RAM: one write port and one read port, a library cell of the rearrangement engine.
    \item CRC = Cyclic Redundancy Check, an error-detecting checksum (CRC-32 on the CoaXPress link and on each move list sent to the controller).
    \item CSR = Compressed Sparse Row, a storage format for sparse graphs (the decoder's adjacency); also Control and Status Register, a host-accessible FPGA register.
    \item clk = Clock.
    \item CPU = Central Processing Unit.
    \item CMAC = AMD's integrated 100\,Gb/s Ethernet MAC/PCS block.
    \item CNN = Convolutional Neural Network.
    \item CoaXPress = A high-speed interface standard for transmitting image data and control signals between a camera and a host.
    \item CXP-6 = A CoaXPress rate class with a line rate of \SI{6.25}{Gb/s} per lane.
    \item DDR = "Double Data Rate" is technology used in modern memory chips with SDRAM.
    \item DGX Spark = NVIDIA desktop system with a GB10 chip (Grace CPU and integrated GPU) and a ConnectX-7 network card.
    \item DMA = Direct Memory Access.
    \item DOCA GPUNetIO = NVIDIA library that lets GPU code send and receive network packets directly.
    \item DSP = Digital signal processing; in the FPGA resource tables, a DSP slice is a dedicated arithmetic block supporting operations such as multiplication and accumulation.
    \item FF = "Flip-flop" FPGA resource.
    \item FIFO = "First in, first out" FPGA concept for memory handling.
    \item FMC = "FPGA Mezzanine Card" provides fixed I/O for FPGA board.
    \item FPGA = Field-Programmable Gate Array.
    \item fps = Frames per Second.
    \item GAN = Generative Adversarial Network.
    \item GbE = 'Gigabit Ethernet" port.
    \item GEMM =  General Matrix Multiplication.
    \item GPU = Graphics Processing Unit.
    \item GraphMP = The learned message-passing decoder introduced in the companion paper~\cite{companion2026graphmp}.
    \item GTY = A family of AMD high-speed serial transceivers.
    \item HLS = High Level Synthesis.
    \item HPC = High Performance Computing.
    \item HSB = Holoscan Sensor Bridge, NVIDIA's open-source FPGA IP and host software that move sensor data into CPU or GPU memory over RoCE; the FPGA side of NVQLink's reference transport.
    \item II = Initiation Interval, the number of clock cycles between the starts of successive loop iterations or inputs in a pipelined design.
    \item IP = Intellectual Property.
    \item I/O = Input/output.
    \item LP$[[1122,148,\leq 20]]$ = A lifted-product code with 1122 physical qubits, 148 logical qubits, and code distance at most 20; the distance is not the number of correctable errors.
    \item LSB = Least Significant Bit.
    \item LUT = "Look-up Table" FPGA resource.
    \item LUTRAM = LUT as RAM.
    \item MAC = Multiply--accumulate in arithmetic and resource counts; Media Access Control in Ethernet descriptions.
    \item MLP = Multi-Layer Perceptron.
    \item NAND = "NOT AND" boolean logic gate.
    \item NIC = "Network Interface Card" converts data to electrical signal and transfers to classical computer chips.
    \item NVQLink = NVIDIA's interconnect platform for integrating quantum systems with GPU computing.
    \item OPC = OP-Central, \emph{Quantum Machines'} validated host server for the OPX1000.
    \item OPNIC = Quantum Machines' PCIe interface card connecting the host server to the OPX1000 controller.
    \item OPX1000 = \emph{Quantum Machines'} controller device for large scale quantum computers.
    \item ORCA-Quest 2 qCMOS = Camera device used to image atoms.
    \item OSD = Ordered Statistics Decoding.
    \item PCIe = Peripheral Component Interconnect Express.
    \item PL = Programmable Logic.
    \item PLIO = Programmable Logic I/O, the AXI4-Stream interface between the programmable logic and the AI Engine array on Versal devices.
    \item PPVM = "Programmable Pauli Virtual Machine" simulation framework.
    \item PTP = Precision Time Protocol, used to lock the FPGA clock to the host's.
    \item QAT = Quantization-aware training.
    \item QEC = Quantum Error Correction.
    \item QDPU = Quantum Data Processing Unit.
    \item QDM = Quantum Data Model.
    \item qLDPC = Quantum Low Density Parity Check
    \item QDMA = Queue-based Direct Memory Access.
    \item QPU = Quantum Processing Unit.
    \item QSFP = Quad Small Form-factor Pluggable.
    \item QUA = Pulse-level-control programming language for quantum devices.
    \item RAM = Random Access Memory.
    \item RDMA = Remote DMA.
    \item RF = "Reuse Factor" knob to be set in the HLS code (not to be confused with "RF" in RFSoC)
    \item rms = Root Mean Square.
    \item RoCE = RDMA over Converged Ethernet; RoCEv2 carries it over UDP/IP.
    \item ROI = Region of Interest.
    \item ROM = Read-Only Memory.
    \item RTL = "Register Transfer Level" hardware design.
    \item SDK = Software Development Kit.
    \item SDRAM = "Synchronous Dynamic Random-Access Memory" concept used in modern chips.
    \item SFP28 = Small Form-factor Pluggable cage for one \SI{25}{Gb/s} lane.
    \item $t_{exp}$ = Exposure time: how long the camera collects fluorescence for one image.
    \item $t_{cycle}$ = Duration of one detect--rearrange cycle: exposure, camera readout, detection, rearrangement planning, atom transport and, with a host in the loop, the host round trip.
    \item $\tau_{img}$ = Imaging-loss lifetime: an atom survives one image with probability $\exp(-t_{exp}/\tau_{img})$.
    \item $\tau_{vac}$ = Vacuum-limited trap lifetime: an atom survives a time $t$ in its trap with probability $\exp(-t/\tau_{vac})$.
    \item tetris = The atom rearrangement engine (RTL module \texttt{tetris}), named after the Tetris-style compaction algorithm it implements.
    \item TNS = Total Negative Slack.
    \item TTL = Transistor-Transistor Logic.
    \item Versal AI Edge = AMD adaptive SoC family; the \texttt{xcve2802} is the target device for the PL--AIE decoder implementation in this work.
    \item W12A12 = Weights and activations captured by 12 bits.
    \item WNS = Worst Negative Slack.
    \item Zynq UltraScale+ RFSoC ZCU216 = FPGA board used in this work, part number xczu49dr.
    \item ZZ check = A weight-two parity check that measures the product $Z \otimes Z$ of two data qubits.

\end{itemize}

\bibliographystyle{unsrt}
\bibliography{refs}

\end{document}